\documentclass[12pt]{article}

\usepackage{epsf,epsfig}

\usepackage{a4wide,axodraw}
\def\slash#1{#1 \hskip-0.45em /}
\def\Slash#1{#1 \hskip-0.59em /}

\def\beq{\begin{eqnarray}}
\def\eeq{\end{eqnarray}}

\def\be{\begin{equation}}
\def\ee{\end{equation}}

\def\np{n_+}
\def\nm{n_-}

\def\Wc{W_{c}}

\def\WZdag{W\! Z^\dagger}
\def\ZWdag{Z W^\dagger}

\def\Wus{Z}

\begin{document}
\thispagestyle{empty}

\begin{flushright}
  PITHA 02/09\\
  hep-ph/0206152\\
  June 17, 2002
\end{flushright}

\vspace{\baselineskip}

\begin{center}
\vspace{0.5\baselineskip}
\textbf{\Large Soft-collinear effective theory and heavy-to-light\\[0.5em] 
currents beyond leading power}\\
\vspace{3\baselineskip}
{\sc M. Beneke, A. P. Chapovsky, M. Diehl, Th.~Feldmann}\\
\vspace{2\baselineskip}
\textit{Institut f\"ur Theoretische Physik E, RWTH Aachen\\[0.1cm]
D -  52056 Aachen, Germany} \\
\vspace{3\baselineskip}

\vspace*{1cm}
\textbf{Abstract}\\
\vspace{1\baselineskip}
\parbox{0.9\textwidth}{ An important unresolved question in strong
interaction physics concerns the parameterization of power-suppressed
long-distance effects to hard processes that do not admit an operator
product expansion (OPE). Recently Bauer et al.\ have developed an
effective field theory framework that allows one to formulate the
problem of soft-collinear factorization in terms of fields and
operators. We extend the formulation of soft-collinear effective
theory, previously worked out to leading order, to second order in a
power series in the inverse of the hard scale. 
We give the effective Lagrangian and
the expansion of ``currents'' that produce collinear particles in
heavy quark decay. This is the first step towards a theory of power
corrections to hard processes where the OPE cannot be used. We apply this
framework to heavy-to-light meson transition form factors at large
recoil energy.  }
\end{center}

\newpage
\setcounter{page}{1}


\newpage


\section{Introduction}
\label{sec:intro}

An important unresolved question in strong interaction physics
concerns the parameterization of power-suppressed long-distance
effects to hard processes that do not admit an operator product
expansion (OPE). For a large class of processes of this type the
principal difficulty arises from the presence of collinear modes,
i.e. highly energetic, but nearly massless particles.

In a series of papers
\cite{Bauer:2000ew,Bauer:2000yr,Bauer:2001ct,Bauer:2001yt,Bauer:2002nz}
Bauer et al.\ have developed an effective field theory framework that allows
one for the first time to formulate the problem of soft-collinear
factorization in terms of fields and operators, rather than the
factorization properties of Feynman diagrams. Assuming that the field
content of the effective theory correctly includes all long-distance
modes, the effective field theory formulation simplifies or renders
more transparent the factorization proofs that form the conceptual
basis of high-energy QCD phenomenology
\cite{Collins:1988gx,Brodsky:-240pv}.  The effective theory has been
formulated only to leading order in an expansion in the hard scale up
to now, as have been the diagrammatic factorization methods.

In this paper we extend the formulation of soft-collinear effective
theory to second order in a power series in the inverse of the 
hard scale.  We give
the effective Lagrangian and the expansion of ``currents'' that
produce collinear particles in heavy quark decay. This is the first
step towards a theory of power corrections to hard processes that do
not admit an OPE. Previous approaches to this problem include the
renormalon approach (reviewed in \cite{Beneke:1998ui}), where power
corrections have been identified through the asymptotics of the
perturbative series at leading power. This approach is
severely limited since only those power corrections can be included
that are related through renormalization to operators appearing 
already at leading power.  The effective theory approach is free from
this limitation.

The outline of this paper is as follows: in Section~\ref{sec:fields}
we set up the notation, power counting rules, fields and gauge
transformations of the effective theory. In
Sections~\ref{sec:lagrangian} and \ref{sec:currents} we derive the
effective Lagrangian of soft-collinear effective theory (SCET) and the
effective heavy-to-light 
currents, respectively, to second order in the expansion
parameter of the effective theory. We then discuss in
Section~\ref{sec:htol} power-suppressed effects that break the
symmetries between heavy-to-light meson transition form factors at
large recoil, extending previous work of two of us
\cite{Beneke:2000wa} on perturbative symmetry-breaking corrections.
We conclude in Section~\ref{sec:conclude}. 



\section{Power counting and fields} 
\label{sec:fields}

We begin with detailing the power counting rules and fields from which
the effective theory will be constructed. This discussion is a
repetition of previous work \cite{Bauer:2000yr}, using our preferred
notation, except for two aspects. Bauer et al.\ use a hybrid
momentum-position space representation in their formulation of the
effective theory, where collinear fields are labelled by the large
part of the collinear momentum of the particles they create and
destroy. The corresponding derivatives are replaced by ``label
operators''. We find it more convenient, especially for the extension
to power corrections, to stay within the conventional position space
representation, where the effective field theory can be formulated
in more familiar field theory terminology.  The second modification
concerns the definition of gauge transformations and Wilson lines, and
is necessary to extend the theory beyond leading order. The position 
space representation also requires that one expands ultrasoft fields 
in their position arguments. 
This will be explained in Section~\ref{sec:lagrangian}.
 
\subsection{Kinematical definitions}

\begin{figure}[t]
   \vspace{0.2cm}
   \epsfxsize=12cm
   \centerline{\epsffile{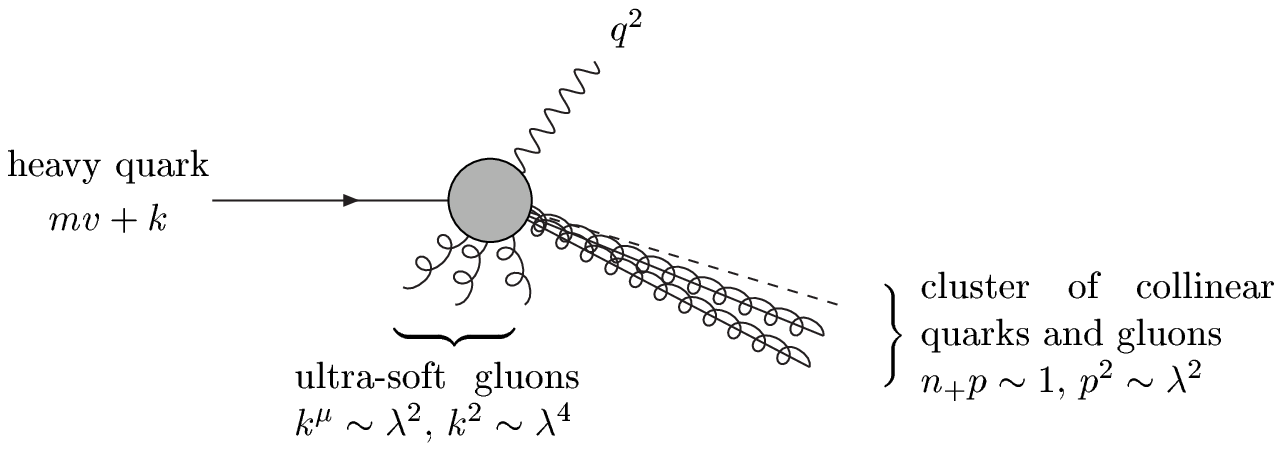}}
   \vspace*{-0.3cm}
\caption[dummy]{\label{setup} Kinematics of heavy quark decay into 
a single cluster of collinear and ultrasoft particles.}
\end{figure}

The effective theory is designed to expand a physical process in
powers of a small ratio of scales, $\lambda$. In perturbative
computations the relevant scales are determined by specifying the
external momenta of certain Green functions. In a real hadronic
process these external momentum configurations are enforced by
hadronic wavefunctions or the selection of specific final states by
the observer.  The situation we are interested in here consists of a
cluster of nearly collinear particles moving with large momentum of
order $m$ into a
direction $\nm$, such that the total invariant mass is of order
$(m\lambda)^2$. This acquires a Lorentz-invariant meaning if there is
a source for collinear particles. We may think for instance of a heavy
quark of mass $m$ at rest, which decays semileptonically into a set of
partons with large energy and small invariant mass, as is relevant to
exclusive decays, see Figure~\ref{setup}. 
More complicated situations often arise in jet
physics, where the final state can consist of several clusters moving
into different well-separated directions. However, in this paper we 
restrict ourselves to the case of a single direction $\nm$.

We decompose the momentum $p$ of a {\em collinear} particle as 
\begin{eqnarray}
  p^\mu = (n_+p) \, \frac{n_-^\mu}{2} + p_\perp^\mu + (n_-p)
  \frac{n_+^\mu}{2},
\end{eqnarray}
where $n_\pm^\mu$ are two light-like vectors, $n_+^2=n_-^2=0$ with
$\np\nm = 2$. The individual momentum components are assumed to have
the following scaling behaviour,
\begin{equation}
n_+ p \sim m, \quad
p_\perp \sim m\lambda, \quad
n_- p \sim m \lambda^2,
\label{mom/scaling}
\end{equation}
such that $p^2\sim m^2 \lambda^2$ as required. In the following we
shall put $m=1$ in scaling relations such as (\ref{mom/scaling}), 
understanding that the dimension of any quantity is
restored by inserting the appropriate power of $m$.

The invariant mass of the final state is not altered if we add to it a
particle with momentum $k$ whose components are all of order
$\lambda^2$. The effective theory therefore also contains such {\em
ultrasoft} particles. The effective theory does {\em not} contain {\em
soft} particles, whose momentum components are all of order $\lambda$,
as external states, since adding such a particle (with momentum $k$) 
to a cluster of
collinear particles with momentum $p$ implies an invariant mass
$(p+k)^2\sim \lambda$ in contradiction with our kinematical
assumption.


\subsection{Fields}
\label{sec/scaling}

We construct an effective field theory by introducing separate collinear 
and ultrasoft fields that destroy and create collinear and 
ultrasoft (anti-)particles, respectively. It follows that an ultrasoft
field $\phi_{\rm us}(x)$ varies significantly only over distances 
of order $x\sim 1/\lambda^2$, so derivatives acting on these fields 
scale as 
\be
\partial_\mu \phi_{\rm us} \sim \lambda^2 \phi_{\rm us}.
\ee
Collinear fields have significant variations over $\nm x\sim 1$, 
and $x_\perp\sim 1/\lambda$. Derivatives acting on collinear fields 
$\phi_{c}$ therefore scale as 
\be
n_+\partial \phi_{c} \sim \phi_{c}, \quad
\partial_\perp \phi_{c} \sim \lambda \phi_{c}, \quad 
n_-\partial \phi_{c} \sim \lambda^2\phi_{c}.
\label{scaling/part}
\ee
The effective theory is constructed to reproduce the Green functions 
of QCD under the kinematical assumption above as an expansion in
$\lambda$. With a large scale $m$ present, quantum fluctuations 
with virtuality $m^2$ also contribute. These {\em hard} modes 
can be integrated out perturbatively. In practice, this amounts 
to Taylor-expanding the loop integrals in $\lambda$, assuming that 
all loop momenta scale as $m$. The effective theory accounts 
for those loop momentum regions where this expansion breaks down, 
i.e. for those configurations where some propagators become on-shell
as $\lambda\to 0$. For the given kinematics this leads to the 
collinear, soft and ultrasoft modes defined above. Since soft 
modes do not appear as external states, they can be integrated 
out.\footnote{This is similar to non-relativistic effective theory 
where soft modes with momentum $k\sim m v$ give rise to 
instantaneous effective interactions in the effective theory of 
potential heavy quarks and ultrasoft gluons and light 
quarks \cite{Beneke:1998zp}.} 
This will be understood in the following, but since (with the
exception of Section~\ref{subsec:norenorm}) this paper 
is concerned with power corrections in $\lambda$ at tree level, 
the details of integrating out soft modes will not be 
important.

In the case of QCD 
we introduce collinear and ultrasoft quark and gluon fields 
and assign a scaling rule to them. A collinear quark field has 
two large and two small components. The decomposition of the 
spinor is as follows:
\begin{equation}
\label{coll/projections}
  \psi_{c}(x) = \xi(x) + \eta(x) ,
  \qquad
  \xi(x) \equiv \frac{\not\!n_-\!\!\not\!n_+}{4} \,\psi_{c}(x) ,
  \quad
  \eta(x) \equiv \frac{\not\!n_+\!\!\not\!n_-}{4}\,\psi_{c}(x) ,
\end{equation}
where $(\slash n_\mp \slash n_\pm)/4$ are projection operators, and
$\slash n_- \xi = \slash n_+ \eta = 0$.  The scaling of these
components is obtained from the projection of the 
QCD quark propagator. For $\xi$ fields 
\be
\label{xiprop}
 \langle 0 | T \xi(x) \bar\xi(y) |0\rangle =
   \frac{\not\!n_-\!\!\not\!n_+}{4}\,
 \langle 0 | T \psi_{c}(x) \bar\psi_{c}(y) |0\rangle
   \frac{\not\!n_+\!\!\not\!n_-}{4} = 
 \int \frac{d^4p}{(2\pi)^4}\, \frac{\slash{n}_-}{2}
   \frac{i n_+ p}{p^2 + i\epsilon}
\, e^{-i p (x-y)}  . 
\ee
For collinear momentum $p$ the right-hand side is of order
$\lambda^2$. From this and the analogous equation for $\eta$ 
with $\nm$ and $\np$ interchanged, we conclude 
\begin{equation}
\label{collquarks}
\xi \sim \lambda , \qquad \eta \sim \lambda^2 .
\end{equation}
The small $\eta$ field will later be integrated out. 
The corresponding argument for an ultrasoft quark field, which we
denote by $q$, dictates
\begin{equation}
q \sim \lambda^3
\end{equation}
for all components of the spinor. Here $k^2\sim \lambda^4$ and 
$d^4k\sim\lambda^8$ for
ultrasoft momentum has been used.

The scaling of the collinear gluon field is determined by projecting 
the gluon propagator (in a covariant gauge with gauge fixing parameter
$\alpha$)
\begin{equation}
\langle 0 | T A_{c}^\mu(x) A_{c}^\nu(y) |0\rangle = 
\int \frac{d^4p}{(2\pi)^4}\, \frac{i}{p^2 + i\epsilon} 
\left[-g^{\mu\nu} + (1-\alpha)\, \frac{p^\mu p^\nu}{p^2}\right]
e^{-i p (x-y)} 
\end{equation}
on its various components. This requires the gluon fields to scale 
identically to a collinear momentum, 
\be
n_+A_{c} \sim 1, \quad     
A_{\perp c} \sim \lambda, \quad     
n_-A_{c} \sim  \lambda^2.
\ee
For an ultrasoft gluon field, 
\be
A_{\rm us}^\mu \sim \lambda^2
\label{coll/pc}
\ee
for all components. Comparing the scaling properties of derivatives
with those of gluon fields we see that collinear and ultrasoft
covariant derivatives have homogeneous scaling rules: the terms in a
covariant collinear derivative $i D_{c}^\mu \phi_{c} = 
(i \partial^\mu + g
A_{c}^\mu) \phi_{c}$ acting on a collinear field all scale in the same
manner. A corresponding statement holds for $i D_{\rm us}^\mu \phi_{\rm us} =
(i \partial^\mu + g A_{\rm us}^\mu) \phi_{\rm us}$ acting on ultrasoft fields.

In cases where collinear particles are produced in the decay of a
heavy quark, we also need an effective heavy quark field. When
collinear modes interact with a nearly on-shell heavy quark, the heavy
quark will subsequently be off-shell by an amount 
of order 1. These off-shell modes
are removed from the effective theory, and heavy quarks interact
only with ultrasoft gluons. It is therefore appropriate to remove the
large component $m v$ of a heavy quark momentum $p=m v+k$ just as in
heavy quark effective theory by introducing the field
\be 
h_v(x) = \frac{1+\slash v}{2}\,e^{i m \, v\cdot x} \, Q(x)
\ee
which varies significantly only over distances $x\sim 1/\lambda^2$, 
since $k$ is ultrasoft. The momentum space 
propagator of $h_v$ is $1/(v k)$, so 
$h_v\sim \lambda^3$ just as the massless ultrasoft quark field.

We finally need to assign a scaling to the integration element $d^4x$
in each term of the effective action, which must reproduce the power
counting for Feynman graphs in momentum space.  Expressing each field
through its Fourier transform, $\phi_i(x) = \int d^4p_i\, e^{-i xp_i}
\tilde{\phi}_i(p_i)$, we see that the integral over $x$ in the action
eliminates one momentum integral
\begin{equation}
\int d^4x\, \left( \prod_{i=1}^n d^4 p_i \right)
\exp(-i x \sum_{i=1}^n p_i)=  
  \int \left( \prod_{i=1}^{n-1} d^4 p_i \right) d^4 p_n\,
     (2\pi)^{4} \delta^{(4)}(p_n + \sum_{i=1}^{n-1} p_i)
\end{equation}
in a term with $n$ fields.  The scaling of $d^4 x$ is thus
$1/\lambda^4$ if the eliminated integral $\int d^4 p_n$ is over a
collinear momentum, and $1/\lambda^8$ if it is ultrasoft.  For
products involving both collinear and ultrasoft fields, one must
choose the eliminated momentum to be collinear and thus count $d^4x$
as $1/\lambda^4$.  This is because eliminating the ultrasoft
integration 
in an integral such as $\int d^4p_{1c}\, d^4p_{2c}\, d^4p_{3 us}$ would
no longer ensure that the momentum $p_3 = - (p_{1c} + p_{2c})$ is
ultrasoft.


\subsection{Gauge transformations}

The effective theory must have a remnant gauge symmetry from gauge
functions $U(x)$ that can themselves be classified as collinear or
ultrasoft.

It is clear that ultrasoft fields cannot transform under collinear
gauge transformations, since this would turn the ultrasoft field into
a collinear field. Under ultrasoft gauge transformations ultrasoft
fields transform in the usual way. The collinear quark field $\xi$ is
also multiplied by $U$ under both collinear and ultrasoft gauge
rotations. Since ultrasoft gluon fields can be added to collinear
gluon fields without changing the character of the collinear field,
ultrasoft gluons can be viewed as slowly varying 
background fields for the collinear
ones. Then $A_{c}$ transforms covariantly under ultrasoft gauge
transformations, while it transforms inhomogeneously under collinear
gauge transformations, but with the derivative replaced by the
covariant derivative with respect to the background field. These
properties are summarized by:
\begin{eqnarray}
&& 
\begin{array}{lll}
\mbox{collinear: }\quad & \displaystyle
   A_{c}\to U_{c} \, A_{c} \, U^{\dagger }_{c}
   + \frac{i}{g} \, U_{c}
   \left[D_{\rm us}, U^{\dagger }_{c}\right], \qquad & 
   \xi \to U_{c} \, \xi, \\[0.2cm] 
& A_{\rm us} \to A_{\rm us}, &
q \to q,\\[0.3cm] 
\mbox{ultrasoft: }\quad & 
   A_{c} \to  U_{\rm us} \, A_{c} \, U^{\dagger}_{\rm us}, & 
   \xi \to U_{\rm us} \, \xi, \\[0.1cm] 
&  \displaystyle
A_{\rm us} \to U_{\rm us}\, A_{\rm us} \, U^{\dagger}_{\rm us}
  + \frac{i}{g} \, U_{\rm us}
  \left[\partial, U^{\dagger}_{\rm us}\right], &
  q \to U_{\rm us} \, q ,
\end{array}
\label{gaugetrafos}
\end{eqnarray}
where $D^\mu_{\rm us} = \partial^\mu - i g A^\mu_{\rm us}$ is an
ultrasoft covariant derivative.  The heavy quark field $h_v$ has
transformation properties identical to $q$. Note that the sum
$A_{c}+A_{\rm us}$ transforms in the standard way under both types of
gauge transformations.

Under collinear transformations the components
$\np A_{c}$ and $A_{c\perp}$ acquire a component of order $\lambda^2$
due to the ultrasoft gluon field in the transformation rule. 
Ultrasoft gauge transforma\-tions are homogeneous in $\lambda$, i.e.
all terms in the transformation law have the same scaling in
$\lambda$, but operators such as $iD_{\perp c}\xi$ transform
inhomogeneously in $\lambda$ under ultrasoft transformations as well. 
A consequence of this is that operators that have a definite scaling
with $\lambda$ are in general not covariant with respect to 
gauge transformations as defined above, but should be combined with
higher-order terms in the $\lambda$ expansion into a gauge-covariant
object.  In the following, we will always keep gauge
invariance manifest as far as possible, before we switch to the strict
$\lambda$ expansion of the effective operators.


\subsection{Wilson lines}
\label{sec:wilson}

The effective theory makes extensive use of various Wilson lines.  We
collect the relevant definitions and properties here. The basic Wilson
line is defined as
\begin{equation}
\label{wline/def}
  W(x) = P\exp\left(ig\int_{-\infty}^0 \!ds \,n_+ A(x+s
    n_+)\right),
\end{equation} 
where $P$ denotes path-ordering defined as $P [A(x)
A(x+sn_+)]=A(x) A(x+sn_+)$, if $s<0$. Here $A=A_{c}+A_{\rm us}$ 
denotes the sum of the
collinear and ultrasoft gluon fields. The Wilson line
is a unitary operator, $W W^\dagger = W^\dagger W =1$, and has the
following useful properties
\begin{equation}
  i n_+ D \, W = W \, i n_+\partial,
   \qquad
  (i n_+ D + i\epsilon)^{-1}  =
  W \,(i n_+ \partial+i\epsilon)^{-1} \,W^\dagger.
 \label{useful-Wilson}
\end{equation}
Under general gauge transformations $U(x)$, $W(x)$ transforms as 
\begin{equation}
  W(x)\to U(x) W(x) U^{\dagger}(x -\infty \, n_+). 
\end{equation} 
The particular form of the transformation matrix at $n_- x \to \infty$
turns out to be irrelevant, since Wilson lines always appear in
combinations where the transformation at infinite distance drops out.
An example is the finite-distance Wilson line $W(x)W^\dagger(x+sn_+)$,
which corresponds to a line integral between $x+sn_+$ and $x$, and
which transforms as
\begin{equation}
    W(x)W^\dagger(x+sn_+)
    \to
    U(x)W(x)W^\dagger(x+sn_+)U^\dagger(x+sn_+).
\label{Wtot/gauge/qcd}
\end{equation} 
It is therefore consistent to choose $U(x -\infty \, n_+)=1$ and this
choice will be adopted in the following. Under collinear and
ultrasoft gauge transformations we then have
\begin{eqnarray}
 &&W \to U_{c}\, W,
\qquad
 W \to U_{\rm us} \, W.
\label{wtrafo}
\end{eqnarray}

We also introduce the ultrasoft Wilson line $\Wus$ that is
defined as in (\ref{wline/def}) but with the full gauge field replaced
by the ultrasoft one, $A_{\rm us}$. This Wilson line is invariant under
collinear gauge transformations and transforms as
\be
Z \to U_{\rm us} \, Z
\ee
under ultrasoft ones.  A particularly important quantity is the
product of Wilson lines, $\WZdag$.  It fulfills the equalities
\begin{equation}
 i n_+ D \, \WZdag = \WZdag \, i n_+ D_{\rm us},
   \qquad
  (i n_+ D + i\epsilon)^{-1} \WZdag =
  \WZdag \,(i n_+ D_{\rm us} +i\epsilon)^{-1} 
\label{coll/wzdag}
\end{equation}
and transforms as 
\begin{eqnarray}
\WZdag \to U_{c} \, \WZdag,\qquad 
\WZdag \to U_{\rm us} \, \WZdag \,U^\dagger_{\rm us}
\end{eqnarray}
under collinear and ultrasoft gauge transformations.  Furthermore we
specify the two limiting cases
\begin{eqnarray}
  && \WZdag \to 1 \qquad \mbox{for $A_{c} \to 0$},
 \nonumber\\[0.1cm]
  && \WZdag \to W_{c} \equiv  
   P\exp\left(ig\int_{-\infty}^0 \!ds \,n_+ A_{c}(x+s
    n_+)\right) \qquad \mbox{for $A_{\rm us} \to 0$}.
\end{eqnarray}
Note that the purely collinear Wilson line $W_{c}$ defined above does
not have a simple behaviour under gauge transformations as
soon as ultrasoft gauge fields are taken into account.

The integration measure $ds$ in the definition of the Wilson line
counts as $1/\lambda^2$ if the integrand contains ultrasoft
fields only, but it counts as $1$, if the integrand  contains a collinear
field. This can be seen in the momentum space representation, where 
$\int ds$ corresponds to $1/(\np p)$.  
It follows that all three Wilson lines, $W$, $Z$, $W_{c}$ count
as order $1$. In contrast to $Z$ and $W_{c}$, however, 
$W$ is not homogeneous
in $\lambda$ due to the occurrence of both, collinear and ultrasoft
gauge fields. Due to the case-dependent counting rule for $ds$, $W$
does not even have a useful expansion in $\lambda$, where ultrasoft
fields should appear as corrections. What we actually need below is
the $\lambda$ expansion of the product $\WZdag$, which is
well-behaved. To construct the expansion we rewrite (\ref{coll/wzdag})
as
\begin{equation}
i n_+ D_{c} \,\WZdag = \Big [\WZdag, g n_+ A_{\rm us} \Big],
\end{equation}
where the derivative acts only on $\WZdag$, and solve this equation
iteratively with the ``boundary condition'' $\WZdag \to 1$ when
$A_{c}\to 0$.  We obtain
\begin{eqnarray}
&& \hspace*{-1.5cm}\WZdag(x) = W_{c}(x)
 + W_{c}\, (i n_+\partial+i\epsilon)^{-1}\, W_{c}^\dag \Big[W_{c}, 
g n_+ A_{\rm us} \Big] + {\cal O}(\lambda^4) 
\nonumber \\
&& \hspace*{0.0cm} =  W_{c}(x) \left(1 -i \int_{-\infty}^0 ds\,
 W_{c}^\dag(x+s n_+)
             \Big[W_{c}, g n_+ A_{\rm us} \Big](x+s n_+)\right)
   + {\cal O}(\lambda^4),
\label{expand-WZ}
\end{eqnarray}
where we used that the inverse of $i\np\partial$, here defined with a 
$+i\epsilon$-prescription, acts as 
\be
\frac{1}{i n_+ \partial+i\epsilon} \,\phi(x) =
  -i \int_{-\infty}^0\!ds\, \phi(x+s n_+).
\label{inversion}
\ee
Note that the commutator term in (\ref{expand-WZ}), $[W_{c}, g n_+
A_{\rm us}] = [W_{c}-1, g n_+ A_{\rm us}]$, has at least one collinear
field when the exponential is expanded, and therefore counts as a
collinear field.  The integration element $ds$ then counts as $O(1)$,
which is crucial to ensure that the second term in (\ref{expand-WZ})
is $\lambda^2$ suppressed. In a similar manner one sees that $(1 -
\WZdag)$ is also a collinear field. On the other hand, the
Wilson lines $W_{c}$ and $\WZdag$ without the unity subtracted do not
qualify as collinear fields and expressions such as
\be
\int_{-\infty}^0 ds\, [\WZdag f_{\rm us}](x+s n_+)
\ee
where $f_{\rm us}(x)$ denotes a product of only ultrasoft fields, 
do not have a well-defined $\lambda$ scaling behaviour and must be avoided. 


\section{Effective Lagrangian}
\label{sec:lagrangian}

In this section we construct the effective Lagrangian for collinear
and ultrasoft quarks and gluons including all interactions suppressed
by one or two powers of $\lambda$. 
We shall first derive the Lagrangian in a form that is
manifestly gauge-invariant, but where the operators do not have a
homogeneous $\lambda$ scaling. Each operator generates a series in
$\lambda$ and is counted by its largest term in the $\lambda$
expansion. In a second step each operator is expanded in a series of
terms each of which has a definite $\lambda$ scaling. We then show
that the effective Lagrangian obtained by tree level matching is not
renormalized to any order in the coupling $\alpha_s$, provided a
Poincar\'{e}-invariant regularization procedure such as dimensional
regularization can be chosen. This remarkable non-renormalization
property extends to all orders in $\lambda$.


\subsection{Collinear Lagrangian}
\label{coll:lagrangian}

We begin with deriving the Lagrangian for collinear quarks interacting
with collinear and ultrasoft gluons, neglecting the ultrasoft quark
field for now. This will be corrected for in the next subsection. The
QCD Lagrangian for massless quarks is
\begin{equation}
\label{qcd_lagrangian}
  {\cal L}_{c} = \bar\psi_{c} 
\left(i \!\not\!\!D +i\epsilon\right)\psi_{c},
\end{equation} 
where $\psi_{c}$ is a four-component spinor, assumed to describe a
nearly on-shell particle with large momentum in the $\nm$ direction.
The covariant derivative is defined as $D=\partial-ig A$, where $A$ is
the sum of the collinear and ultrasoft gluon fields.  Inserting the
decomposition (\ref{coll/projections}) into the QCD Lagrangian, we
obtain
\begin{equation}
\label{lag1}
  {\cal L}_{c} = \bar\xi\,\frac{\not\!n_+}{2} \, i n_- D \,\xi  +
  \bar\eta\,\frac{\not\!n_-}{2} \, i n_+ D \, \eta +
  \bar\xi \left(i\!\not\!\!D_\perp +i\epsilon\right)\eta+
  \bar\eta \left(i\!\not\!\!D_\perp +i\epsilon\right)\xi.
\end{equation}
Here we indicated explicitly that the $i\epsilon$ prescription of the
original Lagrangian is attached to the transverse derivatives. Having
this in mind, we will suppress it in the following.

We proceed by integrating out the small field $\eta$ from the path
integral. Since the Lagrangian is quadratic this amounts to solving
the classical equation of motion, which gives
\begin{equation}
  \eta(x) = - \frac{\not\!n_+}{2}
  \left(i n_+ D+i\epsilon\right)^{-1}
  i\!\not\!\!D_\perp \, \xi(x) 
 = i \,  \frac{\not\!n_+}{2}\, W(x) \int_{-\infty}^0 ds \, \left[
   W^\dagger \, i\!\not\!\!D_\perp \, \xi\right](x+s n_+),
\label{etasol}
\end{equation}
where we have used (\ref{useful-Wilson}) and 
(\ref{inversion}) to obtain the second equality.
This is consistent with $\eta\sim \lambda^2$ as stated in
(\ref{collquarks}). In writing the solution for $\eta$ we have {\em
defined}\/ the inverse of $i n_+ D$ with a $+i\epsilon$
prescription. This prescription is not given by the QCD Lagrangian, and
we could have chosen the opposite sign or a principal value
prescription. The ambiguity in this prescription cancels out in
physical quantities, but the prescription is needed to define individual
terms that appear in the calculation. This is similar to the need to
define a treatment of spurious singularities of gluon propagators in
light-cone gauge. 
The functional determinant of the operator $in_+D$,
which appears upon integrating out $\eta$, amounts to an irrelevant
gauge-field independent constant and can be dropped. To see this, note
that the determinant is gauge-invariant so that we can evaluate it in
light-cone gauge $n_+ A=0$, where $i n_+ D=i
n_+\partial$.\footnote{In a general gauge the functional determinant
reproduces the quark loop diagrams with only $\eta$ fields and any
number of external $\np A$ gluon fields. The loop integral vanishes,
because the propagator for the quark lines,  
$(i n_+\partial+i\epsilon)^{-1}$, implies that
all poles lie on the same side of the real axis.}
The result for the collinear quark Lagrangian is then simply given by
inserting (\ref{etasol}) into (\ref{lag1}), and reads
\begin{eqnarray}
\label{lag2}
  {\cal L}_{c} &=&  \bar\xi\, i n_- D \,\frac{\slash n_+}{2} \,\xi  +
  \bar\xi \,
  i \Slash D_\perp
  \frac{1}{i n_+ D+i\epsilon}\,
  i\Slash D_\perp\frac{\slash n_+}{2}\,
  \xi
\nonumber\\
&=&  \bar\xi(x) \, i n_-D\, \frac{\slash n_+}{2} \,\xi(x)  +
  i \int_{-\infty}^0\!ds\,
  \bigg[\bar\xi \,i \overleftarrow{\Slash D}_\perp
  W\bigg](x)\,
  \bigg[
  W^\dagger\,i\Slash D_\perp \frac{\slash n_+}{2} \,\xi\bigg](x+s
  n_+),
\label{coll/lagrangian}
\end{eqnarray}
where $\overleftarrow{D}=
\overleftarrow{\partial}+igA$ is the covariant derivative
acting to the left.
Note that the effective Lagrangian is formally not hermitian 
because of the 
$+i\epsilon$ prescription on $(i\np D)^{-1}$, which also implies 
that products of fields $\phi(x_1) \cdots \phi(x_n)$ are ``ordered'' 
along the $n_+$ direction, $n_- x_1 \geq \ldots \geq n_- x_n$. The 
hermitian conjugate of the Lagrangian corresponds to the 
$-i\epsilon$ prescription. The non-hermiticity is of no consequence, 
since it is related to modes with $\np p=0$, while collinear modes 
have $\np p\sim 1$. Concretely, the non-local quark-gluon 
vertices that follow from expanding the Wilson line in the gauge 
field individually contain factors $1/(\np p +i\epsilon)$, but these  
unphysical poles cancel in the sum of all diagrams.

The Lagrangian (\ref{coll/lagrangian}) 
is equivalent to the Lagrangian derived in
\cite{Bauer:2000yr} in the hybrid momentum-position space
representation, when $W$ is replaced by the collinear Wilson line
$W_{c}$ and when the ultrasoft gluon field is neglected in $i\Slash
D_\perp$. Using the power counting rules developed in
Section~\ref{sec:fields} it is easy to see that this corresponds to
the leading term in the expansion in $\lambda$.  The result 
(\ref{coll/lagrangian}) is the exact collinear Lagrangian to all
orders in $\lambda$. It is manifestly collinear and ultrasoft
gauge-invariant, which follows from the transformation properties 
(\ref{wtrafo}) of $W$, or directly from the first line of 
(\ref{coll/lagrangian}). The
expansion of (\ref{coll/lagrangian}) in $\lambda$ will be discussed
below, after ultrasoft quarks have been included.

The derivation of the Lagrangian did in fact not require the
assumption that $\psi_{c}$ or $\xi$ are collinear fields. It 
can also be viewed as integrating out two of the components of the
quark spinor field in full QCD, with $\nm$ and $\np$ chosen
arbitrarily. This makes it clear that the Lagrangian
(\ref{coll/lagrangian}) is equivalent to the quark Lagrangian of full
QCD, including all hard and soft modes (and ultrasoft quarks),
provided we consider $\xi$ as a field that destroys and creates all
these modes, not only collinear ones.  This peculiar result follows
because the notion of a collinear particle has no Lorentz-invariant
meaning in the absence of sources that create such particles. The
collinear Lagrangian must therefore exactly reproduce the Green
functions of full QCD (projected appropriately as in (\ref{xiprop})),
when they are evaluated in a frame where all particles have large
momentum in the $\nm$ direction. This interpretation is also confirmed
by the observation that the equation of motion that follows from
(\ref{coll/lagrangian}) and the expression
for $\eta$ is identical to what one obtains in a treatment of full QCD
in the infinite momentum frame \cite{Kogut:1970xa} or in light-cone
quantization \cite{Brodsky:1998de,Srivastava:1999gi}. 
In the following we shall not
pursue this interpretation further since the Lagrangian will
eventually be coupled to sources of collinear particles and we are
interested in performing an expansion in $\lambda$, which implies that
fields must be classified as collinear and ultrasoft.


\subsection{Including ultrasoft quarks}
\label{sec/with/us}

We now incorporate the ultrasoft light quark field, $q$, into 
the effective Lagrangian. The QCD Lagrangian gives rise to the 
following tree-level interactions among collinear and ultrasoft 
fields:
\begin{eqnarray}
 {\cal L} &=& \bar\xi\,\frac{\slash n_+}{2} \, i n_- D \,\xi  +
  \bar\eta\,\frac{\slash n_-}{2} \, i n_+ D \, \eta +
  \bar\xi \, i\Slash D_\perp \, \eta + 
  \bar\eta \, i\Slash D_\perp \, \xi 
\nonumber\\[0.1cm]
&& + \, \bar \xi \, g \Slash A_{c} \, q + \bar \eta \, g \Slash
A_{c} \, q + \bar q \, g \Slash A_{c} \, \xi +  \bar q \,
g \Slash A_{c} \, \eta + \bar q \, i \Slash D_{\rm us} \, q,
\label{Ltmp}
\end{eqnarray}
where $i D_{\rm us}=i\partial + g A_{\rm us}$.  Note that terms such
as $\bar \xi \, i \Slash D_{\rm us} \, q$ vanish by momentum conservation
enforced by the $d^4x$ integral in the action. Collinear gauge
invariance is not manifest in (\ref{Ltmp}).  
We now integrate out the $\eta$ field as before and order the terms in
an expansion in $\lambda$. In this process we use the field equation
for $\xi$, which is equivalent to a field redefinition such that the
final Lagrangian will be manifestly gauge-invariant under the
transformations~(\ref{gaugetrafos}).

Since the Lagrangian (\ref{Ltmp}) is still quadratic in the $\eta$
field, it can be integrated out straightforwardly. The $\eta$ field is
now given by
\begin{eqnarray}
  \eta &=& - \frac{1}{i n_+ D + i \epsilon} \, 
       \frac{\slash n_+}{2} \left( i \Slash
    D_\perp \xi + g \Slash A_{c} \, q \right).
\label{etasol2}
\end{eqnarray}
In the following $(i\np D)^{-1}$ will always be understood with a
$+i\epsilon$ prescription. Inserting (\ref{etasol2}) into (\ref{Ltmp})
results in
\begin{eqnarray}
  {\cal L} &=& {\cal L}_{c} + {\cal L}_{\rm us} 
+ \bar \xi \, g \Slash A_{c} \, q + \bar q \, g \Slash A_{c}
\, \xi - \bar q \,  g \Slash A_{c} \,  \frac{1}{i n_+ D} 
  \frac{\slash n_+}{2} \, g \Slash A_{c} \, q
 \nonumber \\[0.2em] &&
- \,\bar \xi \, i \Slash D_\perp \, \frac{1}{i n_+ D} 
  \frac{\slash n_+}{2} \, g \Slash A_{c} \, q 
- \bar q \, g \Slash A_{c} \,  \frac{1}{i n_+ D} 
  \frac{\slash n_+}{2} \, i \Slash D_\perp \xi ,
\label{Ltmp2}
\end{eqnarray}
with ${\cal L}_{c}$ given by (\ref{coll/lagrangian}), and ${\cal
L}_{\rm us} = \bar q \, i\Slash D_{\rm us} \, q$ the purely ultrasoft
Lagrangian. Note that both ${\cal L}_{c}$ and ${\cal L}_{\rm us}$
are order one terms in the action, since the integration measure
$d^4x$ counts as $\lambda^{-4}$ in $\int d^4x \, {\cal L}_{c}$ but
counts as $\lambda^{-8}$ in $\int d^4x \, {\cal L}_{\rm us}$, where
the integrand involves only ultrasoft fields. The mixed terms in the
Lagrangian are subleading contributions in $\lambda$ to the action,
beginning at order $\lambda$.

The new terms at order $\lambda$ involving the ultrasoft quark field 
are
\begin{eqnarray}
  {\cal L}^{(1)} &=&  
  \bar \xi \left( g \Slash A_{\perp c}  - i \Slash D_\perp \, 
   \frac{1}{i n_+ D} \, g n_+ A_{c} \right)  q 
  + \bar q \left( g \Slash
  A_{\perp c} -  g n_+  A_{c} \,  \frac{1}{i n_+ D} 
   \, i \Slash D_\perp   \right)  \xi. 
\end{eqnarray} 
This can be simplified with the identity
\begin{eqnarray}
\frac{1}{i n_+D}\, g n_+ A_{c} \, \phi_{\rm us} &=& 
(1-\WZdag) \, \phi_{\rm us} 
 - \frac{1}{i n_+ D}\,(1-\WZdag)\,
i n_+ D_{\rm us} \, \phi_{\rm us},
\label{T0sol}
\end{eqnarray}
which follows from (\ref{coll/wzdag}) and is easily verified by acting
with $in_+ D$ from the left. 
The key point to note about this identity is that the first
term on the right-hand side is of order~$1$, but the second is of
order $\lambda^2$ due to the ultrasoft derivative. For this to be
correct it is essential that $(1-\WZdag)$ is a collinear field, which
ensures that $(i n_+ D)^{-1}$ counts as 1 and not as $1/\lambda^2$.
Thus, neglecting terms of order $\lambda^3$, we can write 
\begin{equation}
 \left( g \Slash A_{\perp c}  - i \Slash D_\perp \, 
   \frac{1}{i n_+ D} \, g n_+ A_{c} \right)  q  = 
\left( g \Slash A_{\perp c}  - i
    \Slash D_\perp \, (1-\WZdag) \right) q 
= \left( i\Slash D_\perp \WZdag -  i\Slash D_{\perp\rm us}\right) q. 
\end{equation}
The last term gives $\bar \xi  i\Slash D_{\perp\rm us} \,q$, which 
vanishes in the action by momentum conservation, so 
${\cal L}^{(1)}$ takes the final form
\begin{eqnarray}
{\cal L}^{(1)} &=& 
\bar \xi i \Slash D_\perp \WZdag q + \bar q 
Z W^\dagger  i \Slash D_\perp \xi + {\cal O}(\lambda^3 {\cal L}_c),
\label{L1eff/final}
\end{eqnarray}
which is manifestly gauge-invariant.

With a similar reasoning we can bring the terms of order 
$\lambda^2$ in (\ref{Ltmp2}) into a manifestly gauge-invariant form. 
Consider first the term involving two ultrasoft quark fields, 
which using (\ref{T0sol}) can be approximated as 
\be 
 - \bar q \,  g \Slash A_{c} \,  \frac{1}{i n_+ D} 
  \frac{\slash n_+}{2} \, g \Slash A_{c} \, q \approx
 - \bar q \, g  n_+ A_{c} \,  (1-\WZdag) \,\frac{\slash n_-}{2} \,q .
\ee
Writing $ \,g n_+ A_{c} = i n_+ D
- i n_+ D_{\rm us}$, using the property (\ref{coll/wzdag}) of the 
Wilson line and momentum conservation, we see that this 
term is in fact of order $\lambda^4$ and can be neglected. This leaves 
us with the remaining terms at order $\lambda^2$, 
\be
\label{ll2}
  {\cal L}^{(2)} = 
\bar \xi  \left(
    g n_- A_{c} + i \Slash D_\perp \, \frac{1}{in_+ D} \, g\Slash
  A_{\perp c} \right) \frac{\slash n_+}{2} \, q 
+
 \bar q \, \frac{\slash n_+}{2} \left(g n_- A_{c}
+ g \Slash A_{\perp c}  \, \frac{1}{in_+ D}\, i \Slash D_\perp
\right) \xi .
\ee
Consider the term with $\bar q$ to the left. Using momentum conservation 
$g \nm A_{c}$ can be replaced by $i\nm D$. Then we can rewrite this term 
as 
\be
\label{ll21}
  \bar q \, Z W^\dagger \,i\nm D \,\frac{\slash n_+}{2} \xi 
+ \bar q \, (1-Z W^\dagger) \,i\nm D \,\frac{\slash n_+}{2} \xi 
+ \bar q \, g \Slash A_{\perp c}  \, \frac{1}{in_+ D}\, i \Slash D_\perp
\frac{\slash n_+}{2}\xi.
\ee 
The point of introducing the Wilson lines is that we can now use 
the leading-order equation of motion for the $\xi$ field, 
\be
i n_- D \,\frac{\slash n_+}{2} \,\xi  = 
  - i \Slash D_\perp
  \frac{1}{i n_+ D}\,
  i\Slash D_\perp\frac{\slash n_+}{2}\,
  \xi,
 \label{eom-xi}
\ee
in the second term. This would not be correct in a term with 
$Z W^\dagger$ or $1$ in place of $1-Z W^\dagger$. 
The field equation cannot be used when multiplied from the 
left by a field which is not collinear, because it has been 
obtained by functional differentiation of the action 
with respect to the collinear field $\bar \xi$. Multiplying the 
field equation with an ultrasoft field would project out the
ultrasoft component of the operators in (\ref{eom-xi}) which is 
not consistent with their original appearance as collinear operators 
in the action.
Now (\ref{ll21}) reads
\be 
\label{ll22}
  \bar q \, Z W^\dagger \,i\nm D \,\frac{\slash n_+}{2} \xi 
- \bar q \, \left[(1-Z W^\dagger) i \Slash D_\perp - 
   g \Slash A_{\perp c}\right]  \, \frac{1}{in_+ D}\, i \Slash D_\perp
\frac{\slash n_+}{2}\xi.
\ee
The square bracket equals
\be
-\left(Z W^\dagger i \Slash D_\perp -  i \Slash D_{\perp\rm us} Z 
W^\dagger \right) +  i \Slash D_{\perp\rm us}\,(1-Z W^\dagger).
\ee
We can integrate the ultrasoft derivative by parts in the last term 
to see that it is higher-order in $\lambda$. Here it is again crucial 
that $(1-Z W^\dagger)$ is a collinear field. On the contrary, the 
first two terms individually are not collinear fields and only the 
difference of the two ensures that $(i\np D)^{-1}$ in (\ref{ll22}) 
does not count as $1/\lambda^2$. This gives the final result 
\be 
\label{ll23}
  \bar q \, Z W^\dagger \,i\nm D \,\frac{\slash n_+}{2} \xi 
+\bar q \,  \left(Z W^\dagger i \Slash D_\perp -  i \Slash D_{\perp\rm us} Z 
W^\dagger \right)\, \frac{1}{in_+ D}\, i \Slash D_\perp
\frac{\slash n_+}{2}\xi,
\ee
for one of the two terms of (\ref{ll2}), which is now manifestly 
gauge-invariant. The other term can be manipulated in a similar way, where we must now
use the leading-order equation of motion for $\bar\xi$, obtained by
taking the functional derivative of ${\cal L}_c$ with respect to
$\xi$,\footnote{For this we use $\int d^4\!x\, g(x)\, (i n_+ \partial +
i\epsilon)^{-1} f(x) = - \int d^4\!x\, g(x)\, (i n_+ \!
\overleftarrow{\partial} - i\epsilon )^{-1} f(x)$, which 
follows from (\protect\ref{inversion}) and
(\protect\ref{left-inversion}).}
\begin{equation}
\bar \xi\, i\nm \overleftarrow{D}\,\frac{\slash n_+}{2} = 
{}- \bar \xi \,i\overleftarrow{\Slash{D}}_\perp 
\left(i n_+ \! \overleftarrow{D} - i\epsilon \right)^{-1}
i\overleftarrow{\Slash{D}}_\perp\frac{\slash n_+}{2} ,
\end{equation}
where we have introduced $(i n_+ \! \overleftarrow{D} -
i\epsilon)^{-1} = W^\dag\, (i n_+ \! \overleftarrow{\partial} -
i\epsilon)^{-1}\, W$ and
\begin{equation}
  \label{left-inversion}
\phi(x) \left(i n_+ \! \overleftarrow{\partial} - i\epsilon
        \,\right)^{-1} \equiv 
i \int^{\infty}_0\!ds\, \phi(x+s n_+) .
\end{equation}

Collecting all terms in the expansion we find our final result 
for the soft-collinear effective Lagrangian,
\begin{eqnarray}
\label{collfinal}
{\cal L} &=&  
\bar\xi\, \left(i n_- D + i \Slash D_\perp \frac{1}{i n_+ D}\,
         i\Slash D_\perp \right)  \frac{\slash n_+}{2} \xi 
+ \bar{q}\, i \Slash{D}_{\rm us}\, q 
+ \bar \xi i \Slash D_\perp \WZdag q + \bar q 
Z W^\dagger  i \Slash D_\perp \xi 
\nonumber\\[0.2cm]
&& +\,\bar \xi\,\frac{\slash n_+}{2} \,i\nm D \,\WZdag \,q 
+ \bar\xi \,\frac{\slash n_+}{2}\,i \Slash D_\perp\, \frac{1}{in_+ D}\,
 \left(i \Slash D_\perp \WZdag -  \WZdag i \Slash D_{\perp\rm us}
\right) q 
\\[0.2cm]
&& +\,  \bar q \, Z W^\dagger \,i\nm D \,\frac{\slash n_+}{2} \xi 
+\bar q \,  \left(Z W^\dagger i \Slash D_\perp -  i \Slash D_{\perp\rm us} Z 
W^\dagger \right)\, \frac{1}{in_+ D}\, i \Slash D_\perp
\frac{\slash n_+}{2}\xi 
+ {\cal O}(\lambda^3 {\cal L}_c). 
\nonumber
\end{eqnarray}
The terms containing only $\xi$ and only $q$ are leading in $\lambda$, 
where one must remember that the integration element $d^4x$ in 
the effective action
counts as $\lambda^{-8}$ for a purely ultrasoft term 
and as $\lambda^{-4}$ for all others. The remaining two terms in the 
first line are order $\lambda$ interactions and the remaining two lines 
are order $\lambda^2$. The effective Lagrangian is manifestly 
invariant under collinear and ultrasoft gauge transformations. 
The quark-gluon Lagrangian (\ref{collfinal}) 
is supplemented by the pure Yang-Mills 
Lagrangian, which at this first step of the $\lambda$ expansion 
retains its original form.


\subsection{Expansion in $\lambda$}

\label{lagrangian:lambda}

In deriving (\ref{collfinal}) we counted the operators by their 
smallest power of $\lambda$. Each term in the effective Lagrangian 
(\ref{collfinal}) gives rise to a tower of operators which 
have a homogeneous scaling behaviour, meaning that the insertion 
of such operators gives a contribution of order $\lambda^n$ to 
the scattering matrix element, where $n$ is a certain fixed number that 
can be determined from the power counting rules. In a typical 
application of the effective theory we calculate the scattering 
matrix element to some order in $\lambda$, so we must determine the 
Feynman rules that follow from the Lagrangian order by order 
in $\lambda$. The expansion in terms with a definite (homogeneous) scaling 
in $\lambda$ is also necessary to define effective theories with 
multiple scales in dimensional regularization \cite{Beneke:1998zp}.

We now discuss this further expansion of the effective Lagrangian 
(\ref{collfinal}). We must first split     
$i D_\perp = i D_{\perp c} + g A_{\perp\rm us}$, the
first term being of order $\lambda$ and the second of order $\lambda^2$, and 
expand the Wilson lines in $\lambda$. The expansion of $\WZdag$ 
has already been discussed in Section~\ref{sec:wilson}. 
The combination $W(x) W^\dag(x+s
n_+)$ of Wilson lines along the finite distance from $x+s n_+$ to 
$x$, which appears in the Lagrangian, also has to be expanded 
in the ultrasoft gluon field. Contrary to $W(x)$ itself, this 
combination {\em can} be expanded in $\lambda$, since it stands 
between collinear fields so that  $s$ counts as order 1. 
We obtain with $|s| \sim 1$ 
\begin{eqnarray}
&& W(x) W^\dag(x+s n_+) = W_{c}(x) W_{c}^\dag(x+s n_+) 
\nonumber \\[0.2cm]
&& \hspace*{1cm}
+\,i \int_s^0 dt\, W_{c}(x) \Big[W_{c}^\dag\, g n_+ A_{\rm us}\, 
       W_{c}\Big](x+t n_+) W_{c}^\dag(x+s n_+) 
  + {\cal O}(\lambda^4).
\label{expand-Wilson}
\end{eqnarray}
We need this to expand the inverse differential operator
as 
\begin{equation}
\frac{1}{i n_+ D} = \frac{1}{i n_+ D_{c}}
- \frac{1}{i n_+ D_{c}} g n_+ A_{\rm us} \frac{1}{i n_+ D_{c}} +
{\cal O}(\lambda^4),
\label{expand-Wilson-nice}
\end{equation}
where $i D_{c}=i\partial + g A_{c}$. 
This expansion holds if collinear fields appear both to the
left and to the right of the operator.

In addition we must expand the position arguments of 
ultrasoft fields in the transverse direction and the direction 
of $\np$, since the ultrasoft fields vary more slowly than collinear
fields in these directions. Defining $x_- = \frac{1}{2} (n_+ x)\, n_-$,
the relevant expansion is
\begin{eqnarray}
\phi_{\rm us}(x) &=& 
  \phi_{\rm us}(x_-) + \Big[x_{\perp} \partial
\phi_{\rm us}\Big](x_-)
\nonumber \\
&& + \,\frac{1}{2} \,n_- x \Big[n_+ \partial \,\phi_{\rm us}\Big](x_-)
 + \frac{1}{2} \Big[x_{\mu \perp} x_{\nu \perp}
      \partial^\mu\partial^\nu \phi_{\rm us} \Big](x_-)
 + {\cal O}(\lambda^3 \phi_{\rm us}).
  \label{taylor}
\end{eqnarray}
The collinear field
multiplying this expansion varies over distances $x_\perp \sim
1/\lambda$ and $n_- x \sim 1$, hence the term with
$\partial_\perp \phi_{\rm us}$ is of relative order $\lambda$ and the
terms with $n_+ \partial \phi_{\rm us}$ and $\partial_{\perp}
\partial_{\perp} \phi_{\rm us}$ are of relative order $\lambda^2$, 
since all derivatives on ultrasoft fields scale as $\lambda^2$.  
In terms of momentum space Feynman 
rules this Taylor-expansion corresponds to the fact that 
in a collinear-ultrasoft vertex the momentum transfer by the 
ultrasoft lines to collinear lines 
is small in the transverse direction and the 
direction of $\np$ compared to the momentum carried by collinear lines. 
To obtain a homogeneous expansion in $\lambda$ the momentum space 
diagrams must be expanded in these small momentum components. 
As a consequence of this momentum is not conserved at the interaction 
vertices. In coordinate space this corresponds to a breaking of 
manifest translation invariance due to the Taylor-expansion of 
ultrasoft fields around an arbitrary point in the transverse and 
$\np$ direction. 
Of course translation invariance is recovered order by order 
in $\lambda$. 

Performing this ``multipole'' expansion up to order $\lambda^2$, the 
Lagrangian reads 
\be
\label{fivet}
{\cal L} = {\cal L}_\xi^{(0)} + {\cal L}_\xi^{(1)} + 
{\cal L}_\xi^{(2)} + {\cal L}_{\xi q}^{(1)} + {\cal L}_{\xi q}^{(2)}
+ \cal{L}_{\rm us},
\ee
where the various terms are given by
\begin{eqnarray}
\label{expand2}
{\cal L}^{(0)}_{\xi} &=& 
   \bar{\xi} \left(  i n_- D
 +  i \Slash{D}_{\perp c}
         \frac{1}{i n_+ D_{c}}\, i\Slash{D}_{\perp c} \right)
                                \frac{\slash{n}_+}{2} \, \xi,
\nonumber\\[0.1em]
{\cal L}^{(1)}_{\xi} &=& 
  \bar{\xi} \left(i \Slash{D}_{\perp c}
         \frac{1}{i n_+ D_{c}}\, g\Slash{A}_{\perp\rm us} 
  + g\Slash{A}_{\perp\rm us}
         \frac{1}{i n_+ D_{c}}\,  i \Slash{D}_{\perp c}
  + \left[ (x_\perp\partial) \, (g n_- A_{\rm
      us})\right] \right)
\frac{\slash n_+}{2} \, \xi , 
\nonumber\\[0.1em]
{\cal L}^{(2)}_{\xi} &=& 
  \bar{\xi}\,g\Slash{A}_{\perp\rm us}
         \frac{1}{i n_+ D_{c}}\,g\Slash{A}_{\perp\rm us} 
                                \frac{\slash{n}_+}{2} \xi 
 - \bar{\xi}  \,i \Slash{D}_{\perp c}
         \frac{1}{i n_+ D_{c}}\, gn_+ A_{\rm us}\, 
            \frac{1}{i n_+ D_{c}}\, i\Slash{D}_{\perp c} 
                                \frac{\slash{n}_+}{2} \xi
\nonumber\\[0.1cm]
  && + \,\bar \xi \left( i \Slash{D}_{\perp c}
  \frac{1}{i n_+ D_{c}}\, \left[(x_\perp \partial)\, 
  g\Slash A_{\perp
           \rm us}\right] +  \left[(x_\perp \partial)\, g\Slash A_{\perp
           \rm us}\right]  \frac{1}{i n_+ D_{c}}\,i \Slash{D}_{\perp c}
     \right) \frac{\slash n_+}{2} \, \xi
\nonumber\\[0.2cm]
  && + \,\bar \xi \Bigg( \frac{1}{2}\,(n_- x) \left[
      (n_+\partial)(gn_- A_{\rm us})\right] + \frac{1}{2} \,
    x_\perp^\mu x_\perp^\nu
    \left[\partial_\mu \partial_\nu \, (gn_- A_{\rm
        us})\right] \Bigg) \frac{\slash n_+}{2} \, \xi
\label{Lxi2:prelim}
\eeq
for the part of the Lagrangian without ultrasoft quark fields, and
\beq
{\cal L}^{(1)}_{\xi q} &=& 
    \bar{\xi} \,i\Slash{D}_{\perp c} W_{c}\, q + 
    \bar q \,W_{c}^\dagger i\Slash{D}_{\perp c} \,\xi,
\nonumber\\[0.1cm]
{\cal L}^{(2)}_{\xi q} &=& 
     \bar{\xi}\, g\Slash{A}_{\perp \rm us} W_{c}\, q
   + \bar{\xi}\,\frac{\slash{n}_+}{2}\left( i n_- D 
   +  i \Slash{D}_{\perp c}\,
      \frac{1}{i n_+ D_{c}}\,i \Slash{D}_{\perp c} \right)
                        W_{c} \,q
\nonumber \\
&&  + \, \bar \xi \, i \Slash D_{\perp c} \, W_{c} \left[(x_\perp
  \partial) \,  q\right] +  \mbox{``h.c.''}
\label{12xiq}
\end{eqnarray}
for the interactions between collinear and ultrasoft quarks.
In the last line ``+$\,$h.c.'' refers to the terms of the 
form $\bar q[\ldots]\xi$. In these expressions, and in the remainder of this 
subsection, all ultrasoft fields are understood to be evaluated at position 
$x_-$. For instance, $i\nm D$ at point $x$ equals $i\nm\partial + 
g \nm A_c(x)+g\nm A_{\rm us}(x_-)$. The square bracket means that the 
derivative operates only inside the bracket,
and that all ultrasoft fields or their derivatives are evaluated at $x=x_-$ 
after the derivatives are taken. On the other hand, in an expression 
such as $i \Slash{D}_{\perp c} W_{c} \,q$, the field $q$ is evaluated at $x_-$ 
so that the transverse derivative effectively acts only on $W_c$. Note  
that we cannot use the equation of motion for $\bar \xi$ to drop the 
second term in ${\cal L}^{(2)}_{\xi q}$, since $W_c \,q$ to the right 
of the operator is not a collinear field. The terms in the 
Lagrangian without  
the ultrasoft quark field and without factors of $x$ from the 
multipole expansion appear to be equivalent to the 
corresponding terms in the hybrid momentum-position space
representation of the Lagrangian  
given in \cite{Manohar:2002fd} on the basis of 
reparameterization invariance. The derivation presented here shows
that (leaving aside the terms generated by the multipole expansion 
and those involving the ultrasoft quark field which have not been 
considered in \cite{Manohar:2002fd}) these are the {\em only} terms 
that appear in the effective Lagrangian 
at order $\lambda$ and $\lambda^2$. 

The individual terms ${\cal L}^{(i)}$ of the effective Lagrangian
(\ref{fivet}) are no longer invariant under ultrasoft and collinear 
gauge transformations. The transformations (\ref{gaugetrafos}) mix
terms of different order in
the $\lambda$ expansion, so that only the sum $\sum_{i=0}^n {\cal L}^{(i)}$
is gauge-invariant up to higher-order corrections. 
This is related to the inhomogeneous
transformation of collinear fields under multipole-expanded ultrasoft
gauge transformations,
\beq
  \xi(x) &\to& U_{\rm us}(x) \, \xi(x) = U_{\rm us}(x_-)\,\xi(x) +
    [(x_\perp \partial) \, U_{\rm us}](x_-) \,\xi(x)  + \ldots,
\nonumber 
\\[0.2em]
 A_c(x) &\to& 
    U_{\rm us}(x_-)  \,A_c(x) \,U_{\rm us}^\dagger(x_-) +
    [(x_\perp \partial) \, U_{\rm us}](x_-) \,A_c(x)\,U_{\rm
      us}^\dagger(x_-)
\nonumber\\[0.0cm]
&& +\,U_{\rm us}(x_-)  
\,A_c(x)\,[(x_\perp \partial) \, U^\dagger_{\rm us}](x_-)
    + \ldots,
\label{taylor/gauge}
\eeq
and similarly for $W_c(x)$. 
Furthermore, under collinear gauge transformations (\ref{gaugetrafos})
the collinear gluon projections $(n_+ A_c)$ and $A_{\perp c}$ 
receive $\lambda$ suppressed contributions from ultrasoft gluon
fields. The only gauge invariance that remains manifest for 
every term in 
the $\lambda$ expanded effective Lagrangian (\ref{fivet}) 
is for ultrasoft gauge transformations $U_{\rm us}(x_-)$ that depend only 
on $x_-$. In this case collinear fields
transform homogeneously with $U_{\rm us}(x_-)$. The same is true for partial
derivatives $[\partial_\perp q]$, $[(n_+\partial) \,q ]$ and 
ultrasoft gluon field components $A_{\perp \rm us}$, $n_+ A_{\rm us}$.
Although the lack of manifest gauge invariance causes no problem for
the calculation of on-shell Green functions 
in the effective theory, it would be desirable to cast the Lagrangian
into a form which is manifestly gauge-invariant under the full class
of gauge transformations (\ref{gaugetrafos}). 
This can be achieved by making use of the
equations of motion for collinear quarks and gluons, which is
equivalent to a redefinition of the corresponding fields in the
effective Lagrangian. In the following we demonstrate how this 
can be done for the simpler case of {\em abelian} gauge
fields. 

In the abelian theory ultrasoft and collinear gauge fields
do not interact, and the collinear gluon field does not
transform at all under ultrasoft gauge transformations. Using the 
equation of motion for the
collinear quark field is then sufficient 
to rewrite the effective Lagrangian in the desired form.
The terms ${\cal L}_{\xi}^{(0)}$ and ${\cal L}_{\xi q}^{(1)}$
do not contain ultrasoft gluon fields and already take their
final form. The remaining order $\lambda$ corrections to the soft-collinear 
Lagrangian can be rewritten with the help of
the commutator identity
\be
i \left[x_\perp^\mu, \,  i \Slash{D}_{\perp c}\,
\frac{1}{i n_+ D_{c}}\,  i \Slash{D}_{\perp c}\right]
= \gamma_\perp^\mu \,\frac{1}{i n_+ D_{c}}\,  i \Slash{D}_{\perp c} + 
 i \Slash{D}_{\perp c}\,\frac{1}{i n_+ D_{c}}\,
\gamma_\perp^\mu.
\ee
This results in
\begin{equation}
  {\cal L}^{(1)}_{\xi} =
  \bar{\xi} \left(
     i \left[ (g x_\perp A_{\rm us}) , \, i \Slash{D}_{\perp c}
         \frac{1}{i n_+ D_{c}} i \Slash{D}_{\perp c}\right]
  +
  \left[ (x_\perp\partial) \, (gn_- A_{\rm
      us})\right] \right)
\frac{\slash n_+}{2} \, \xi. 
\label{L1:prelim}
\end{equation}
We can now use the leading-order equations of motion for $\xi$ 
and $\bar \xi$ to obtain the final result for ${\cal L}^{(1)}_{\xi}$ 
\be
{\cal L}^{(1)}_{\xi} =
  \bar{\xi} \left( x_\perp^\mu n_-^\nu \, gF_{\mu\nu}^{\rm
      us}\right) 
\frac{\slash n_+}{2} \, \xi \qquad\mbox{(abelian gauge field)},
\label{1xi}
\ee
where $F$ denotes the field-strength tensor. This interaction term is 
the light-front 
equivalent of the electric dipole interaction $\vec{x}\cdot \vec{E}$ 
of ultrasoft photons 
with non-relativistic electrons. 

Similar manipulations bring the second-order terms ${\cal L}^{(2)}_\xi$, 
${\cal L}^{(2)}_{\xi q}$ into a form in which $A_{\rm us}$ appears only 
in $i \nm D$ or the field-strength tensor. To derive the new form of the 
Lagrangian, one has to keep track of the changes of the Lagrangian at 
order $\lambda^2$ induced after every application of the equation of motion 
at order $\lambda$. The details of this derivation are given in 
Appendix~\ref{app:abelian}. The result is that (for abelian gauge
fields) 
${\cal L}^{(2)}_\xi$, 
${\cal L}^{(2)}_{\xi q}$ take the form
\beq
{\cal L}^{(2)}_{\xi} &=&
  \frac{1}{2} \, \bar \xi \left(
  (n_-x) \, n_+^\mu n_-^\nu \, gF_{\mu\nu}^{\rm us}  
  + x_\perp^\mu n_-^\nu \left[(x_\perp \partial) \,
   g F_{\mu\nu}^{\rm us}\right]  \right) \frac{\slash  n_+}{2} \, \xi
\nonumber\\[0.1cm]
&& + \,\frac{1}{2} \, \bar \xi \left(
    i \Slash D_{\perp c}  \,
 \frac{1}{i n_+ D_{c}} \, x_\perp^\mu \gamma_\perp^\nu \,
 g F_{\mu\nu}^{\rm us}  +   x_\perp^\mu \gamma_\perp^\nu \,
 g F_{\mu\nu}^{\rm us}  \,
 \frac{1}{i n_+ D_{c}} \, i \Slash D_{\perp c}
 \right) \frac{\slash n_+}{2}
\, \xi,
\label{2xi}\\[0.1cm]
{\cal L}^{(2)}_{\xi q} &=&  
\bar{\xi}\,\frac{\slash{n}_+}{2} \left( i n_- D
   + i \Slash{D}_{\perp c}\,
     \frac{1}{i n_+ D_{c}}\, i \Slash{D}_{\perp c}\right)
     \Wc \,q
+ \, \bar \xi \, i \Slash D_{\perp c} W_{c} \left[(x_\perp
  D_{\rm us}) \,  q\right] \nonumber\\[0.1cm] 
&& +  \mbox{ ``h.c.''}.
\label{2q}
\eeq
The various
applications of the equations of motion for $\xi$ are equivalent
to the change of variables
\be
   \xi \to \left( 1 + i g x_\perp A_{\rm us} - \frac12 \, (g x_\perp
     A_{\rm us})^2 + \frac{i}{2} \, (n_- x) \, (g n_+ A_{\rm us}) + 
     \frac12 \, x_\perp^\mu x_\perp^\nu \, [i g \,\partial_\nu A_{\mu, 
       \rm us}] \right) \xi.
\label{Taylor/xinew}
\ee
The new collinear field $\xi$ on the right-hand side 
of (\ref{Taylor/xinew}) can be shown to transform as
\be
  \xi(x) \to  U_{\rm us}(x_-) \, \xi(x) +  {\cal O}(\lambda^3 \xi),
\ee
i.e.\ up to higher orders in $\lambda$ 
it transforms homogeneously with $U_{\rm us}(x_-)$ under 
the original transformation $U_{\rm us}(x)$ that depends on {\em all} 
space-time components of $x$. The new effective Lagrangian is now 
manifestly invariant under the full set of gauge transformations,
because ultrasoft gluon fields and space-time derivatives
appear in covariant derivatives or field strength tensors only. 
Note that invariance of the Lagrangian under collinear and
ultrasoft gauge transformations forces covariant derivatives in
the $x_-$ directions to always appear as $in_-D$ or $in_-D_{\rm
  us}$ but not as $in_-D_c$.

The derivation of the corresponding results for non-abelian gauge fields 
is more complicated because of the interactions between ultrasoft and 
collinear gluons. This requires a careful treatment of the 
$\lambda$ expansion of the 
Yang-Mills Lagrangian which is left for future work.


\subsection{No renormalization}

\label{subsec:norenorm}

The effective Lagrangian constructed so far reproduces the propagation
and interaction of collinear and ultrasoft particles, but it does not 
yet include the effect of hard and soft fluctuations (loops) that 
contribute to a scattering process in full QCD. These effects are
included by matching perturbatively the QCD Lagrangian to the
effective Lagrangian.\footnote{We assume here that the strong coupling
  is small at the scales $m$ and $m\lambda$. When $m\lambda$ is of 
order of the strong interaction scale, only hard fluctuations can be
integrated out perturbatively.} In general, this modifies the 
coefficients of the various operators in the effective Lagrangian, 
and may induce new operators, so that 
\be
{\cal L}_{\rm tree} = \sum_i {\cal O}_i \quad\longrightarrow\quad 
{\cal L} = \sum_{i^\prime} C_{i^\prime} {\cal O}_{i^\prime}.
\ee
We now show that the tree-level ultrasoft-collinear effective Lagrangian 
(\ref{collfinal}) or (\ref{fivet}) 
is not renormalized at any order in the strong coupling, and at 
any order in the $\lambda$-expansion, provided a factorization
scheme can be used that does not break Lorentz invariance, such as 
dimensional regularization.\footnote{It is assumed here without proof that 
dimensional regularization regularizes all hard and soft integrals.} 
In other words, no new operators are induced in higher orders 
in the matching and all coefficient functions retain their tree-level 
values, while the strong coupling evolves as in full QCD.

Consider an $N$-loop diagram in full QCD, where all loop momenta $l_i$ 
are hard, which means that all components are of order $\lambda^0$, 
and where all linear combinations of the $l_i$ are hard. The external 
momenta (not necessarily on-shell)  
of the diagram consist of a number of collinear momenta 
$p_j$, soft momenta $k_{m,\rm s}$ and ultrasoft momenta $k_{n,\rm us}$. 
We include soft momenta here, because such hard diagrams can appear 
as subdiagrams in a larger diagram with soft loop momenta. Included 
in the definition of a hard subgraph is the instruction to Taylor-expand 
the {\em integrand} in all quantities that are subleading in $\lambda$. 
All integrals are assumed to be defined in dimensional regularization. 
The resulting structure is a polynomial in external momenta except 
for the large components of collinear momenta. This precisely matches the 
structure of the operators that appear in the effective Lagrangian 
before integrating out also soft modes. The procedure outlined here 
defines a factorization scheme for the Wilson coefficients 
$C_i$ and is similar to the scheme used to define non-relativistic 
effective theories in dimensional regularization 
\cite{Beneke:1998zp,Manohar:1997qy}.

The initial $N$-loop diagram before Taylor-expansion takes the form 
\be
\label{initialdiagram}
\int\prod_{i=1}^N d^dl_i\,\prod_k \frac{1}{(\sum l_i + \sum p_j 
+\sum k_{m,\rm s} +\sum k_{n,\rm us})^2}\times {\rm polynomial},
\ee
where the product over $k$ refers to the internal lines and the 
sums denote line-specific linear combinations of momenta with 
coefficients $\pm 1, 0$, which we denote by $L_{k,\rm h}$, $P_{k,c}$, 
$K_{k,\rm s}$ and $K_{k,\rm us}$. Using the scaling rules for 
the various momenta, we find
\be 
(L_{k,\rm h} + P_{k,c} + K_{k,\rm s} + K_{k,\rm us})^2 = 
L_{k, \rm h}^2 + (\np P_{k,c}) (\nm L_{k,\rm h}) + {\cal O}(\lambda).
\ee
The integrand of (\ref{initialdiagram}) must be Taylor-expanded 
in all the terms of order $\lambda$ and higher, and we obtain a 
sum of terms of the form 
\be
\label{expandeddiagram}
\int\prod_{i=1}^N d^dl_i\,\prod_k 
\frac{1}{(L_{k,\rm h}^2 + (\np P_{k,c}) (\nm L_{k, \rm h}))^{a_k}}
\times {\rm polynomial}
\ee
with integers $a_k$. But all these integrals vanish in dimensional 
regularization, because they could only depend on scalar products 
of the vectors $(\np P_{k, c}) \,\nm^\mu$, which vanish because 
$\nm^2=0$.\footnote{A more formal way to derive this goes as follows: 
perform a rescaling of all loop momenta,
$$
      n_-l\to \theta^2 n_-l, \qquad
      l_{\perp}^\mu \to \theta l_\perp^\mu.
$$
Under this scaling the loop integral transforms into 
itself times an irrational power of $\theta$ for general $d$ 
and integer $a_k$. Hence the integral must vanish.}
It follows that the soft-collinear Lagrangian is not corrected by 
hard loops in the factorization scheme defined above. This unusual 
non-renormalization property is connected to the fact that, before 
fields are restricted to describe particular momentum modes, the 
collinear Lagrangian is in fact equivalent to full QCD as noted 
earlier, and acquires a Lorentz-invariant meaning only once 
it is coupled to a source that produces collinear particles. 
These sources {\em are} renormalized by hard fluctuations, but 
the Lagrangian is not. 

A similar reasoning can be used to show that soft fluctuations 
do not correct the soft-collinear effective Lagrangian either. 
Note that soft modes have virtualities of order $\lambda^2$, 
equal in magnitude to the virtuality of collinear modes. Nonetheless, 
soft modes 
can be integrated out, because they cannot appear as external 
modes under our basic kinematical assumption. A general 
soft diagram has all loop momenta $l_i$ soft, 
which means that all components are of order $\lambda^1$, 
and all linear combinations of the $l_i$ are also soft. The external 
momenta of the diagram consist of a number of collinear momenta 
$p_j$ and ultrasoft momenta $k_{n}$. The matching onto 
the soft-collinear Lagrangian is again defined by Taylor-expanding 
the loop integrand in all small quantities and by evaluating 
all diagrams in dimensional regularization. With the same definitions 
as above the propagator denominators of a soft loop integral 
are 
\beq
(L_{k, \rm s}+P_{k, c} + K_{k, \rm us})^2 &=&  
(\np P_{k, c})(\nm L_{k, \rm s}) 
\nonumber\\
&&\hspace*{-3cm}+ \,\Big[L_{k, \rm s}^2 + L_{\perp
  k, \rm s}
P_{\perp k, c} + P_{k, c}^2 + (\np P_{k,c}) (\nm K_{k,
  \rm us})\Big] + {\cal O}(\lambda^3),
\eeq
where the first term on the right-hand side is of order $\lambda$ 
and all others explicitly given are of order $\lambda^2$. For generic 
collinear momenta $\np P_{k, c} = {\cal O}(1)$, or else 
$P_{m, c}= 0$ if no collinear external momentum flows through 
line $m$. The soft matching correction then only involves 
integrals of the form 
\beq
\label{expandedsoft}
&& \int\prod_{i=1}^N d^dl_i\,
\prod_{k} 
\frac{1}{((\np P_{k, c}) (\nm L_{k, \rm s}) + i \epsilon)^{a_{k}}} 
\prod_{m} \frac{1}{( L_{m, \rm s}^2 +
  i\epsilon)^{a_{m}}}
\times {\rm polynomial}
\eeq
with integers $a_{k,m}$.
These integrals are again zero in dimensional regularization. 

This argument does not cover exceptional configurations  
where collinear particles are nearly comoving, such that 
a subset of external collinear momenta sum up to nearly zero
in an internal line $n$, i.e.\ $\np P_{n,c} \sim \lambda$. 
In this case the soft matching correction involves integrals of the form 
\beq
\label{expandedsoft2}
&& \int\prod_{i=1}^N d^dl_i\,
\prod_{n} \frac{1}{( (\np P_{n, c})(\nm L_{n, \rm s})+
L_{n, \rm s}^2+ L_{\perp n, \rm s} 
P_{\perp n,c} + P_{n,c}^2 +i\epsilon)^{a_{n}}}
\nonumber \\[0.2em]
&& \quad \times 
\prod_{k} 
\frac{1}{(
(\np P_{k,c}) (\nm L_{k, \rm s}) + i \epsilon)^{a_{k}}} 
\prod_{m} \frac{1}{(L_{m, \rm s}^2 + i\epsilon)^{a_{m}}}
\times {\rm polynomial}
\eeq
with integers $a_{k,m,n}$, and these integrals 
are not obviously zero in dimensional regularization. We checked 
for some one- and two-loop diagrams that these integrals do vanish, 
because there is always at least one $(n_+l_i)$ integration where all 
poles lie on the same side of the real axis, but we do not have an 
all-order proof of this result.

Under our kinematical assumption there are no tree diagrams involving
soft particles. We 
therefore conclude that (barring the potential contribution 
from exceptional configurations
discussed above) soft modes are not relevant for the 
construction of the effective Lagrangian.

\subsection{Heavy quark Lagrangian}

We will later discuss the production of collinear particles in the
decay of a heavy quark near mass-shell. Heavy quarks near mass-shell
are described by the Lagrangian of heavy quark effective theory
(HQET), ${\cal L}_{\rm HQET}$
\cite{Eichten:1990zv,Georgi:1990um,Grinstein:1990mj}. In our framework
this Lagrangian corresponds to the interaction of heavy quarks with
ultrasoft gluons. An interaction of a near-on-shell massive quark with
a collinear gluon will throw the heavy quark off mass-shell and the
corresponding process would have to be added to the effective theory
as an explicit matching correction. The question we would like to
address here is whether, when we combine the heavy quark with
collinear and ultrasoft modes, the Lagrangian for the combined system
is 
\be
\label{lagcomb}
{\cal L} = {\cal L}_{\rm HQET} + {\cal L}_{\rm SCET},
\ee
where ${\cal L}_{\rm SCET}$ is the soft-collinear Lagrangian, 
or whether the newly present collinear modes require that 
we add terms to the effective heavy-quark Lagrangian not 
present in the HQET Lagrangian with ultrasoft gluons only. Note 
that at this point we do not discuss sources (currents) which 
can convert a heavy quark into collinear particles.

Before turning to this question we briefly review the standard 
HQET formalism for later use. 
Starting from the QCD Lagrangian with a heavy quark field $Q$,
we define the four-component ultrasoft heavy-quark field
$
Q_v(x)=e^{i m \, v\cdot x} \, Q(x)
$,
which varies only over distances $x\sim 1/\lambda^2$, $m$ being
the heavy quark mass. 
In terms of this new field the heavy quark Lagrangian reads
\begin{eqnarray}
  {\cal L} &=& \bar Q_v
  \left(i \Slash D - m \, (1-\slash v)\right) Q_v
  \ = \ \bar Q_v
  \left(i \Slash D_{\rm us} - m \, (1-\slash v)\right) Q_v,
\label{Qvlagr}
\end{eqnarray}
where to obtain the second equality we dropped the term containing 
the collinear gluon field, which vanishes by momentum conservation.
In other words, the ultrasoft heavy quark does not interact 
with collinear gluons in the effective theory. 
One then performs the projection onto large and small components,
\begin{eqnarray}
&&  h_v = \frac{1+\slash v}{2} \, Q_v, \qquad
    H_v = \frac{1-\slash v}{2} \, Q_v,
\end{eqnarray} 
and integrates out the small field $H_v$. This gives the tree-level 
effective Lagrangian
\begin{eqnarray}
 {\cal L}_{\rm HQET} &=& \bar h_v \left[i v D_{\rm us} + 
 \frac{(i \Slash D_{\rm us})^2}{2 m} + {\cal O}(\lambda^6) \right] h_v,
\label{hqet}
\end{eqnarray} 
and the leading-order equation of motion,
\begin{eqnarray}
  i v  D_{\rm us} \, h_v = {\cal O}(\lambda^4) \, h_v.
\label{hqet/eom}
\end{eqnarray}
Using (\ref{hqet/eom}) $Q_v$ can be expressed as
\begin{eqnarray}
Q_v = h_v+H_v &=& 
\left( 1 + \frac{i \Slash D_{\rm us}}{2m} + {\cal O}(\lambda^4)
\right) h_v.
\label{Qv}
\end{eqnarray}
One also derives from the propagator of $h_v$ that $h_v\sim \lambda^3$ and 
$H_v\sim \lambda^5$. Under ultrasoft gauge-transformations, $h_v \to
U_{\rm us}\, h_v$, but $h_v\to h_v$ under collinear gauge 
transformations. 

\begin{figure}[t]
   \vspace{-3cm}
   \epsfysize=27cm
   \epsfxsize=18cm
   \centerline{\epsffile{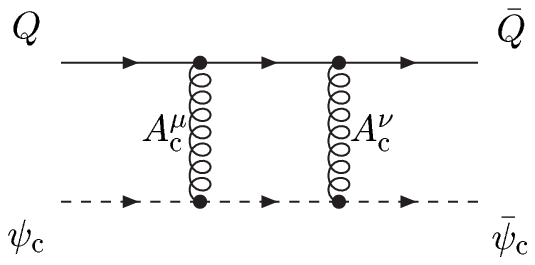}}
   \vspace*{-21.2cm}
\caption[dummy]{\label{fig1} Collinear interaction of the heavy 
quark.}
\end{figure}

Now consider the interaction of a heavy quark with a collinear 
quark through two collinear gluons as in Figure~\ref{fig1}, where 
it is implied that both external heavy quark lines are near mass-shell. 
As a consequence of this the large components of the 
collinear gluon momenta must add to something of order 
$\lambda^2$. The intermediate heavy quark propagator is off-shell, 
and at first sight it might appear that this results in a new 
effective interaction of $h_v$ with collinear gluons or quarks 
which is not present in the HQET Lagrangian. 
However, diagrams 
of this type do in fact not represent permitted configurations 
for the physical process to which the effective theory applies. 

To see this, recall that the effective theory diagrams correspond
exactly to those configurations of momenta where the QCD diagram
develops on-shell singularities.  According to a theorem by Coleman
and Norton \cite{Coleman:1965aa} these singularities correspond to
classically allowed scattering processes.  A heavy quark cannot emit
or absorb two collinear particles without going off-shell, so that in
such a classical scattering process one collinear particle must be in
the initial and the other in the final state.  In the effective theory
for the decay of a $B$ meson into collinear particles through a weak
interaction current, there are no initial-state collinear
particles.  The invariant mass of the initial state is close to the
mass of the heavy quark, and this excludes collinear particles in the
initial state -- $B$ mesons consist only of ultrasoft heavy quarks and
ultrasoft light quarks and gluons.  We note that the same is true for
soft particles: even if they appeared as degrees of freedom in the
effective theory, there would be no vertex between a heavy-quark line
and two or more soft quarks or gluons.

The conclusion of this discussion is that the combined heavy quark and
soft-collinear system (in the absence of external sources, i.e.\
heavy-to-light currents) is described by the
sum of the two Lagrangians (\ref{lagcomb}) with no additional
interaction terms.


\section{Heavy-to-light current to order $\lambda^2$}
\label{sec:currents}

In this section we discuss the representation of weak currents 
$\bar \psi\Gamma Q$, where $\Gamma$ denotes a Dirac matrix, in 
the effective theory. We consider the kinematic region where large 
momentum is transferred from the heavy quark to the final state 
light quarks and gluons.

\subsection{Derivation of the effective current}
\label{sec/current}

\begin{figure}[t]
   \vspace{-3cm}
   \epsfysize=27cm
   \epsfxsize=18cm
   \centerline{\epsffile{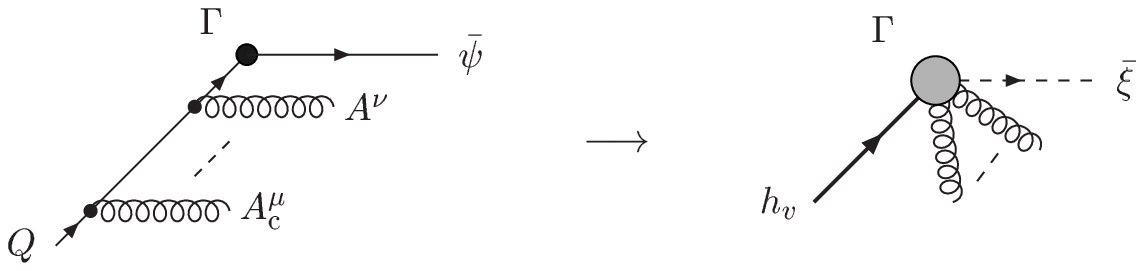}}
   \vspace*{-21cm}
\caption[dummy]{\label{fig/current}
Tree-level matching of the heavy quark current.}
\end{figure}

The purpose of this subsection is to derive the effective heavy-to-light 
current at tree level, including corrections of order $\lambda$ and 
$\lambda^2$. This extends previous work \cite{Bauer:2000yr}, 
where the effective 
current has been derived at order $\lambda^0$. 
The emission of a collinear gluon from the heavy quark 
puts the heavy quark off mass-shell by an amount of order $m^2$. The 
corresponding diagrams with off-shell heavy quark lines are not part
of the effective theory, but must be reproduced by the effective
current. Once a collinear gluon has been emitted from the heavy 
quark, the heavy quark stays off-shell when further collinear 
{\rm or ultrasoft} gluons are radiated until the heavy quark decays
into a light quark.\footnote{At tree level, we do not have
  to consider the emission of soft gluons, since soft gluons do not 
  appear as external lines for a physical process.} The 
infinite number of tree diagrams that correspond to the emission of 
collinear and ultrasoft gluons has to be summed into the effective 
current as sketched in Figure~\ref{fig/current}. 

In position space we obtain  
\begin{eqnarray}
J_{\rm QCD}(x) &=& 
\bar \psi(x) \, \Gamma \, 
\sum\limits_{n=0}^\infty \int dz_1 \cdots dz_n \, D_F(x-z_1) \,
 ig\Slash A(z_1) \cdots D_F(z_{n-1}-z_n) \,
 ig\Slash A_{c}(z_n) \, Q(z_n)
\nonumber \\[0.2em]
&=&
\bar \psi(x) \, \Gamma \, 
\sum\limits_{n=0}^\infty \, \frac{1}{i\slash
  \partial - m} \,
 \Big(-g\Slash A(x)\Big) \cdots \frac{1}{i\slash
  \partial - m} \,
 \Big(-g\Slash A_{c}(x)\Big) \, Q(x)
\nonumber \\[0.2em]
&=&
e^{-i m \, v\cdot x} \
\bar \psi \,  \Gamma 
\left(1-\frac{1}{ i\Slash D - m \, (1-\slash v)}\,
g\Slash A_{c} \right) Q_v,
\label{start}
\end{eqnarray}
where $D_F$ denotes the heavy-quark propagator and the 
covariant derivative $\partial_\mu -i g A_\mu$ contains collinear
and ultrasoft gluons, $A=A_{c}+A_{\rm us}$. The collinear 
gluon field next to the heavy quark field arises because the 
gluon at the beginning of the quark line must be collinear 
to put the quark off-shell. We omitted the $i\epsilon$ prescription
in the heavy quark propagators, since they are far off-shell by 
construction. The field $Q_v$ is defined in (\ref{Qv}).  
The light-quark field $\bar\psi$ will be discussed below. 
In the expression ~(\ref{start}) gauge invariance of the
current is not evident, since the
collinear gluon field $A_{c}^\mu$ appears explicitly. Formally,
gauge invariance can be made explicit by writing 
\be 
g \, \Slash A_{c} \, Q_v
= \left[ i \Slash D - m \, (1-\slash v)\right] Q_v -
  \left[ i \Slash D_{\rm us} - m \, (1-\slash v)\right] Q_v 
= \left[ i \Slash D - m \, (1-\slash v)\right] Q_v
\label{80}
\ee
where we used the equation of motion following from
(\ref{Qvlagr}).  

Our first task is to expand 
\be
{\cal Q} \equiv \left(1-\frac{1}{ i\Slash D - m \, (1-\slash v)}\,
g\Slash A_{c} \right) Q_v
\label{81}
\ee
in powers of $\lambda$. As before, we organize this expansion such that
every term is covariant under the gauge transformations 
(\ref{gaugetrafos}). For the discussion of reparameterization 
invariance it will be convenient {\em not} to use the 
canonical definition $v^\mu \equiv (n_+^\mu + n_-^\mu)/2$
but to keep $v$ as an independent vector. Consistency with our power 
counting scheme requires only that 
\begin{equation}
n_+ v \sim 1 , \qquad n_- v \sim 1 , \qquad  v_\perp \sim \lambda
\end{equation}
together with $v^2=1$, $n_\pm^2=0$, and $n_+n_- = 2$. 
To expand ${\cal Q}$, we decompose the collinear gluon field into 
its different components and expand the inverse differential operator 
in $i \Slash D_\perp$ and $i\nm D$. We do not (yet) separate the 
smaller ultrasoft gluon field in 
expressions such as $(i n_+ A)$ or $A_\perp^\mu$, since this would 
not allow us to maintain invariance under the transformations 
(\ref{gaugetrafos}). The leading term in the inverse differential operator 
can be written as
\be
S_0\equiv \frac{1}{(i n_+ D) \, \slash n_-/2 - m(1 - \slash v)} 
= \frac{1}{n_- v} 
\left(\frac{1+\slash v}{i n_+ D}+\frac{\slash n_-}{2m}\right),
\ee
where here and in the following the inverse $(i n_+ D)^{-1}$ is
understood to be defined with
the $+i\epsilon$ prescription. We then obtain 
\begin{eqnarray}
{\cal Q}&=& 
\left(1-S_0 \,g\,\np A_c\,\frac{\!\not\!n_-}{2}\right) 
\left(1+\frac{i\!\!\not\!\! D_{\rm us}}{2 m}\right)h_v 
\nonumber\\
&&\hspace*{0cm} 
-\,S_0\left[\left(-i\!\!\not\!\! D_\perp\right)S_0 
\,g\,\np A_c\,\frac{\!\not\!n_-}{2} + g\!\not\!\!A_{\perp c}\right] h_v
\nonumber\\
&&\hspace*{0cm} 
-\,S_0\left[\left(-i n_- D\,\frac{\!\not\!n_+}{2}\right)S_0
\,g\,\np A_c\,\frac{\!\not\!n_-}{2} + g\,\nm A_c\,
\frac{\!\not\!n_+}{2} \right] h_v
\nonumber\\
&&\hspace*{0cm} 
-\,S_0\left(-i\!\!\not\!\! D_\perp\right)
S_0\left[\left(-i\!\!\not\!\! D_\perp\right)S_0 
\,g\,\np A_c\,\frac{\!\not\!n_-}{2} + g \!\not\!\!A_{\perp c}\right] h_v
+ {\cal O}(\lambda^3 h_v).
\end{eqnarray}
Defining 
\begin{equation}
T_0=\frac{1}{i n_+ D}\,g\,\np A_c,
\end{equation}
and using some Dirac algebra, the above 
expansion is rewritten as
\begin{eqnarray}
{\cal Q}&=& 
\left(1-T_0\right) h_v +\left(1-\frac{1}{\nm v}\,(1+\!\not\!v)
\,\frac{\!\not\!n_-}{2} T_0\right) 
\frac{i\!\!\not\!\! D_{\rm us}}{2 m} h_v 
\nonumber\\[0.2cm]
&&\hspace*{0cm} 
-\,\frac{1}{\nm v}\,\frac{\!\not\!n_-}{2 m}
\Big(\left(-i\!\!\not\!\! D_\perp\right) T_0 + 
g\!\not\!\!A_{\perp c}\Big) h_v
-\,\frac{1}{\nm v}\,\frac{\!\not\!n_-}{2m}\frac{\!\not\!n_+}{2}
\Big(\left(-i n_- D\right) T_0 + g\,\nm A_c \Big) h_v
\nonumber\\[0.2cm]
&&\hspace*{0cm} 
-\, \frac{1}{\nm v}\Big(2 v_\perp^\mu+\,(\np v) \,n_-^\mu\Big) 
\frac{1}{i n_+ D}
\Big(\left(-i D_\mu\right) T_0 + g A_{\mu,c} \Big) h_v
\nonumber\\[0.2cm]
&&\hspace*{0cm} 
+\,\frac{1}{\nm v}\,\frac{1}{m}\,\frac{1}{i n_+ D}
\left(-i\!\!\not\!\! D_\perp\right)
\Big(\left(-i\!\!\not\!\! D_\perp\right) T_0 
+ g \!\not\!\!A_{\perp c}\Big) h_v
+ {\cal O}(\lambda^3 h_v).
\label{exp1}
\end{eqnarray}
This simplifies considerably when one uses (\ref{T0sol}) to expand 
$T_0 h_v$ in $\lambda$, and then the identity
\begin{equation}
\left(-i D^\mu\right) (1-W Z^\dagger) + 
g A_{c}^\mu =
i D^\mu W Z^\dagger - i D^\mu_{\rm us}.
\end{equation}
The term involving $v_\perp$ can be eliminated because 
\be
\frac{1}{\nm v}\Big( 2 v_\perp+(\np v) \nm\Big) = \frac{2 v}{\nm
  v}-\np.
\ee
Combining various terms of similar structure 
and making use of the property (\ref{coll/wzdag}) of $\WZdag$, 
we find 
\begin{eqnarray}
{\cal Q} &=&
\WZdag \left(1+ \frac{i\!\!\not\!\! D_{\rm us}}{2 m} \right) h_v 
-\, \frac{1}{\nm v}\,\frac{\!\not\!n_-}{2m}
\Big(i\!\!\not\!\! D \WZdag
- \WZdag\,i\!\!\not\!\! D_{\rm us}\Big) h_v
\nonumber\\[0.2cm]
&&
-\, \frac{1}{\nm v}\,\frac{1}{i n_+ D}
\left(\left\{2i v  D + \frac{i\!\!\not\!\! D_\perp
i\!\!\not\!\! D_\perp}{m} \right\}\WZdag
- 2i v  D_{\rm us} -  \frac{i\!\!\not\!\! D_{\perp }
i\!\!\not\!\! D_{\perp\rm us}}{m}\right) h_v
\nonumber\\[0.2cm]
&&
+ \,{\cal O}(\lambda^3 h_v).
\end{eqnarray}
Neglecting corrections of order $\lambda^3$ or smaller, 
the last two terms involving ultrasoft covariant derivatives 
can be modified by 
inserting $\WZdag$ to the left and by replacing $i\!\!\not\!\!
D_{\perp }$ by $i\!\!\not\!\! D_{\perp\rm us}$. This gives the final
result 
\begin{eqnarray}
{\cal Q} &=&
\WZdag \left(1+ \frac{i\!\!\not\!\! D_{\rm us}}{2 m} \right) h_v 
-\, \frac{1}{\nm v}\,\frac{\!\not\!n_-}{2m}
\Big(i\!\!\not\!\! D \WZdag
- \WZdag\,i\!\!\not\!\! D_{\rm us}\Big) h_v
\nonumber\\[0.2cm]
&&
-\, \frac{1}{\nm v}\,\frac{1}{i n_+ D}
\Bigg(\left\{2i v  D + \frac{i\!\!\not\!\! D_\perp
i\!\!\not\!\! D_\perp}{m} \right\}\WZdag
-\WZdag \left\{ 2i v  D_{\rm us} +
\frac{i\!\!\not\!\! D_{\perp\rm us}
i\!\!\not\!\! D_{\perp\rm us}}{m}\right\}\Bigg) h_v
\nonumber\\[0.2cm]
&&\hspace*{0cm}
+ \,{\cal O}(\lambda^3 h_v).
\label{finalQ}
\end{eqnarray}
Note that the term with the derivative  
$2 i v D = (n_- v) (i n_+ D) + 2i v_\perp
D_\perp + (n_+ v) (i n_- D)$ is of order $\lambda^2$,
since the apparent order $\lambda^0$ term, $i n_+ D \WZdag$, 
cancels in the combination  
$i n_+ D \WZdag - \WZdag i n_+ D_{\rm us}=0$. Hence (\ref{finalQ}) 
does give a valid $\lambda$ expansion of ${\cal Q}$. It is also 
manifestly collinear and ultrasoft gauge-covariant. In the absence 
of the collinear gluon field ${\cal Q}$ reduces to $Q_v$ as it
should.\footnote{Although the last term in curly brackets in
  (\ref{finalQ}) is of order $\lambda^4$ when $A_c$ is non-zero, 
it becomes of order $\lambda^2$ as $A_c\to 0$, because then 
$(i\np D)^{-1}\to (i\np D_{\rm us})^{-1}\sim 1/\lambda^2$. 
However, this term exactly cancels against the first curly bracket
when $A_c\to 0$.}

Before we continue the discussion of $\bar\psi\Gamma {\cal Q}$ 
using (\ref{finalQ}), we briefly mention an equivalent derivation 
of (\ref{finalQ}). Rather than summing tree diagrams we solve the 
Dirac equation
\be
(i\Slash D-m)Q = 0
\label{dirac}
\ee
in the presence of collinear gluon fields $A_{c}^\mu$,
together with the boundary condition that in the absence of a
collinear field the solution reduces to the ultrasoft case (\ref{Qv}),
\beq
  Q(x) &\to& e^{-imvx} \, Q_v(x) \qquad \mbox{for $A_{c}^\mu \to 0$} \ .
\label{bound}
\eeq
This equivalence can be seen from inserting (\ref{80}) into (\ref{81}), which
implies that $e^{-i m \, v\cdot x} \, {\cal Q}(x)$ 
is determined by the solution of the Dirac equation (\ref{dirac}). 
We expand 
$  {\cal Q} = {\cal Q}^{(0)}+ \lambda \, {\cal Q}^{(1)}+\lambda^2
       \, {\cal Q}^{(2)} +\ldots $ and 
solve the Dirac equation
\be
\left[
 \left(\frac{\slash n_-}{2} \,in_+D - m \, (1-\slash v)\right)
    +i\Slash D_\perp
     +\frac{\slash n_+}{2} \,in_-D
     \right]
     \left[{\cal Q}^{(0)}+ \lambda \, {\cal Q}^{(1)}+\lambda^2
       \, {\cal Q}^{(2)} +\ldots \right] = 0
\label{eom/expand}
\ee
iteratively in $\lambda$ subject to 
the boundary condition (\ref{bound}).
The advantage of this procedure is that collinear and ultrasoft
gauge invariance is manifest at every step of the derivation.
For instance, to order $\lambda^0$ the equation is
\be
    \left(\frac{\slash n_-}{2} \, (in_+D) - m \, (1-\slash v) \right)
     {\cal Q}^{(0)} = 0,
\ee
which is solved by ${\cal Q}^{(0)} = \WZdag h_v$ up to terms 
of order $\lambda^2$ and also satisfies the boundary condition up to
terms of this order. Substituting this into (\ref{eom/expand}) 
yields the equation for ${\cal Q}^{(1)}$. We solve this equation 
and satisfy the boundary condition by adding terms of the form 
$\WZdag f_{\rm us}$, where $f_{\rm us}$ is an ultrasoft field. 
When this program is carried out to order $\lambda^2$ 
one reproduces the result (\ref{finalQ}) for ${\cal Q}$. In this 
approach the particular form 
$f(D^\mu) \, \WZdag - \WZdag \, f(D^\mu_{\rm us})$ of many 
terms in (\ref{finalQ}) can be traced back to the requirement 
that ${\cal Q}\to Q_v$ is demanded in the absence of collinear 
fields. 

Returning to the derivation of the effective current, we now have 
\be
\label{jstart}
 e^{i m \, v\cdot x}\,J_{\rm QCD} \equiv J = 
\left(\bar\xi+\bar\eta+\bar q\right)\Gamma {\cal Q}
\ee
with ${\cal Q}$ given by (\ref{finalQ}). Inserting  
for $\bar \eta$ the expression that corresponds to (\ref{etasol2}) and
expanding this to the required order in $\lambda$ does not give the 
correct result. The reason for this is that the effective current is
constructed such that it reproduces the on-shell matrix elements 
with a current insertion in QCD. In the effective theory this amounts
to calculating insertions of the effective current together with 
time-ordered products with terms of higher order in $\lambda$ in the 
effective Lagrangian. Different forms of the effective Lagrangian 
which are equivalent by the equations of motion give different 
results for the effective current beyond leading order in $\lambda$. 
The correct current is automatically determined by matching the 
matrix elements in the full and in the effective theory. In order to avoid
the explicit calculation of the time-ordered product terms, we couple 
the current to a weak external field $B$ and add the terms 
\be
{\cal L}_j =
e^{-i m \, v\cdot x} \,
\bar\psi \Gamma {\cal Q} B + \mbox{``h.c.''} 
\equiv (\bar \xi+\bar\eta+\bar q) j + \mbox{``h.c.''} 
\label{lj}
\ee
to the Lagrangian (\ref{Ltmp}). We now repeat the derivation of the effective 
Lagrangian (\ref{collfinal}) 
in Section~\ref{sec/with/us} and follow the modifications 
induced by the presence of the additional source term. 
The solution for the field $\eta$ in (\ref{etasol2}) now reads
\be
 \eta = - \frac{1}{i n_+ D} \, 
       \frac{\slash n_+}{2} \left( i \Slash
    D_\perp \xi + g \Slash A_{c} \, q +j\right).
\ee
Inserting this into (\ref{Ltmp}) and (\ref{lj}) results in 
\beq
{\cal L}_j &=& 
 \left( \bar{\xi} +\bar q - 
\left(\bar\xi \ i\!\overleftarrow{\Slash D}_{\perp}\,+\bar q g  
\Slash A_{c}\right)
     \left(i n_+ \! \overleftarrow{D}\right)^{-1}
     \,\frac{\slash n_+}{2}
     \right)\, j 
+ \mbox{``h.c.''} 
\nonumber\\
&=& 
 \left( \bar{\xi} 
 - \bar\xi \ i\!\overleftarrow{\Slash D}_{\perp}\,
     \left(i n_+ \! \overleftarrow{D}\right)^{-1} \,\frac{\slash n_+}{2}
+\bar q - \bar q \,
\frac{\not\!\nm\slash n_+}{4} \,(1-Z W^\dagger)\right)\, j 
+ \mbox{``h.c.''} 
\eeq
where we expanded in $\lambda$ up to second order, used (\ref{T0sol}) and 
neglected terms quadratic in the weak field $B$. 
In Section~\ref{sec/with/us} we used the $\xi$ equation of motion to
pass from (\ref{ll21}) to (\ref{ll22}). 
With the term ${\cal L}_j$ added, 
the leading-order equation of motion
(\ref{eom-xi}) for the collinear quark field $\xi$ is modified
to
\begin{equation}
i n_- D \,\frac{\slash n_+}{2} \,\xi  = 
  - i \Slash D_\perp
  \frac{1}{i n_+ D}\, 
  i\Slash D_\perp\frac{\slash n_+}{2}\,
  \xi - \frac{\slash{n}_+ \slash{n}_-}{4} j  .
\end{equation}
so that in passing to (\ref{ll22}) one picks up an additional
contribution 
\be
-\bar q\,(1-Z W^\dagger) \frac{\not\!\np\slash n_-}{4} j + 
\mbox{``h.c.''}
\ee
to ${\cal L}_j$. We then obtain the result 
\begin{equation}
{\cal L}_j = 
 \left( \bar{\xi} - \bar\xi \ i\!\overleftarrow{\Slash D}_{\perp}
     \left(i n_+ \! \overleftarrow{D}\right)^{-1}
     \frac{\slash n_+}{2} + \bar{q}\ZWdag\right) j 
+ \mbox{``h.c.''} 
\label{source-term}
\end{equation}
for the source Lagrangian in the effective theory with the Lagrangian 
(\ref{collfinal}), which implies the 
effective current
\be 
J = \left( \bar{\xi} - \bar\xi \ i\!\overleftarrow{\Slash D}_{\perp}
     \left(i n_+ \! \overleftarrow{D}\right)^{-1}
     \frac{\slash n_+}{2} + \bar{q}\ZWdag
     \right)\Gamma{\cal Q},
\label{currfin}
\ee
where it is understood that derivatives and their inverse do not
act on the overall phase factor $e^{-imvx}$. The field redefinition 
\be 
\label{redef}
\xi\to \xi - (1-\WZdag) 
\frac{\not\!\nm\slash n_+}{4}\,q
\ee 
implicit in using 
the field equation for $\xi$ in (\ref{ll21}) 
has rendered this expression for 
$J$ manifestly
invariant under the gauge transformations (\ref{gaugetrafos}),
contrary to the original expression (\ref{jstart}). Under a  gauge
transformation  the external current transforms as $j\to U j$, which 
is compensated by the transformation of the round bracket in 
(\ref{source-term}). This is similar to the structure of the effective
Lagrangian, where (\ref{collfinal}) is invariant but the starting 
expression (\ref{Ltmp}) is not. In other words, it is the field 
$\xi$ after the (implicit) field redefinition (\ref{redef}) that
transforms according to the gauge transformations 
(\ref{gaugetrafos}) but not the original field $\xi$.

The ultrasoft quark field does in fact not contribute to the
heavy-to-light current at relative order $\lambda^2$. 
Since $\bar{q}$ scales as $\lambda^3$, we have 
\be 
\bar{q}\ZWdag \Gamma{\cal Q} \approx \bar{q}\ZWdag \Gamma \,\WZdag h_v 
= \bar q\,\Gamma \,h_v \simeq 0
\ee
for colour-singlet currents. The final term $\bar q\Gamma h_v$ can be 
set to zero, because we assume that there is large momentum transfer
to the final state, so that the operator must contain 
at least one collinear field in order to 
contribute to such final states. Inserting
the result for ${\cal Q}$ from (\ref{finalQ}) into (\ref{currfin}), 
we obtain for the current including 
corrections of order $\lambda^2$:
\be
 \left[\bar \psi(x) \, \Gamma \, Q(x) \right]_{\rm QCD} 
=
e^{-i m v\cdot x}\, \Big[J^{(0)} + J^{(1)} + J^{(2)}+\ldots  \Big]
\ee
with
\begin{eqnarray}
J^{(0)} &=& \bar{\xi}\, \Gamma\, \WZdag h_v,
\nonumber \\[0.2cm]
J^{(1)} &=& - \bar\xi \, i\!\overleftarrow{\Slash D}_{\perp}
     \left(i n_+ \! \overleftarrow{D}\right)^{-1} \frac{\slash n_+}{2} \,
     \Gamma\, \WZdag h_v 
            - \frac{1}{\nm v} \,\bar\xi\, \Gamma\, \frac{\not\!n_-}{2m} \,
            \Big[\Big[i\!\!\not\!\! D_\perp \WZdag\Big]\Big] \,h_v ,
\label{finalJ}
\\[0.2cm] 
J^{(2)} &=& \bar{\xi}\, \Gamma\, \WZdag 
    \frac{i\!\!\not\!\! D_{\rm us}}{2 m}\, h_v 
    + \frac{1}{\nm v} \,\bar\xi \, i\!\overleftarrow{\Slash D}_{\perp}
     \left(i n_+ \! \overleftarrow{D}\right)^{-1}\frac{\slash n_+}{2} \,
     \Gamma\,  \frac{\not\!n_-}{2m} \,
         \Big[\Big[i\!\!\not\!\! D_\perp \WZdag\Big]\Big] \,h_v
\nonumber \\[0.2cm]
 && \hspace*{-1.3cm}- \,\frac{1}{\nm v} \,\bar{\xi}\, \Gamma\, 
\frac{\slash{n}_-\slash{n}_+}{4m}\,
\Big[\Big[i n_- D \, \WZdag\Big]\Big] \,h_v
- \frac{1}{\nm v} \,\bar{\xi}\, \Gamma\, \frac{1}{i n_+ D}\,
\left[\left[\left\{2 i v D + \frac{i\!\!\not\!\! D_\perp
i\!\!\not\!\! D_\perp}{m} \right\} \WZdag\right]\right] h_v,
\nonumber 
\end{eqnarray}   
where the double-bracket is defined by
\be
\Big[\Big[f(D)\WZdag\Big]\Big]\equiv f(D)\WZdag-\WZdag f(D_{\rm us}).
\label{doublebracket}
\ee
The double-bracket structures generically appear in (\ref{finalJ}),
because they guarantee that the power counting of the various terms 
is preserved when the collinear gluon field is taken to vanish. Each 
double-bracket vanishes separately in this limit, so that terms with 
$1/(i\np D)$ are never enhanced. 


\subsection{Reparameterization constraints}

Contrary to the soft-collinear 
effective Lagrangian, the effective heavy-to-light current 
does receive radiative corrections, which may induce non-trivial
Wilson coefficients multiplying the operators that appear at
tree level, as well as new operators with Wilson coefficients whose 
expansion begins at higher orders in $\alpha_s$. We discuss in the 
following how the various operators are related by reparameterizations
of the velocity vector $v$ and the light-like vectors $n_\pm$, 
which can be used to restrict the number of independent Wilson
coefficients. Such reparameterizations of the soft-collinear effective 
theory have been used in 
\cite{Manohar:2002fd,Chay:2002vy,Chay:2002mw} to constrain the 
renormalization of some of the power-suppressed operators in the 
effective Lagrangian. In Section~\ref{subsec:norenorm} we have seen 
that none of the possible operators in the effective Lagrangian 
can acquire a Wilson coefficient different from its tree-level 
value, so reparameterization invariance does not provide further 
information on the Lagrangian. In the following we show how 
reparameterizations relate the various terms of the current 
(\ref{finalJ}). We do not attempt a complete analysis, which would 
have to treat the set of all possible operators at order
$\lambda$ and $\lambda^2$, not only those that appear at tree-level.

The effective current depends on the arbitrary velocity vector $v$ 
defining the projection $h_v$ of the heavy quark field. As in 
heavy quark effective theory the current must be invariant under 
a velocity transformation $v \to v'=v + \delta v$ with $(v')^2 =1$,
where $\delta v \sim \lambda^2$, since $\delta v$ is related to the 
ambiguity in splitting a heavy quark momentum in its kinematical piece 
$m v$ and a residual momentum of order $\lambda^2$. Invariance of 
the QCD current implies that $J(x)$ or, equivalently, ${\cal Q}(x)$ 
should transform as 
\be
{\cal Q}(x) \to e^{i m\delta v \cdot x}\,{\cal Q}(x). 
\ee
Working out the transformations of the various terms in (\ref{finalQ}), we 
find that velocity reparameterization invariance ties 
\be 
\WZdag \, \frac{i\Slash D_{\rm
      us}}{2m} \, h_v,
\ee
which appears at order $\lambda^2$, to the leading-order term 
$\WZdag \, h_v$ just as in heavy quark effective theory, but no other 
relations are enforced at order $\lambda^2$. Invariance under velocity
reparameterizations implies, however, that the two
terms in the definition of the double-bracket (\ref{doublebracket})  
should not be separated.

Manifest reparameterization invariance 
is achieved by expressing all operators in
terms of the field $Q_v$ defined in (\ref{Qv}) and
the differential operator  
\be
  {\cal V}^\mu \equiv v^\mu + \frac{iD^\mu}{m}
\label{Vdef}
\ee
rather than $h_v$, $v^\mu$ and  $iD^\mu$ \cite{Luke:1992cs}, since 
\be 
Q_v(x) \to e^{i m\delta v \cdot x}\,Q_v(x), \qquad
f({\cal V}^\mu)\,Q_v(x) \to e^{i m\delta v \cdot x}\,f({\cal V}^\mu)\,Q_v(x).
\ee
We can indeed cast the expression (\ref{finalQ}) for ${\cal Q}$ into  
this form, the result being  
\begin{eqnarray}
{\cal Q} &=&
\WZdag Q_v 
-\, \frac{1}{\nm {\cal V}}\,\frac{\!\not\!n_-}{2}
\Big[\Big[\Slash{{\cal V}} \,\WZdag\Big]\Big] Q_v
\nonumber\\[0.2cm]
&&
-\,\frac{1}{{\cal V}^2-1}\,
\Big[\Big[\left\{\nm {\cal V}\,\np {\cal V} + \Slash{{\cal V}}_\perp 
\Slash{{\cal V}}_\perp \right\}\WZdag\Big]\Big] Q_v
+ {\cal O}(\lambda^3 Q_v).
\label{finalQreparam}
\eeq
Here we have used that up to higher orders in $\lambda$ we can
approximate 
$1/(n_- v)\approx 1/(n_- {\cal V})$ and $m /(n_- v \, i n_+ D) \approx
1/({\cal V}^2 - 1)$. This can be written even more concisely by noting
that \be 
\nm {\cal V}\,\np {\cal V} + \Slash{{\cal V}}_\perp 
\Slash{{\cal V}}_\perp  = 
\Slash{{\cal V}} \,\Slash{{\cal V}} - i\np D\,\frac{\Slash{n}_-}{2 m}
\left(\frac{\Slash{n}_+}{2}\,\nm {\cal V} +\Slash{{\cal V}}_\perp 
\right) + {\cal O}(\lambda^3)
\ee
under the double-bracket, so that 
\be 
{\cal Q} = 
\WZdag Q_v-\frac{1}{{\cal V}^2-1}\,
\Big[\Big[\Slash{{\cal V}} \,\Slash{{\cal V}}\,\WZdag\Big]\Big] Q_v
+ {\cal O}(\lambda^3 Q_v).
\label{finalQreparam2}
\ee

By the same line of reasoning 
the light-like vectors $n_+$ and $n_-$ that characterize collinear
particles in the effective theory are not uniquely determined.
The allowed reparameterizations that are consistent with the
power-counting scheme for fields and derivatives, and leave 
$n_+ n_- =2$ and $n_\pm^2=0$ invariant, can be divided into three classes
that correspond to particular Lorentz transformations of the
vectors $n_+$ and $n_-$ \cite{Brodsky:-240pv,Kogut:1970xa,
Manohar:2002fd}:
\be
\begin{array}{lll}
 \mbox{a) longitudinal boosts: } 
&        n_+ \to \alpha \, n_+, \quad &n_- \to \alpha^{-1} \, n_-,  \
\nonumber \\[0.3em]
  \mbox{b) ``minus-transverse'' boosts: } 
&  n_+ \to n_+, \qquad&
    n_- \to n_- + 2 e_\perp - e_\perp^2 \, n_+, 
\\[0.3em]
\mbox{c) ``plus-transverse'' boosts: }
&  n_- \to n_- , \qquad&
    n_+ \to n_+ + 2 f_\perp - f_\perp^2 \, n_- .
\end{array}
\label{tboosts/1}
\ee
The velocity vector is considered as independent of $n_\pm$ and 
remains fixed under all three transformations.
Here $e_\perp$ and $f_\perp$  are arbitrary vectors in the transverse
plane, and power-counting for $n_-D$, $D_\perp$ and $v_\perp$ 
 requires $\alpha \sim 1$, and $e_\perp, f_\perp \sim \lambda$ or smaller. 
Of course, each single operator in the effective theory must still be
manifestly invariant under rotations within the transverse plane
(which leave $n_+$ and $n_-$ unaffected). 

The requirement that a vector $a$ remains invariant under the above 
transformations allows us to deduce the transformation of its 
transverse, plus- and minus-components. Under longitudinal boosts a)
\be
\np a \to \alpha \,\np a, \qquad \nm a\to \alpha^{-1} \,\nm a,
\qquad a_\perp\to a_\perp.
\ee
Under minus-transverse boosts b)
\beq
\np a &\to& \np a,
\nonumber\\
\nm a &\to & \nm a + 2 e_\perp a- e_\perp^2 \,\np a,
\nonumber\\
a_\perp^\mu &\to & a_\perp^\mu -\np a \,e_\perp^\mu - e_\perp a\,
\np^\mu+e_\perp^2 \np a \,\np^\mu, 
\eeq
while under plus-transverse boosts c)
\beq
\np a &\to & \np a + 2 f_\perp a- f_\perp^2 \,\nm a,
\nonumber\\
\nm a &\to& \nm a,
\nonumber\\
a_\perp^\mu &\to & a_\perp^\mu -\nm a \,f_\perp^\mu - f_\perp a\,
\nm^\mu+f_\perp^2 \np a \,\nm^\mu.
\eeq
The transformation of $\xi$ follows from the invariance of the full 
spinor $\psi_c$ and the definition $\xi=(\Slash{n}_-\Slash{n}_+) \,\psi_c/4$, 
which gives
\be
\xi \stackrel{a)}{\to}  \xi,\qquad
\xi \stackrel{b)}{\to}  \xi + \frac{\not\!e_\perp\Slash{n}_+}{2}\,\xi,\qquad
\xi \stackrel{c)}{\to}  \xi + \not \hspace*{-0.11cm} f_\perp\frac{1}{i\np D}\, 
i\Slash{D}_\perp\xi.
\ee
The Wilson line $\WZdag$ is invariant under the transformations a) and
b) and transforms under plus-transverse boosts as 
\be 
  \WZdag \to   \WZdag - \frac{2}{in_+D} \,\Big[\Big[\,if_\perp D_\perp
    \WZdag \Big]\Big] + {\cal O}(\lambda^3).
\label{WZtraf}
\ee

We now discuss the
implications of the different reparametrizations of collinear fields
in turn. 
Longitudinal boosts constrain the allowed operators in the 
effective theory, but do not relate different operators. It is easy
to see that for each operator
in the effective heavy-to-light current (\ref{finalJ}) 
the powers of $\alpha$
induced by the rescaling of $n_+$ and $n_-$ compensate 
each other, so each term is invariant.

In contrast, transverse boosts lead to non-trivial relations 
between different operators in the heavy-to-light current. In
particular, the leading order current $\bar{\xi}\, \Gamma\, \WZdag
h_v$ is not invariant under transverse boosts. It is easy to 
see that the transformation of $\bar\xi$ is compensated by 
that of 
\be
  - \bar \xi \, i\!\overleftarrow{\Slash D}_{\perp}\,
     \left(i n_+ \!\overleftarrow {D}\right)^{-1}
     \,\frac{\slash n_+}{2},
\ee
which suggests that we express the current in terms of 
\be
\bar\Xi\equiv 
\bar \xi - \bar \xi \, i\!\overleftarrow{\Slash D}_{\perp}\,
     \left(i n_+ \!\overleftarrow {D}\right)^{-1}
     \,\frac{\slash n_+}{2},
\ee
which is simply $\bar \xi+\bar \eta$ in the absence of the ultrasoft quark
field. This relates the first term of $J^{(1)}$ in (\ref{finalJ}) to 
the leading-order current and the second term in $J^{(2)}$ to 
the second term 
in $J^{(1)}$. To exploit transverse boost invariance further, we 
write (\ref{finalJ}) as 
\begin{eqnarray}
J  &=& \bar{\Xi}\, \Gamma\, \WZdag Q_v
-\frac{1}{\nm v} \,\bar\Xi\, \Gamma\, \frac{\not\!n_-}{2m} \,
            \Big[\Big[i\!\!\not\!\! D_\perp \WZdag\Big]\Big] \,Q_v
- \,\frac{1}{\nm v} \,\bar{\Xi}\, \Gamma\, 
\frac{\slash{n}_-\slash{n}_+}{4m}\,
\Big[\Big[i n_- D \, \WZdag\Big]\Big] \,Q_v
\nonumber\\
&& \hspace*{-0cm}
- \,\frac{1}{\nm v} \,\bar{\Xi}\, \Gamma\, \frac{1}{i n_+ D}\,
\Big[\Big[2 i v D \, \WZdag\Big]\Big]\,Q_v  
- \,\frac{1}{\nm v} \,\bar{\Xi}\, \Gamma\, \frac{1}{i n_+ D}\,
\left[\left[\frac{i\!\!\not\!\! D_\perp
i\!\!\not\!\! D_\perp}{m}  \WZdag\right]\right] Q_v  +\ldots 
\nonumber\\
&\equiv& \sum_{i=0}^4 J_i,
\label{newJ}
\end{eqnarray}   
where the $J_i$ are defined by the successive 
five terms in the expression for $J$.  

We consider first the plus-transverse boosts c). The leading-order term 
$J_0$ is not invariant due to the transformation (\ref{WZtraf}) of $\WZdag$. 
The non-invariant term is cancelled by the transformation of 
the term $J_3$ containing the derivative $2 i v D$, so the Wilson coefficients 
of these two operators must be related. We note that if we had 
written $2 i v D$ in terms of $i\np D$, $i\nm D$ and $2 i v_\perp D$, 
reparameterization invariance would have tied all these 
terms together into $2 i v D$. Plus-transverse boosts yield no 
further relations up to order $\lambda^2$, since the variations of 
$J_{1,2,4}$ are all of order $\lambda^3$.

The leading term $J_0$ is invariant under minus-transverse boosts b). 
The order-$\lambda$ term $J_1$ is not invariant. Its variation reads 
\be
\delta_{b)} J_1 = 
-\frac{1}{\nm v} \,\bar\Xi\, \Gamma\, \frac{1}{2m} \,
\Big[\Big[\left\{2\!\not\! e_\perp i\!\!\not\!\! D_\perp -
\Slash{n}_-\Slash{n}_+ i e_\perp D\right\}\WZdag\Big]\Big] \,Q_v 
+\ldots. 
\ee
The first term in curly brackets is cancelled by the transformation 
of $J_4$, the second by the transformation of $J_2$. This implies a 
relation among the Wilson coefficients of these terms. We note that 
if we had decomposed $i\!\!\not\!\! D_\perp
i\!\!\not\!\! D_\perp$ into its symmetric and anti-symmetric part, 
both parts would have taken part in the cancellation of 
$\delta_{b)} J_1$.

We conclude from this that for a given
Dirac structure $\Gamma$ the operators in the effective 
heavy-to-light current that appear at tree level
are divided into two sets that correspond to $\{J_0,J_3\}$ and 
$\{J_1,J_2,J_4\}$ in (\ref{newJ}). The Wilson coefficients of the 
operators in each set are related by reparameterization 
invariance. The fact that the operator structure $J_1$ is not related
to $J_0$ under transverse boosts has already been noted in 
\cite{Bauer:2002uv}.

Upon renormalization we may also encounter new
power-suppressed operators, which are not present at tree level. 
These operators will also be related by reparameterization 
symmetries, but the derivation of the complete set of 
relations is beyond the scope of the present paper. 
Note that the terms $J_3$ and $J_4$ involving $2 i v D$ and 
$i\Slash{D}_\perp i\Slash{D}_\perp/m$, respectively, 
are not separately invariant under heavy quark 
velocity reparameterizations, but their sum is. This suggests 
that there may be a further relation that connects the two sets 
of operators defined above, but since the non-invariant term 
in $J_{3,4}$ is of order $\lambda^3$, one would need to analyze the 
set of higher-order operators to draw a definite 
conclusion.\footnote{This relation becomes even more suggestive when the 
current is represented as $\bar\Xi\,\Gamma {\cal Q}$ with 
${\cal Q}$ expressed by (\ref{finalQreparam2}). Then the 
only quantity not manifestly invariant under transverse 
boosts is $\WZdag$, and the variations of the two terms in 
(\ref{finalQreparam2}) compensate each other.}
Having exploited the reparameterization invariances 
we will from here on present our results with the
particular choice $v = (n_+ + n_-)/2$ 
for the heavy quark velocity vector.  


\subsection{Expansion in $\lambda$}

The expansion of the effective heavy-to-light current in terms of operators
with definite (homogeneous) $\lambda$ scaling
is performed in the same manner as for the collinear Lagrangian in 
Section~\ref{lagrangian:lambda}. The derivatives and inverse
derivatives in the current 
that operate on $\bar \xi$ to the left are expanded as
\begin{equation}
\bar{\xi} \,i\!\overleftarrow{\Slash D}_{\perp}
     \left(i n_+ \overleftarrow{D}\right)^{-1}
     \frac{\slash n_+}{2} 
= \bar{\xi} \left\{i\!\overleftarrow{\Slash D}_{\perp c}
     \left(i n_+ \overleftarrow{D_{c}}\right)^{-1} \frac{\slash n_+}{2}
  - g \Slash{A}_{\perp \rm us}
     \left(i n_+ \overleftarrow{D_{c}}\right)^{-1} \frac{\slash n_+}{2}
  + \ldots \right\},
\end{equation}
where $A_{\rm us}$ is now evaluated at $x=x_-$.  
Similarly, the quantity ${\cal Q}$ in (\ref{finalQ}) is expanded 
using (\ref{expand-WZ}) for $\WZdag$, and 
all ultrasoft fields are multipole-expanded according to 
(\ref{taylor}).
Multiplying everything together we obtain the equivalent of 
(\ref{finalJ}), where now every term has a definite scaling 
in $\lambda$. To present the result 
we find it convenient to split the heavy-to-light current in the
effective theory into two terms (denoted $J^{(A)}$ and $J^{(B)}$)
which are separately
invariant under the reparameterizations of $n_\pm$ discussed in the 
previous subsection.  Neglecting 
terms beyond order $\lambda^2$, we obtain
\beq
\label{J:expansion}
 \left[\bar \psi(x) \, \Gamma \, Q(x) \right]_{\rm QCD} 
& \to & e^{-i m v\cdot x}\, \Big\{J^{(A)} + J^{(B)} \Big\}
\cr &= & e^{-i m v\cdot x}\, \Big\{J^{(A0)} + J^{(A1)} + J^{(A2)} + J^{(B1)}
+  J^{(B2)}\Big\}
\eeq
with 
\begin{eqnarray}
J^{(A0)} &=& \bar{\xi}\, \Gamma\, W_{c} \, h_v,
\nonumber \\[0.2em]
J^{(A1)} &=& \bar \xi \, \Gamma \, \Wc \, \left[ x_\perp
  \partial \, h_v \right] 
 - \bar \xi \, i\!\overleftarrow{\Slash D}_{\perp c}
     \left(i n_+ \overleftarrow{D}_c\right)^{-1} \frac{\slash n_+}{2} \,
     \Gamma \, \Wc \, h_v,
\nonumber \\[0.2em]
J^{(A2)} &=& \bar \xi \, \Gamma \, \Wc \left(\frac12 (n_- x)
\left[ (n_+\partial) \, h_v\right] + \frac12 \, x_{\mu \perp} x_{\nu \perp}
\left[\partial^\mu \partial^\nu \, h_v\right]  + 
  \left[\frac{i\!\!\not\!\! D_{\rm
    us}}{2 m} h_v \right] \right)  
\nonumber \\[0.1em]
  && - \, \bar \xi \, \Gamma \, \frac{ 1}{i n_+ D_{c}}
\,\Big[\Big[ i n_-  D \, \Wc\Big]\Big]\, h_v 
+ \bar \xi \, \Gamma  \,\frac{ 1}{i n_+ D_{c}}
\,\Big[ \Wc, \, g n_+ A_{\rm us}\Big] \, h_v
\label{current:lambda}
\\[0.1em]
  && -  \, \bar{\xi} \,  i\!\overleftarrow{\Slash D}_{\perp c}
     \left(i n_+ \overleftarrow{D_{c}}\right)^{-1} \frac{\slash
       n_+}{2}
     \, \Gamma \, \Wc \, \left[ x_{\perp} \partial \, h_v \right]
    +  \bar \xi \,  g \Slash{A}_{\perp \rm us}
     \left(i n_+ \overleftarrow{D_{c}}\right)^{-1} \frac{\slash
       n_+}{2} \, \Gamma \, \Wc \, h_v,
\nonumber  
\eeq
and
\beq
J^{(B1)} &=& 
  - \bar{\xi}\, \Gamma\, \frac{\slash{n}_-}{2m}\, 
  i\Slash{D}_{\perp c} \, W_{c} \,  h_v,
\label{lambda2}
\\[0.2em]
J^{(B2)} &=& 
- \bar \xi \, \Gamma \, \frac{\slash n_-}{2m} \, x_{\mu\perp}
\, i \Slash D_{\perp c} \, \Wc \left[ \partial^\mu \, h_v\right] 
 - \bar \xi \, \Gamma \, \frac{\slash n_-}{2m} \left(
   \Big[g \Slash A_{\perp \rm us}, \Wc\Big]
 +  \Big[\Big[i n_-  D \, \Wc\Big]\Big] \right) h_v
\nonumber \\[0.1em]
&& - \, \bar \xi \, \Gamma \,
 \frac{1}{in_+D_{c}} \, \frac{i\Slash D_{\perp \rm
       c} \,i\Slash D_{\perp \rm c} }{m} \, \Wc \,  h_v 
+ \bar{\xi}\, i\!\overleftarrow{\Slash D}_{\perp c}
     \left(i n_+ \overleftarrow{D_{c}}\right)^{-1}
     \frac{\slash n_+}{2}\, \Gamma\, 
     \frac{ \slash{n}_-}{2m}\, 
                        i\Slash{D}_{\perp c} \, W_{c} \,  h_v,
\nonumber
\end{eqnarray}
where on the right-hand sides the ultrasoft fields $h_v$ and $A_{\rm us}$
are evaluated at $x=x_-$, and derivatives on ultrasoft fields
only act within square brackets. 
The leading-order current $J^{(A0)}$ coincides with the result derived
in \cite{Bauer:2000yr} by summing momentum space Feynman diagrams. 
When we compare our terms at order $\lambda$ with the result obtained 
in \cite{Chay:2002vy}, we find agreement for the term  
$J^{(A1)}$ (except for the first term
coming from the multipole expansion of the heavy quark field),
but find that the term $J^{(B1)}$ has been missed there. 
The second-order corrections in $J^{(A2)}$ and $J^{(B2)}$ have
not been considered before. 

The $\lambda$ expansion of the heavy-to-light current 
as given by (\ref{current:lambda}) and (\ref{lambda2}) corresponds 
to the effective Lagrangian (\ref{fivet}).
Gauge invariance is therefore manifest only for ultrasoft 
gauge transformations restricted to depend only on $x_-$. 
The same steps that led us (for abelian gauge fields)  
from the expanded Lagrangian (\ref{fivet}) to the 
Lagrangian with terms (\ref{1xi})-(\ref{2q}), 
automatically convert the current into an expression where ultrasoft 
gauge fields appear only in covariant derivatives and field-strength 
tensors as in the corresponding Lagrangian. This is derived most 
conveniently by noting that the manipulations of the Lagrangian 
correspond to the  field redefinition of the collinear quark field
$\xi$ given in (\ref{Taylor/xinew}). Inserting this into 
the source term for the heavy-to-light
current in (\ref{source-term}) we find that  
partial derivatives and ultrasoft gauge fields
in the heavy-to-light current given by (\ref{current:lambda}) and 
(\ref{lambda2}) recombine into covariant derivatives.
The currents $J^{(A0)}$ and $J^{(B1)}$ in
(\ref{current:lambda}) and (\ref{lambda2}) already assume their 
final form. For the other terms we 
obtain (for abelian gauge fields only)
\beq
 J^{(A1)} &=& \bar \xi \, \Gamma \, W_{c} \, 
  \left[ x_{\perp} D_{\rm us}
 \, h_v \right]  - \bar \xi 
   i\!\overleftarrow{\Slash D}_{\perp c}
     \left( i n_+ \overleftarrow{D}_{c}\right)^{-1}
     \, \frac{\slash n_+}{2} \, \Gamma\, W_{c} \,h_v,
\nonumber\\
 J^{(A2)} & = &\bar \xi \, \Gamma \, \Wc \left( \frac12 (n_- x)
\left[(n_+ D_{\rm us}) \, h_v\right]+ 
\frac12 \, x_{\mu \perp} x_{\nu \perp}
\left[ D^\mu_{\rm us} D^\nu_{\rm us} \, h_v\right]
 + \left[\frac{i\!\!\not\!\! D_{\rm us}}{2 m} \, h_v \right] \right)
\nonumber \\[0.1em]
  && - \, \bar \xi \, \Gamma \, \frac{ 1}{i n_+ D_{c}} \,
\Big[\Big[ i n_-  D \, \Wc \Big]\Big] \, h_v 
- \bar{\xi} \,  i\!\overleftarrow{\Slash D}_{\perp c}
     \left(i n_+ \overleftarrow{D_{c}}\right)^{-1} \frac{\slash
       n_+}{2}
     \, \Gamma \, \Wc \,[  x_{\perp } D_{\rm us} \, h_v],
\nonumber\\
 J^{(B2)} & = & 
- \bar \xi \, \Gamma \, \frac{\slash n_-}{2m} \, x_{\mu\perp} \,
  i \Slash D_{\perp c} \, \Wc \,
  \left[ D^\mu_{\rm us} \, h_v \right]
 - \bar \xi \, \Gamma \,  \frac{\slash n_-}{2m}\, \Big[\Big[
   i n_-  D \, \Wc \Big]\Big] \,  h_v
\label{finalJab}\\[0.1em]
&& - \, \bar \xi \, \Gamma \,
 \frac{1}{in_+D_{c}} \, \frac{i\Slash D_{\perp \rm
       c}\, i\Slash D_{\perp \rm
       c}}{m} \, \Wc \,  h_v 
+ \bar{\xi}\, i\!\overleftarrow{\Slash D}_{\perp c}
     \left(i n_+ \overleftarrow{D_{c}}\right)^{-1}
     \frac{\slash n_+}{2}\, \Gamma\, 
     \frac{ \slash{n}_-}{2m}\, 
                       i\Slash{D}_{\perp c}\, W_{c} \,  h_v.
\nonumber
\eeq
This form of the current corresponds to the manifestly gauge-invariant
effective Lagrangian given by (\ref{1xi})-(\ref{2q}) for abelian gauge fields. 
The non-abelian case is left for future work.


\section{Form factors for heavy-to-light meson decays}
\label{sec:htol}

We now apply soft-collinear effective theory to the heavy quark expansion 
of transition form factors for 
$B$ decays into light pseudoscalar or vector mesons in the 
kinematic region where the energy of the final state meson
is of order of the heavy quark mass 
$m$. Because of the large momentum transfer to 
light partons, collinear modes are relevant in this kinematic region. 

The transition form factors encode 
the hadronic effects relevant to semi-leptonic
decays,  but also provide important non-perturbative input to
rare and non-leptonic $B$ decays in the QCD factorization
approach \cite{Beneke:1999br,Beneke:2001at,Bosch:2001gv}.
In QCD one defines three independent form factors 
for $B$ decays into pseudoscalars, and seven independent form
factors for decays into vector mesons.\footnote{
We use the same definitions as in  \cite{Beneke:2000wa}. They are 
summarized in Appendix~\ref{sec:ffdef} for convenience.}
At leading order in an expansion in $1/m$, 
the number of independent form factors for $B$ decays into light 
pseudoscalar (vector) mesons
is reduced to one (two) \cite{Beneke:2000wa,Charles:1998dr}. 
This leads to 
form factor relations which reduce the hadronic uncertainties in
the calculation of exclusive $B$ decays.

In \cite{Beneke:2000wa} two of us have shown (to first order
in the strong coupling $\alpha_s$) that the form factors 
factorize at leading order in $1/m$ into a soft-overlap contribution 
multiplied by a short-distance coefficient $C_i$ and an additive 
hard spectator-scattering correction $\Delta f_i$. For instance, the
three form factors $f_i(q^2)$ 
describing the decay into a light pseudoscalar meson can
be written as 
\be 
f_i(q^2) = C_i^{(0)}(E) \, \xi_P^{(0)}(E) +
\frac{\alpha_s C_F}{4\pi} \, \Delta f_i(E),
\label{fac0}
\ee
where the energy $E$ of the final state meson is related to the
momentum transfer $q^2$ by $E=(m_B^2+m_P^2-q^2)/(2 m_B)$ and 
$\xi_P^{(0)}$ is the universal ``soft'' form factor defined in the 
effective theory which parameterizes the soft-overlap contribution 
at leading order in $1/m$.  The tree-level 
coefficients $C_i^{(0)}$ are given by calculable kinematical
factors.  Radiative corrections to $C_i^{(0)}$ and the hard scattering
corrections $\Delta f_i$ have been obtained in
\cite{Beneke:2000wa} in a ``physical'' factorization scheme 
defining the soft-overlap form factors. The radiative corrections to 
$C_i^{(0)}$ have also been computed in soft-collinear effective 
theory \cite{Bauer:2000yr} with $\overline {\rm MS}$ subtractions. 
The implications of the two calculations coincide for QCD form factor ratios 
from which the soft-overlap form factors drop out. The 
effective theory approach provides the appropriate tools to 
prove a factorization formula such as (\ref{fac0}) to all orders 
in $\alpha_s$. A first step has been taken in 
\cite{Bauer:2000yr}, where it was shown that the relations 
among the soft-overlap contributions to the QCD form factors 
hold to all orders in the absence of hard spectator interactions. 
To complete the proof one would have to demonstrate that the 
infrared contributions that appear in the calculation of the 
hard scattering term can be absorbed into the universal soft-overlap 
form factor to all orders in $\alpha_s$.

Combining the soft-overlap and hard scattering terms requires an 
extension of the effective theory that has not been formulated so
far. In the following discussion we focus on the soft-overlap 
contributions beyond leading order in the $1/m$ expansion. These 
are contributions where the spectator quark in the $B$ meson does 
not undergo a large-momentum transfer interaction, so that the 
final state consists of the 
ultrasoft spectator quark together with other small momentum modes,
and a cluster of partons with large momentum 
that evolves out of the $b\to$ light quark transition. 
The invariant mass of this cluster of partons is of order 
$m \Lambda$, where $\Lambda$ is the strong interaction scale, 
since the interaction of an ultrasoft gluon with momentum $\Lambda$ 
and a nearly on-shell quark or gluon with momentum of order 
$m$ puts the large momentum parton off-shell by an amount
$m \Lambda$.\footnote{From this point of view the configurations 
which are considered in the hard scattering approach are special, 
since they consist of {\em only} large momentum partons, in 
which case their invariant mass must be of order $\Lambda^2$, the light 
meson mass squared.} This implies that we should take the 
scaling $\lambda\sim (\Lambda/m)^{1/2}$ for the expansion parameter 
in soft-collinear effective theory. We may then write 
\be
\label{xi12}
f_i(q^2)|_{\,\rm soft \,\,overlap} = C_i^{(0)} \, \xi^{(0)} +
           \sum_j \left( C_{ij}^{(1)} \, \xi_j^{(1)}
          +  C_{ij}^{(2)} \, \xi_j^{(2)} + \ldots \right),
\ee
where the functions $\xi_j^{(1,2)}$ are defined by matrix elements 
of the terms of order $\lambda$ and $\lambda^2$, respectively, 
in the effective current derived in Section~\ref{sec:currents}. The 
purpose of this section is then to see whether form factor 
relations may survive when one includes power corrections. Note that 
we have to go to second order in $\lambda$ to cover $1/m$ corrections 
because $\lambda\sim (\Lambda/m)^{1/2}$.

Before we begin the discussion of power corrections it is instructive 
to see how the multiplicative structure $C_i(E) \, \xi_i(E)$
in the factorization formula (\ref{fac0}) arises in the effective
theory formalism. With 
hard radiative corrections included, the 
leading-order matching equation
\be
  \left[\bar \psi(x) \, \Gamma \, Q(x)\right]_{\rm QCD} =
  e^{-i m \, v\cdot x} \,  \bar \xi(x) \, \Gamma\, \Wc(x) \, h_v(x_-)
\ee
is modified to \be
  \left[\bar \psi(x) \, \Gamma \, Q(x)\right]_{\rm QCD} =
  e^{-i m \, v\cdot x} \,  
  \sum_{i} \int_{-\infty}^0 ds \ 
  \tilde C_\Gamma^i(s,\mu) \, O_i(x+s\np).
\ee
Here we defined a more general set of 
current operators in the effective theory 
\begin{eqnarray}
  O_i(x) &\equiv & \bar \xi(x) \, \Gamma_i 
  \, \Wc(x) \, h_v(x_-) \ ,
\end{eqnarray}
where $\Gamma_i$ are appropriate Dirac matrices, 
which may be induced by radiative corrections,  
and $\tilde C_\Gamma^i(s)$ are
the corresponding hard coefficient functions. The integral over 
$s$ appears, because the component $\np p$ of collinear momenta is  
of the same order as the hard momenta, so that renormalization in 
coordinate space is non-local.\footnote{The relation between the 
renormalized and bare operators in the effective theory is 
also of this non-local form,  
$$
  {\cal O}_i(x,s,\mu) = 
   \int \, ds' \, \tilde Z(s-s',\mu) \, {\cal O}_i^{\rm bare}(x,s').
$$
The non-local contributions to the $Z$-factor come from the
ultraviolet divergences of collinear loops. Ultrasoft loops give 
local contributions proportional to $\delta(s-s')$.} 
Choosing $x=0$ we obtain for the matrix elements that define the 
heavy-to-light form factors
\beq
\langle M(p')|
[\bar\psi \,\Gamma \,Q]_{\rm QCD}|
\bar B(p)  \rangle &=& 
\sum_{i} \int_{-\infty}^0 ds \ 
  \tilde C_\Gamma^i(s,\mu) \, \langle M(p')| e^{i s\np P }\,
[\bar \xi\, \Gamma_i\, \Wc\, h_v]\,e^{-i s\np P }
|\hat{B}(v)  \rangle
\nonumber\\[0.2cm]
&& \hspace*{-4cm} =\,
  \sum_{i} C_\Gamma^i\Big(\np p'-(m_B-m)-i\epsilon,\mu\Big) \, \langle M(p')| 
\,[\bar \xi\, \Gamma_i\, \Wc\, h_v]\,|\hat{B}(v)  \rangle,
\label{ceff}
\eeq
where we introduced the translation operator $e^{i a P}$ and used 
$P^\mu|\hat{B}(v)\rangle = (m_B-m) v^\mu|\hat{B}(v)\rangle$ on the 
heavy meson state in the effective theory. 
We may replace $\np p'-(m_B-m)$ by $\np p'$ in 
(\ref{ceff}), since the difference is a higher-order effect in the $1/m$
expansion. 
The momentum space 
coefficient functions $C_\Gamma^i(n_+ p')$ defined by 
\begin{eqnarray}
 C_\Gamma^i(n_+ p',\mu) &=& 
\int_{-\infty}^{0} ds \, e^{i s n_+ p'} \,  \tilde C_\Gamma^i(s,\mu)
\end{eqnarray}
correspond to the coefficient functions 
computed in \cite{Bauer:2000yr}. 
The matrix elements on the two sides of (\ref{ceff}) 
define the form factors in QCD and in the effective theory, 
respectively, so that (\ref{ceff}) displays the multiplicative 
structure of the matching equation for the soft-overlap form factors.

\subsection{Leading power}

We first recapitulate the derivation of the form factors at leading
order in $\lambda$, where the effective current is given by 
$\bar\xi\,\Gamma \,W_c \,h_v$ according to (\ref{current:lambda}).  
Because of the projection properties of the
light collinear field, $\slash n_- \, \xi(x)=0$, and the heavy quark
field, $(1-\slash v) \, h_v(x)=0$, the most general decomposition of
the matrix element of the leading-order current between a $\bar B$ meson and 
a light pseudoscalar ($P$) or vector meson ($V$) reads
\begin{eqnarray}
\langle P(p') | \xi \, \Gamma \, \Wc \, h_v |\bar B(p)\rangle
   &=& {\rm tr}\left[ A_P^{(0)}
     \, \frac{\slash n_+ \slash n_-}{4}  \, \Gamma
     \, \frac{1+\slash v}{2}  \right],
\nonumber
\\[0.2em]
\langle V(p',\epsilon) | \bar \xi \, \Gamma \, \Wc \, h_v |\bar B(p)\rangle
   &=& {\rm tr}\left[ A_V^{(0)}
     \, \frac{\slash n_+ \slash n_-}{4} \, \Gamma
     \, \frac{1+\slash v}{2} \right],
\label{APV0}
\end{eqnarray}
where the current operator is evaluated at $x=0$. In
the following we consider only pseudoscalar $B^-$ or $\bar B^0$ mesons.
Note that the dynamics of the fields $\xi$ and $h_v$ in the
effective theory, defined by leading order parts of the 
Lagrangians (\ref{coll/lagrangian}) and (\ref{hqet}),
does not change the projection properties of these fields.  Without loss of
generality we choose the four-vectors $v$, $n_+$, $n_-$ to satisfy $v
=(n_+ + n_-)/2$ as before, and in addition $p = m_Bv$ and $p'
= \frac{1}{2}(n_+ p')\, n_- + \frac{1}{2}(n_- p')\, n_+$.  Up to
relative corrections of order $m_{P,V}^2/ E^2 \sim 
\lambda^4$ we have $p' = E n_-$, where $m_{P,V}$ is the mass of the 
light meson.  The universal
functions $A_P^{(0)}(E,v,n_-)$ and $A_V^{(0)}(E,v,n_-,\epsilon)$ can be
decomposed into their most general independent Dirac structures
allowed by Lorentz invariance and parity,
\begin{eqnarray}
   A_P^{(0)}(E,v,n_-) &=& 2  E \, \xi_P^{(0)}(E), 
     \nonumber\\[0.15em]
   A_V^{(0)}(E,v,n_-,\epsilon) &=& - 2 E \,  \slash
   \epsilon_\perp^\ast \, \gamma_5 \, \xi_\perp^{(0)}(E)
     - 2 E \, (v \cdot \epsilon^\ast) \, 
                   \gamma_5 \, \xi_\parallel^{(0)}(E),
\label{APVdef}
\end{eqnarray}
leaving only one universal form factor $\xi_P^{(0)}(E)$ for decays
into light pseudoscalar mesons, and two form factors $
\xi_\perp^{(0)}(E) $, $\xi_\parallel^{(0)}(E) $ for decays into
transversely and longitudinally polarized light vector mesons.  Here
$\epsilon_\perp =\epsilon - \frac{1}{2} (n_+ \epsilon) \, n_- -
\frac{1}{2} (n_- \epsilon) \, n_+$ is the polarization vector for a
transverse vector meson.  (We slightly changed notation compared to
\cite{Beneke:2000wa} and included the dependence on the polarization
vector and factors $\gamma_5$ into the functions $A_P$ and $A_V$.)
The resulting form factor relations can be found in
\cite{Beneke:2000wa,Charles:1998dr}, or can be read off the
leading contributions (involving only $f^{(++)}_{P,\perp,\parallel}$ 
defined below) 
to the formulae given in
Appendix~\ref{sec:ffdef}.

\subsection{Beyond leading power}

Beyond leading order in $\lambda$ one could continue to write down the
most general decomposition of the matrix element of every term in the 
effective current (\ref{finalJ}), as well as the decomposition of 
time-ordered products of the current with higher-order terms in the 
effective Lagrangian. This would give rise to a large number of 
non-perturbative functions $\xi_j^{(1,2)}$ defined in the effective
theory, and an 
expression such as (\ref{xi12}) for the QCD form factors in terms of
these functions. This approach 
would be useful if the form factors in the effective theory could be
computed in a simpler way than the form factors in QCD, for instance 
in lattice QCD. This is unlikely for soft-collinear effective
theory, where the operators are all non-local in light-like
directions. One must then construct combinations of form factors in 
full QCD from which the unknown non-perturbative functions drop out to
a certain order in $\lambda$. 
For the derivation of such form factor relations a simpler approach is
available that avoids the decomposition of every operator and
time-ordered product. 

We start by writing the
matrix element of the heavy-to-light current in full QCD as
\begin{eqnarray}
\label{ff:general}
\langle P(p') | \bar \psi \, \Gamma \, Q |\bar B(p)\rangle &=& 
{\rm tr}\left[ A_P^{(++)}\, \frac{\slash n_+ \slash
       n_-}{4}  \, \Gamma
     \, \frac{1+\slash v}{2}  \right] 
 + {\rm tr}\left[ A_P^{(+-)} \, \frac{\slash n_+ \slash
       n_-}{4}  \, \Gamma
     \, \frac{1-\slash v}{2}  \right]  
\nonumber \\[0.15em]
  && \hspace*{-2cm}+ \,{\rm tr}\left[ A_P^{(-+)} \, \frac{\slash n_- \slash
       n_+}{4}  \, \Gamma
     \, \frac{1+\slash v}{2}  \right]  
 + {\rm tr}\left[ A_P^{(--)} \, \frac{\slash n_- \slash
       n_+}{4}  \, \Gamma
     \, \frac{1-\slash v}{2}  \right] 
\end{eqnarray}
with an analogous definition for decays into vector mesons. Note that
the decomposition of the right-hand side of (\ref{ff:general}) simply 
uses the projectors $(1\pm \slash v)/2$ and $\slash n_\pm \slash n_\mp/4$ on
the large and small components of the quark fields, labelled by $+$
and $-$, respectively, and is completely general.  
The functions $A_{P}^{(kl)}$ and $A_{V}^{(kl)}$ with $k,l=+,-$ 
can again be decomposed as
\begin{eqnarray}
   A_P^{(kl)}(E,v,n_-) &=& 2  E \, f_P^{(kl)}(E), 
 \nonumber\\[0.15em]
   A_V^{(kl)}(E,v,n_-,\epsilon) &=& - 2 E \,  \slash
   \epsilon_\perp^\ast \, \gamma_5 \, f_\perp^{(kl)}(E)
     - 2 E \, (v \cdot \epsilon^\ast) \, 
                   \gamma_5 \, f_\parallel^{(kl)}(E).
\label{APV-full-def}
\end{eqnarray}
Among the $4+8$ form factors $f_{P}^{(kl)}$ and
$f_{\perp,\,\parallel}^{(kl)}$ only $3+7$ are independent. This follows
from the equations of motion for light and heavy quarks in QCD and
translational invariance,
\begin{eqnarray}
 q^\mu \langle P| \bar \psi \, \gamma_\mu \, b |B\rangle 
&=& (m-m_q) \, \langle P| \bar \psi \, b |B\rangle,
\nonumber \\[0.2em]
 q^\mu \langle V| \bar \psi \, \gamma_\mu\gamma_5 \, b |B\rangle 
&=& - (m+m_q) \, \langle V| \bar \psi \, \gamma_5 \, b |B\rangle 
\end{eqnarray}
with $q=p-p'$. 
This implies
\be
 \frac{m_B- m}{m_B} \, f_{P,\,\parallel}^{(++)} 
  + \frac{q^2}{m_B^2} \Big\{
  f_{P,\parallel}^{(-+)}-f_{P,\,\parallel}^{(--)}  \Big\}
  - \Big\{ 2
  f_{P,\,\parallel}^{(+-)}+f_{P,\,\parallel}^{(-+)}+f_{P,\,\parallel}^{(--)}
  \Big\} = {\cal O}(\lambda^3  f_{P,\,\parallel}^{(++)}),
\label{QCDconstr}
\ee
where we have neglected light quark masses $m_q$ 
 and anticipated that $f_{P,\,\parallel}^{(++)}$ is the leading form
 factor in the $\lambda$ expansion.  
Note that $m_B-m\sim \lambda^2$, $m_{P,V}\sim \lambda^2$, and that 
there is no relation for decays into transversely
polarized vector mesons.
The remaining independent $3+7$ form
factors are in one-to-one correspondence to the conventional form
factors parameterizing heavy-to-light transitions for vector, axial
vector, and tensor currents.  Their definitions and the transformation
between the two sets of form factors are given in
Appendix~\ref{sec:ffdef}.

To find the form factor relations we take the effective current 
(\ref{finalJ}) and determine the order in $\lambda$ where one 
obtains a contribution to the form factors
$f_i^{(kl)}$ (with $i=P,\perp,\parallel$) by decomposing $\Gamma$ in 
each term into its four projections analogous to (\ref{ff:general}). 
For instance, 
the leading-order current $J^{(0)}$ only contributes
to $A_M^{(++)}$ (with $M=P,V$) as seen in the previous subsection, 
so 
\begin{equation}
f_i^{(++)}(E) = \xi_i^{(0)}(E) \, 
             \Big( 1 + {\cal O}(\alpha_s,\lambda) \Big)
\end{equation}
for the leading form factor in the
large-energy/heavy quark mass limit. The order $\lambda$ terms 
in the current contribute only to $A_M^{(++)}$, $A_M^{(+-)}$ and
$A_M^{(-+)}$ but not to $A_M^{(--)}$. Finally, the 
$\lambda^2$ suppressed terms in $J^{(2)}$ contribute to all form
factors. This can be summarized in the relations
\begin{equation}
f_i^{(+-)}(E) \sim  \lambda\, f_i^{(++)}(E) , \qquad
f_i^{(-+)}(E) \sim  \lambda\, f_i^{(++)}(E) , \qquad
f_i^{(--)}(E) \sim  \lambda^2 f_i^{(++)}(E).
\end{equation}
In each case these relations specify the largest term 
in the $\lambda$ expansion. This tells us that there are no form factor 
relations when power corrections of order $\lambda^2$ are included. 
However, at order $\lambda$ we can neglect $f_i^{(--)}(E)$, 
which leaves us with 3+6  form factors for pseudoscalar and vector
mesons, respectively. The relations (\ref{QCDconstr}) approximated 
to order $\lambda$ give two constraints, which implies 2+5
independent form factors. Therefore, we have 1+2 form factor relations
at order $\lambda$.

\subsection{Form factor relations}

The three form factor relations that continue to hold including 
the leading power corrections can be found from the relations between
$f_{i}^{kl}(E)$ and the conventional heavy-to-light
form factors in Appendix \ref{sec:ffdef}, neglecting 
$f_{i}^{--}(E)$. The relations
read
\beq 
R_P &=& \frac{f_+-f_0}{f_T} = \frac{q^2}{m_B \, (m_B+m_P)} 
        \left(1+ {\cal O}(\alpha_s,\lambda^2)\right),
\nonumber\\[0.5cm]
R_\perp &=& 
\frac{\displaystyle \left(1-\frac{q^2}{m_B^2}\right)  T_1 - T_2 +
   \frac{q^2}{m_B^2} \left(1+\frac{m_V}{m_B}\right) A_1}
{\displaystyle \left(1-\frac{q^2}{m_B^2}\right)  V} = 
\frac{q^2}{m_B \, (m_B+m_V)}
 \left(1 +
 {\cal O}(\alpha_s,\lambda^2) \right),
\nonumber\\[0.3cm]
R_\parallel &=& 
\frac{\displaystyle \left(1+\frac{m_V}{m_B}\right)  A_1 -
  \left(1-\frac{m_V}{m_B}\right) \left(1-\frac{q^2}{m_B^2}\right) A_2 - 2
  \, \frac{m_V}{m_B}
  \left(1-\frac{q^2}{m_B^2}\right)  A_0}
{\displaystyle T_2 -
  \left(1-\frac{q^2}{m_B^2}\right) 
   T_3}
\nonumber\\
& =& \frac{q^2}{m_B^2} \left(1 +
 {\cal O}(\alpha_s,\lambda^2)\right)
\label{rel3}
\eeq
for form factors that describe the decays into light pseudoscalar, 
transversely and longitudinally polarized mesons, respectively.
Note that the vanishing
of the right-hand sides of (\ref{rel3}) with the
momentum transfer $q^2$ follows from the absence of massless poles in
hadronic matrix elements for $q^2 \to 0$, and is thus exact in QCD.
Other form factor relations valid at order $\lambda^0$ do receive
$\lambda$ suppressed power corrections in the effective
theory, for instance
\begin{equation}
  \left(1-\frac{q^2}{m_B^2}\right)  V 
- \left(1+\frac{m_V}{m_B}\right)^2 A_1
= {\cal O}(\lambda), 
\qquad
\left(1-\frac{q^2}{m_B^2}\right) T_1 - T_2
= \frac{q^2}{m_B^2}\, {\cal O}(\lambda),
\label{helicity}
\end{equation}
where the first relation is proportional to $f_\perp^{(-+)} +
f_\perp^{(--)}$ and the second to $f_\perp^{(-+)} - f_\perp^{(--)}$.
As observed in \cite{Burdman:2000ku}, the form
factor combinations (\ref{helicity}) describe the transition of a
collinear quark with negative helicity into a vector meson with
positive helicity.  We explicitly see that helicity retention from
quark to meson does not hold beyond leading order in the $\lambda$
expansion. In this respect our conclusion differs from the one 
in \cite{Chay:2002vy}, where it is stated that 
the first equation in (\ref{helicity}), which relates the 
vector and one axial-vector form factor for $B \to V$ decays, 
does not receive order $\lambda$ corrections. (The tensor form
factors $f_T$, $T_{1,2,3}$ were not discussed in~\cite{Chay:2002vy}). 

If we include order $\lambda^2$ corrections, non-trivial
form factor relations do not survive.  However, (\ref{rel3})
may still be useful if we restrict our analysis to small
momentum transfer and impose the scaling $q^2 \sim
\lambda^2$.  In this case the undetermined $\lambda^2$
corrections are always multiplied by $q^2$, and can be neglected to
the considered order.  The same holds for the radiative corrections to
(\ref{rel3}). 

To see how the form factor relations are satisfied in a specific
non-perturbative model, we compare the left-hand sides 
of (\ref{rel3}) with the form factors 
computed in the QCD sum rule approach in  
\cite{Ball:1998tj,Ball:1998kk} to the right-hand sides. To make this 
comparison more accurate we also include on the right-hand sides  
the known $\alpha_s$-corrections to the heavy quark limit, 
which can be obtained from Section 3.3 of \cite{Beneke:2000wa}. 
Interestingly, the hard-scattering contribution drops out from all
three ratios $R_{P,\perp,\parallel}$, and the perturbative correction 
is due to the coefficients $C_i^{(0)}$ in (\ref{fac0}) alone. 
The ratio $R_\perp$ receives no radiative
corrections at all at leading power in $\lambda$ 
because of the helicity argument given above.
On the left-hand side of Fig.~\ref{fig:LCQCD} we plot the
{\em difference}\/ between the functions $R_{P,\perp,\parallel}$ 
and their values in the symmetry limit (defined as the limit where 
power corrections and $\alpha_s$ corrections are neglected). 
The dashed curve gives the QCD sum rule result, and the solid curve
gives  
the result in the effective theory at leading order in $\lambda$,  
including radiative 
corrections. The difference between the two curves is therefore a
measure of $\lambda^2$ power corrections, or corrections of 
order $\alpha_s\lambda$. 
Because of the $q^2$ suppression of the right-hand sides of
(\ref{rel3}) we expect only small corrections.
Indeed, in all three cases the deviations from zero are at most 3\% for $q^2$ 
up to 7~GeV$^2$.

On the right-hand side of Fig.~\ref{fig:LCQCD} we plot the
{\em ratio}\/ of the functions $R_{P,\perp,\parallel}$ 
and their values in the symmetry limit minus 1.
In this case the $q^2$ suppression is divided out, and we expect
deviations from 0 of the order of $\lambda^2$ and/or $\alpha_s$.
In the case of $R_P$ and $R_\parallel$ the effect of radiative
corrections is typically 10\%.
If we take the difference between the light-cone sum rule predictions
and \cite{Beneke:2000wa} as a measure for the power-corrections 
in (\ref{rel3}), we deduce effects of order 15-20\% 
which is compatible with $\lambda^2=\Lambda_{\rm QCD}/m$ or 
$\alpha_s\lambda$. Of course, at this point it is unclear whether 
the difference between the two sets of curves is a real 
power correction or an artefact of the 
QCD sum rule calculation.

\begin{figure}
\begin{center}
  \psfig{file=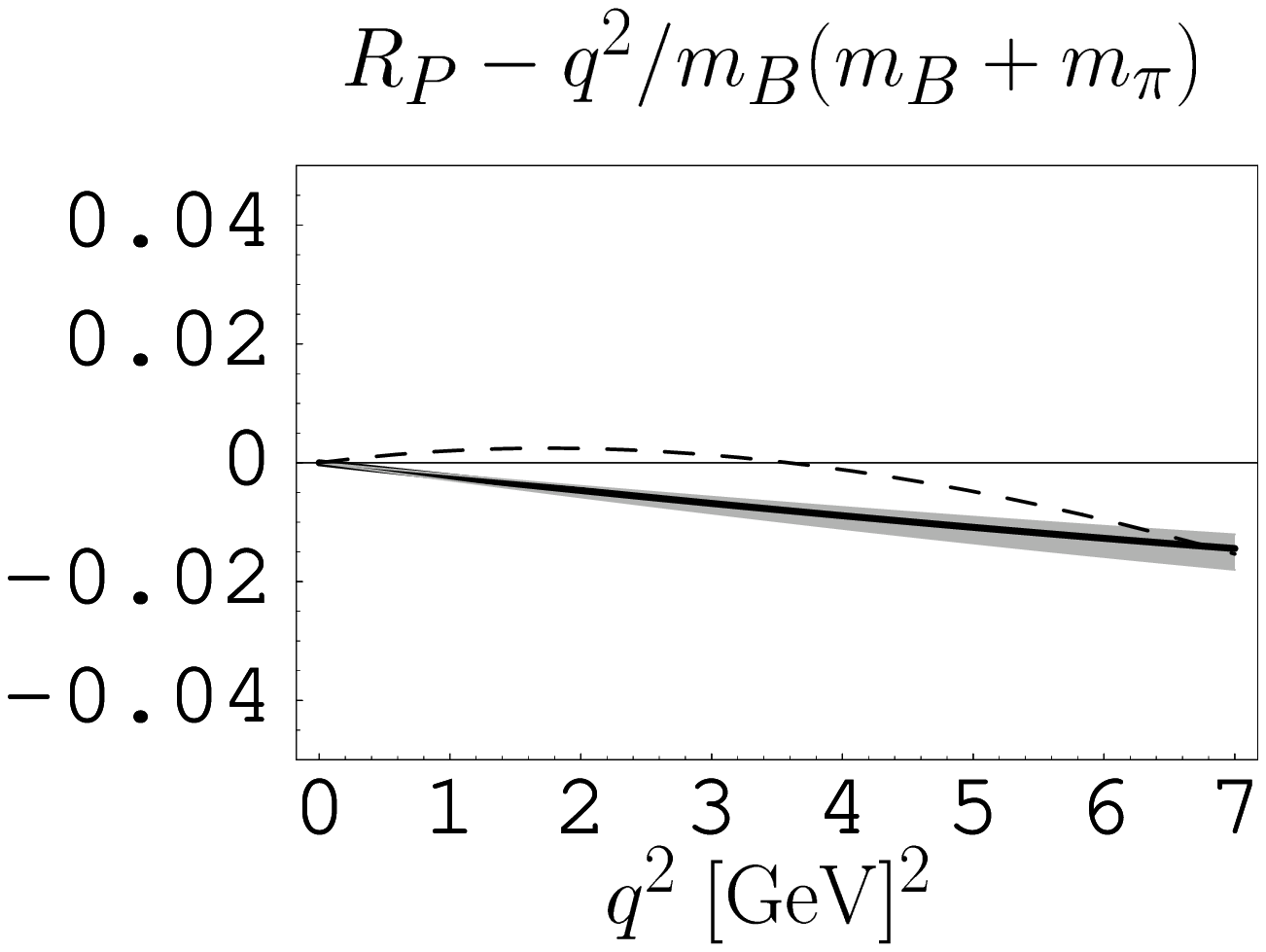, bb=75 380 480 675, width = 0.45 \textwidth}
  \psfig{file=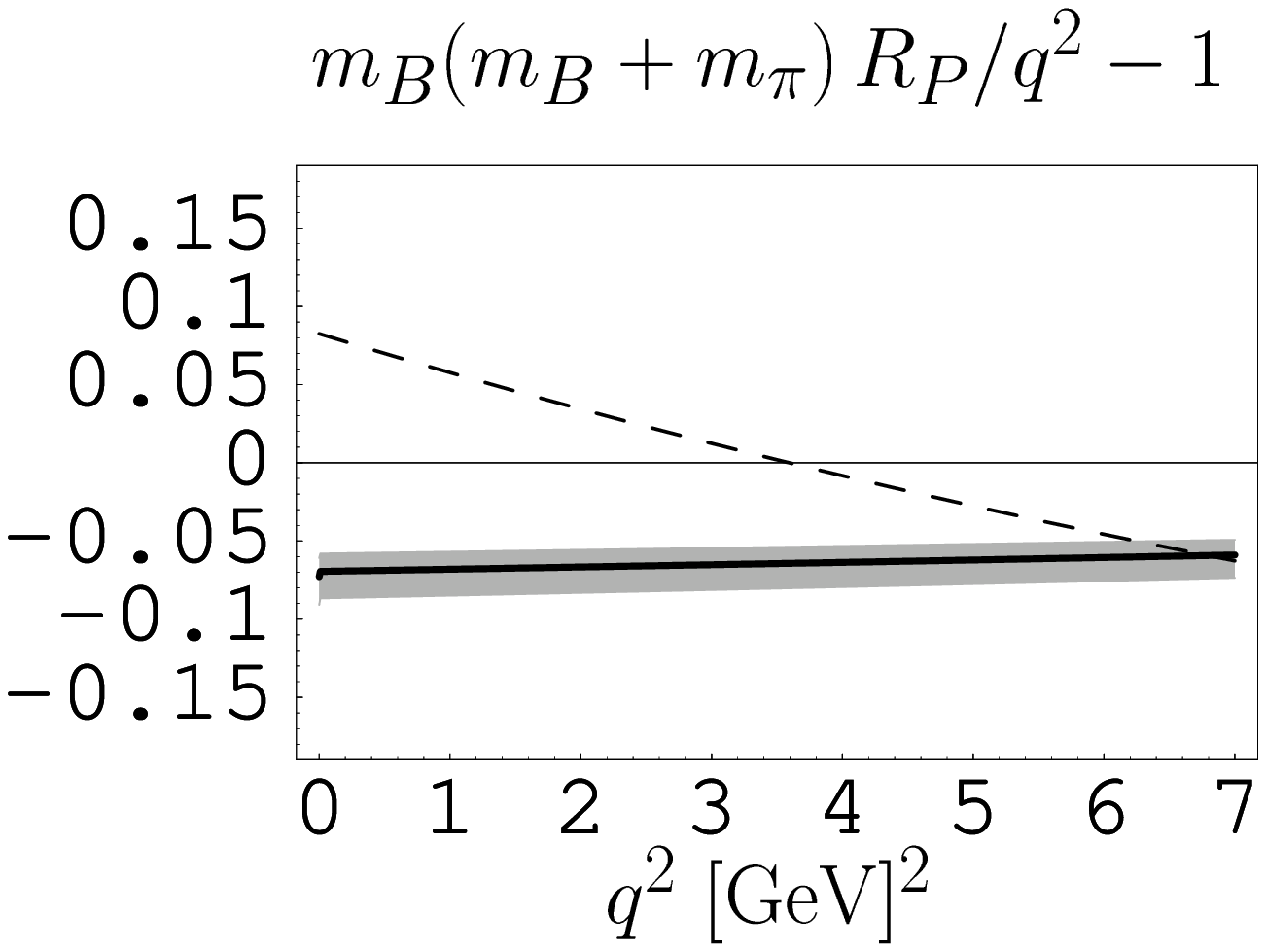, bb=75 380 480 675, width = 0.45
    \textwidth}\\[1.5em] 
  \psfig{file=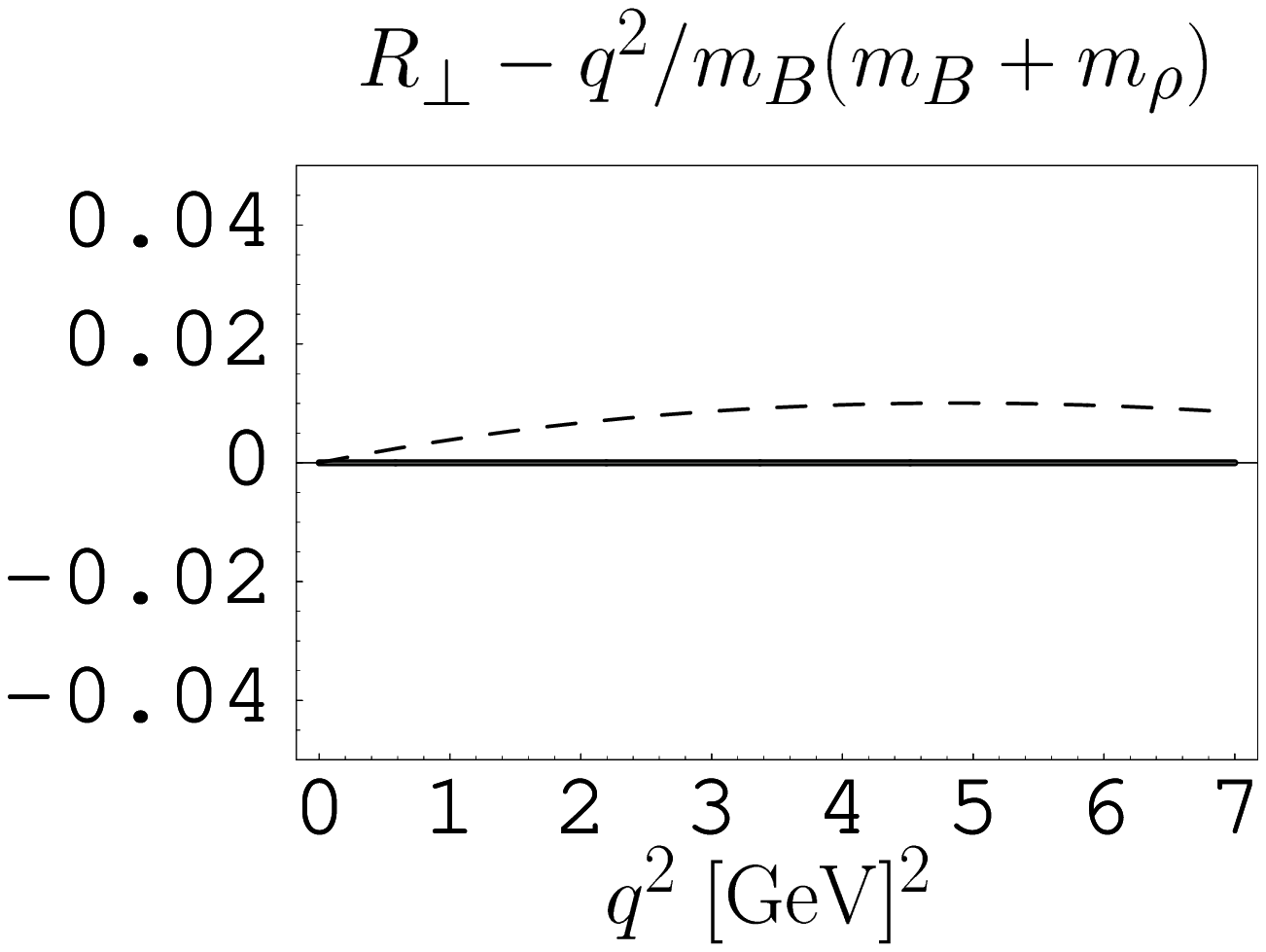, bb=75 380 480 675, width = 0.45 \textwidth}
  \psfig{file=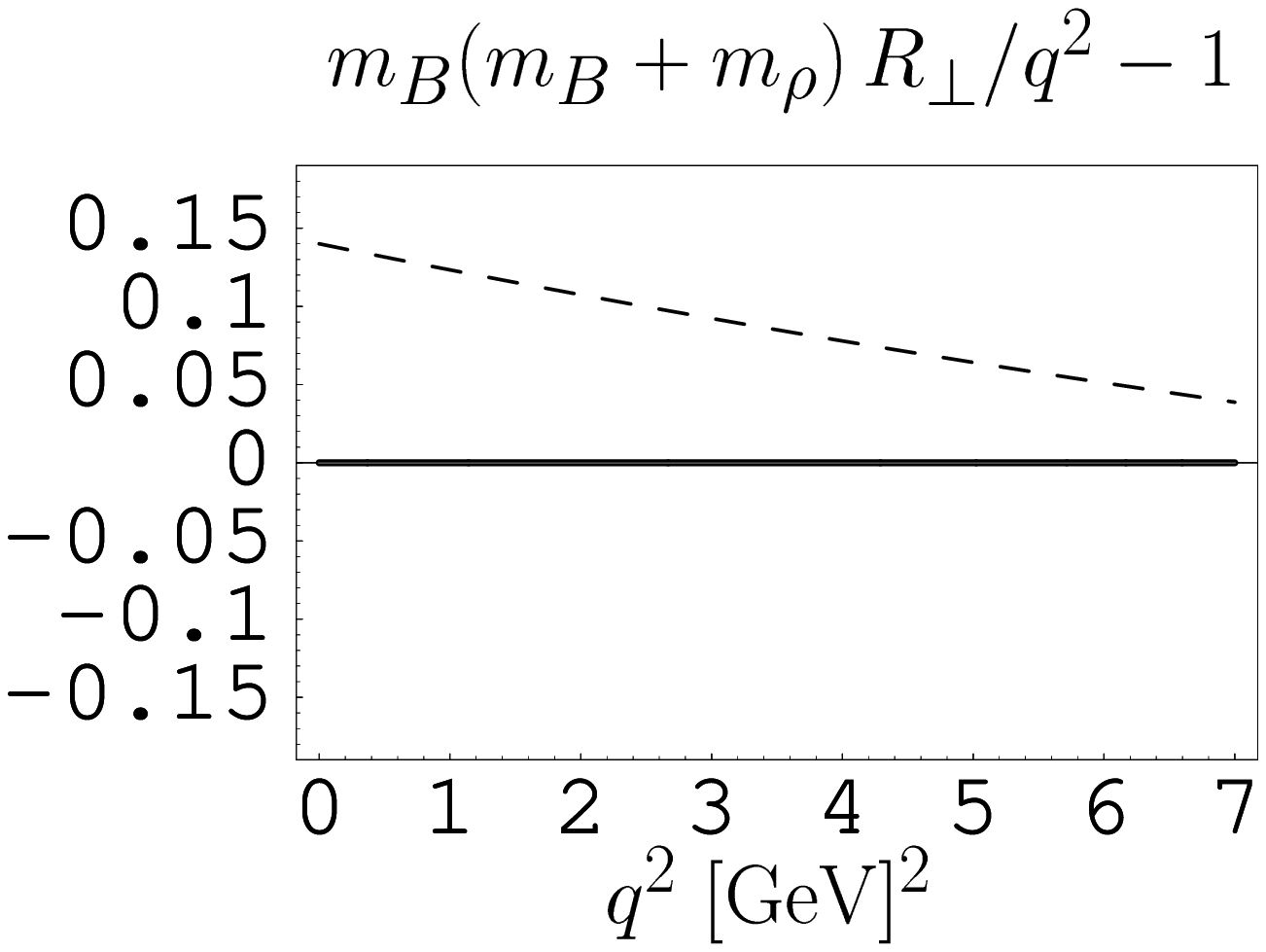, bb=75 380 480 675, width = 0.45
    \textwidth}\\[1.5em] 
  \psfig{file=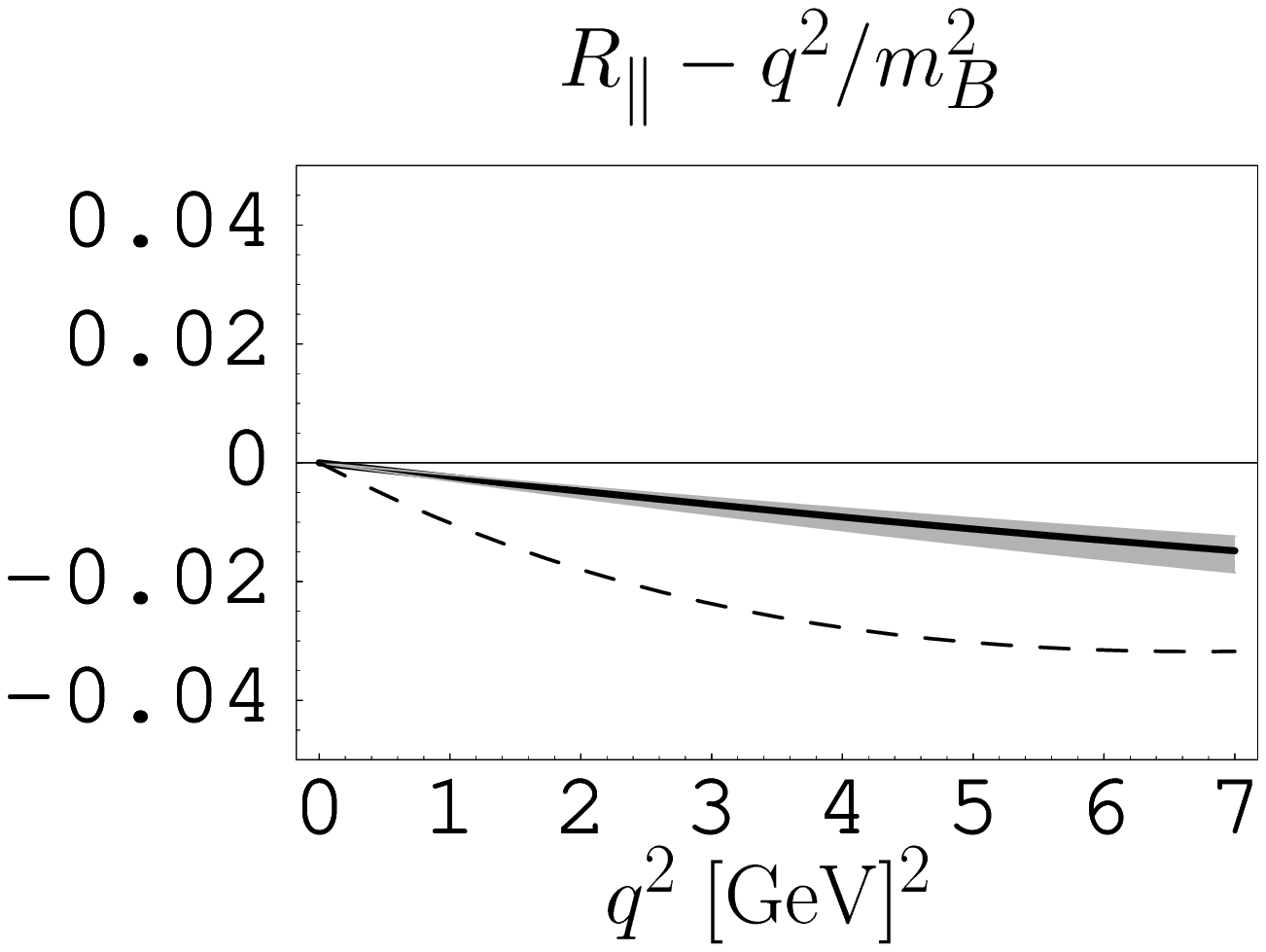, bb=75 380 480 675, width = 0.45 \textwidth}
  \psfig{file=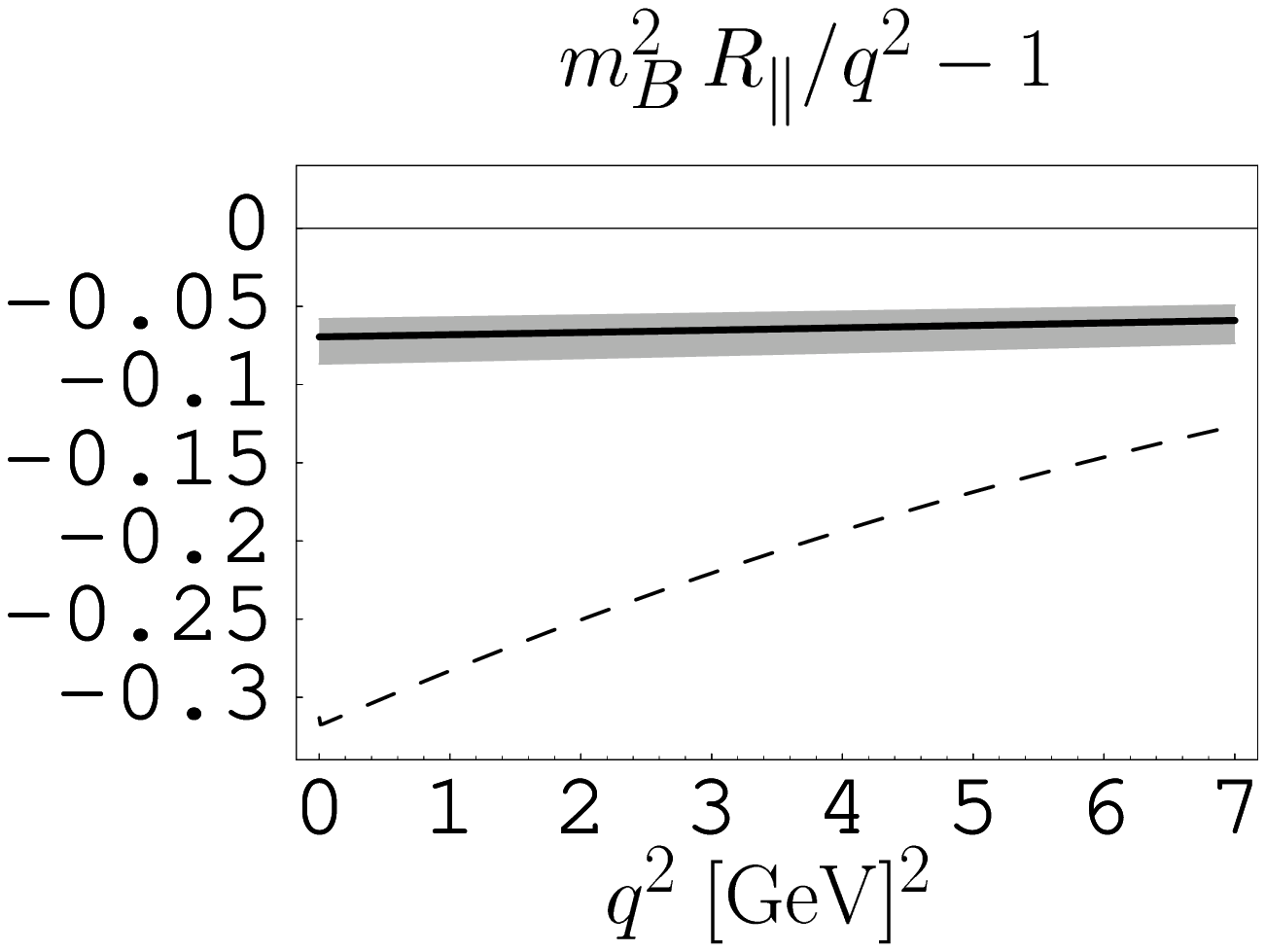, bb=75 380 480 675, width = 0.45 \textwidth}
\end{center}
\caption{\label{fig:LCQCD} Illustration of the relations
(\protect\ref{rel3}) 
for $B\to \pi$ and $B \to \rho$ form factors. On the left-hand
side we plot the ratios $R_{P,\perp,\parallel}$ 
minus  the corresponding values in the symmetry limit. The 
right-hand side shows $R_{P,\perp,\parallel}$ divided 
by their symmetry limits minus 1. 
The dashed line corresponds to the form factors predicted from QCD
light-cone sum rules in \protect\cite{Ball:1998tj,Ball:1998kk}.
The solid line is the order $\lambda^0$ result supplemented by
$\alpha_s$ corrections as calculated in 
\cite{Beneke:2000wa}. The tensor form factors are evaluated at the
scale $m$.  The grey error band reflects the
theoretical uncertainty from varying the scale of $\alpha_s$ from 
$m/2$ to $2 m$.}
\end{figure}

A subtle question concerns the
power-counting of hadronic matrix elements in 
soft-collinear effective theory.  In
contrast to the well-known Isgur-Wise form factor in HQET
\cite{Isgur:1990ed}, which is constrained to scale 
as $\lambda^0$ by its normalization at zero recoil, the 
power counting for the universal form
factors $\xi_{i}^{(0)}(E)$ is not 
determined by the effective theory.  
The scaling behaviour of the quark fields and of the
meson states with $\lambda$ is not sufficient to determine the
corresponding scaling behaviour of form factors. Formally, we have from
the normalization of hadronic states
\begin{eqnarray}
  \langle P(p')| \sim  \lambda^{-2}, \qquad
 |B(m_B v)\rangle \sim \lambda^{-3},
\end{eqnarray}
where we used $p' \approx E n_-$.  Together with $\xi \sim \lambda$
and $h_v \sim \lambda^3$ this results in 
\begin{eqnarray}
  \langle P(p')| \xi \, \Gamma \, \Wc \, h_v|B(m_Bv)\rangle
&\sim & \lambda^{-1} \, X(\lambda).
\label{hadron:scaling}
\end{eqnarray}
Here we introduced the function $X(\lambda)$ which
parameterizes the probability to find the parton configuration 
corresponding to the current operator in the incoming and outgoing
meson states. While the probability to find a nearly on-shell 
$b$-quark in a $B$ meson is of order one, the configuration with
one collinear quark (jet) carrying most of the light meson's energy 
is expected to be suppressed. The suppression factor is usually 
estimated from the end-point behaviour
of the meson's two-particle light cone distribution amplitude 
(see \cite{Beneke:2000wa} and references therein), which gives
\begin{equation}
\left\{\xi_P^{(0)}(E) \, ,\ \xi_\perp^{(0)}(E)\, , \ 
   (v \cdot\epsilon)~\xi_\parallel^{(0)}(E)  \right\} 
\ \propto \ \lambda^{3} \ .
\end{equation}
From this one concludes that $X(\lambda) \sim \lambda^4$. In our 
discussion of power corrections to the transition form factors we
implicitly assumed that the matrix elements of higher-order terms 
in the effective current  have the same $\lambda^4$
suppression as the leading-order matrix element. There might be 
further dynamical suppression of the 
higher-order matrix elements, but an understanding of this 
requires knowledge of the meson wavefunctions that is not provided by
the effective theory. 

In this respect soft-collinear effective theory applied 
to heavy-to-light decays at large recoil is rather different from 
HQET for $B \to D^{(*)}$ form factors. In the latter
case one obtains non-trivial form factors relations beyond
leading order in $\lambda$, because the heavy-quark flavour symmetry
also relates the initial and final {\em hadronic}\/ states. (The same
symmetry is responsible for the absolute normalization of the
Isgur-Wise form factors at zero recoil.) An analogous symmetry
that relates light meson states is absent in soft-collinear effective 
theory.


\section{Conclusions}
\label{sec:conclude}

The effective field theory approach to soft-collinear 
factorization initiated by \cite{Bauer:2000yr,Bauer:2001yt} 
provides a new framework for formulating the factorization theorems 
of QCD for processes with energetic, nearly massless particles. 
Beyond this it opens up the possibility 
to investigate power corrections to such processes
more systematically than in the diagrammatic framework. In this 
paper we have taken a first step in this direction by deriving 
the effective Lagrangian to second order in the expansion parameter 
$\lambda$, and the corresponding power corrections 
to weak interaction quark currents that convert a heavy quark 
into energetic light partons. 

We have formulated the effective theory in a position space 
representation rather than the less familiar hybrid position-momentum space 
representation adopted in \cite{Bauer:2000yr}. Although the physical 
content of the two formulations should be identical, the choice of 
representation implies significant technical differences. In the 
hybrid representation the rapid variations of collinear fields 
are extracted and become field momentum labels, such that a collinear 
field $\xi_p(x)$ has only a residual ultrasoft spatial variation. 
In position space collinear fields $\xi(x)$ are characterized by their
most rapid spatial variation in every direction with no possibility to
extract superimposed slow variations. In this formulation 
the multipole expansion of the argument of ultrasoft fields
accomplishes the equivalent of extracting the ultrasoft variation of 
collinear fields in the hybrid representation. A consequence of this
is that the gauge symmetry of QCD is implemented differently in the 
two formulations of soft-collinear effective theory.

The main results of this paper can be summarized as follows:
\begin{itemize}
\item[--] We have derived the soft-collinear effective Lagrangian to 
order $\lambda^2$ including ultrasoft quark fields. We have also 
found that the tree-level Lagrangian is exact, since Lorentz 
invariance protects every operator (at any order in $\lambda$) 
from acquiring a non-trivial 
coefficient function.
\item[--] We have derived the effective heavy-to-light current 
relevant for exclusive $B$ meson decays to order $\lambda^2$ and 
investigated its properties under reparameterizations of the 
arbitrary vectors that define the momentum decomposition in the 
effective theory. An expression for the current at order $\lambda$ 
has been previously given in \cite{Chay:2002vy}. We differ 
from this result by an additional term that comes from the coupling 
of transverse collinear gluons to the heavy quark.
\end{itemize}
We then investigated whether relations between the transition 
form factors for $B$ meson decay into a light pseudoscalar or vector 
meson survive when one includes power corrections. Including terms of
order $\lambda$ we find one relation among the three form factors for
decays into pseudoscalar mesons and two relations among the 
seven form factors for decays into vector mesons. Including terms of
order $\lambda^2$ no relations survive. This is to be compared to 
the existence of two (five) form factor relations for pseudoscalar
(vector) mesons in the infinite quark mass limit.

\paragraph{Note added in proof.} In the revised version of 
\cite{Bauer:2002uv} (hep-ph/0205289v2) Bauer, Pirjol and Stewart have
also given the tree-level matching coefficient of the operator 
$J_1$, which appears in the expansion of the heavy-light current at
order $\lambda$, see (\ref{newJ}).  Their result agrees with ours. 

\section*{Acknowledgements}

M.B. thanks the Aspen Center of Physics for hospitality when this 
work was begun. The work of A.C. is supported in part by the 
Bundesministerium f\"ur Bildung und Forschung, Project 05 HT1PAB/2.

\begin{appendix}

\section{Derivation of equations (\ref{2xi}) and (\ref{2q})}
\label{app:abelian}

In this appendix we provide the details of the manipulations 
that lead to the final form of the second-order Lagrangian, 
(\ref{2xi}) and (\ref{2q}), for abelian gauge fields. 
We start from (\ref{fivet}) and the explicit expressions 
for the various terms given below (\ref{fivet}). 

To obtain (\ref{1xi}) we use the equation of motion first for 
$\xi$, then for $\bar\xi$. We have to keep the order 
$\lambda$ terms in the equation of motion and account for the fact 
that the Lagrangian changes after the first use of the equation of 
motion. After performing the steps that lead to (\ref{1xi}) 
one has to add to (\ref{Lxi2:prelim}), (\ref{12xiq}) the terms 
\beq
  \Delta {\cal L}_\xi^{(2)} &=& 
  \bar \xi \, (g x_\perp A_{\rm us}) \left( i n_- D + i
    \Slash D_{\perp c}
    \frac{1}{i n_+ D_{c}} i \Slash{D}_{\perp c} \right)
    (g x_\perp A_{\rm us}) \, \frac{\slash n_+}{2} \, \xi  
\nonumber \\
&& + \,\bar{\xi} \left[ i \Slash{D}_{\perp c}
         \frac{1}{i n_+ D_{c}}\, g\Slash{A}_{\perp\rm us} 
  + g\Slash{A}_{\perp\rm us}
         \frac{1}{i n_+ D_{c}}\,  i \Slash{D}_{\perp c},
 \,i g x_\perp A_{\rm us}\right]
\frac{\slash n_+}{2} \, \xi, 
\nonumber \\[0.1em]
   \Delta {\cal L}_{\xi q}^{(2)} &=&  
    - \,\bar{\xi} \, (i g x_\perp A_{\rm us}) \, i\Slash{D}_{\perp c}
    W_{c}\, q
+ \mbox{ ``h.c.''}
\label{new2q}
\eeq
from the equation of motion. 
Formally, using the equations of motion in (\ref{L1:prelim}) 
is equivalent to the change of variables $\xi \to (1+ i g x_\perp
A_{\rm us}) \, \xi$ in ${\cal L}_\xi^{(0)}$ 
and ${\cal L}_\xi^{(1)}$.

The terms in $ \Delta {\cal L}_\xi^{(2)}$ can be simplified by
using the identities (valid only for abelian gauge fields) 
\beq
&&  (x_\perp A_{\rm us}) \, (i n_- D) \, (x_\perp A_{\rm
    us})
= \frac12 \left\{ (in_- D), \, (x_\perp A_{\rm us})^2 \right\}, 
\nonumber\\[0.1em]
&&   (x_\perp A_{\rm us})\,
 i \Slash D_{\perp c}
   \frac{1}{i n_+ D_{c}} i \Slash{D}_{\perp c}\, 
   (x_\perp A_{\rm us})
\cr
&& \qquad = \frac12  \left\{ i \Slash D_{\perp c}
         \frac{1}{i n_+ D_{c}} i \Slash{D}_{\perp c},\, 
    (x_\perp A_{\rm
    us})^2 \right\} + \Slash A_{\perp \rm us} \,  \frac{1}{i n_+ D_{c}}
\, \Slash A_{\perp \rm us} ,
\nonumber\\[0.1em]
&& \left[ i \Slash  D_{\perp c}  \,  \frac{1}{i n_+ D_{c}} \,
   \Slash A_{\rm us} , \, (i x_\perp A_{\rm us})\right]
\cr && \qquad  =
 \left[ \Slash A_{\perp \rm us}  \,  \frac{1}{i n_+ D_{c}} \,i
  \Slash  D_{\perp c} , \, (i x_\perp A_{\rm us})\right]
= - \Slash A_{\perp \rm us} \,  \frac{1}{i n_+ D_{c}}
\, \Slash A_{\perp \rm us}.
\eeq
Using again the leading order equations of motion for $\xi$ and $\bar
\xi$, one obtains  
\be
\Delta {\cal L}_\xi^{(2)} = 
- \, \bar \xi \, g\Slash A_{\perp \rm us} \,  \frac{1}{i n_+ D_{c}}
\, g\Slash A_{\perp \rm us} \, \frac{\slash n_+}{2} \, \xi,
\ee 
which exactly cancels the corresponding term in (\ref{Lxi2:prelim}). 
The remaining terms in ${\cal L}_\xi^{(2)}$ can be simplified by 
commutator relations and the equation of motion. The identity 
\be
\frac{1}{(i\np D_c)^2} = i\left[\frac{1}{i\np D_c},\,\frac{1}{2}\,
\nm x\right]
\ee
allows us to write 
\be 
 \bar \xi \, i\Slash  D_{\perp c}  \,  \frac{1}{i n_+ D_{\rm
      c}} \, (gn_+ A_{\rm us}) \, \frac{1}{i n_+ D_{c}} \,  i\Slash
  D_{\perp c}  \, \frac{\slash n_+}{2} \, \xi 
 =  \frac12 \, \bar \xi \, (n_-x) \left[(n_-\partial) \,
  (gn_+A_{\rm us})\right] \frac{\slash n_+}{2} \xi.
\ee
This term combines with the first term in the third line of 
(\ref{Lxi2:prelim}) into $\np^\mu\nm^\nu F_{\mu\nu}^{\rm us}$. Then we 
use 
\beq
&&\hspace*{-0.5cm}2\left(
\left[(x_\perp \partial)\, \Slash A_{\perp\rm us}\right]
\, \frac{1}{i n_+ D_{c}} \,  i\Slash D_{\perp c} + 
i\Slash D_{\perp c}\, \frac{1}{i n_+ D_{c}} \,
\left[(x_\perp \partial)\, \Slash A_{\perp\rm us}\right]
\right) = 
\\[0.2cm]
&&\hspace*{0.0cm}i\left[x_\perp^\mu x_\perp^\nu 
[\partial_\nu A_{\mu,\rm us}],\,
i\Slash  D_{\perp c}  \, \frac{1}{i n_+ D_{c}} \,  i\Slash
  D_{\perp c}\right]
+ \left(x_\perp^\mu\gamma_\perp^\nu F_{\mu\nu}^{\rm
    us} \, \frac{1}{i n_+ D_{c}} \,  i\Slash D_{\perp c} + 
i\Slash D_{\perp c}\, \frac{1}{i n_+ D_{c}} \,
x_\perp^\mu\gamma_\perp^\nu F_{\mu\nu}^{\rm us}\right)
\nonumber
\eeq
to rewrite the second line of (\ref{Lxi2:prelim}): 
\beq
&& \hspace*{-0.5cm}
\bar \xi \left( 
i \Slash{D}_{\perp c} \frac{1}{i n_+ D_{c}}\, 
\left[(x_\perp \partial)\, g\Slash A_{\perp \rm us}\right] + 
\left[(x_\perp \partial)\, g\Slash A_{\perp \rm us}\right]\,
  \frac{1}{i n_+ D_{c}}\, i \Slash{D}_{\perp c} \right)
\frac{\slash n_+}{2}\xi = 
\nonumber\\[0.2cm]
&&\hspace*{2cm}
\frac{1}{2} \, \bar \xi \left(
    i \Slash D_{\perp c}  \,
 \frac{1}{i n_+ D_{c}} \, x_\perp^\mu \gamma_\perp^\nu \,
 g F_{\mu\nu}^{\rm us}  +   x_\perp^\mu \gamma_\perp^\nu \,
 g F_{\mu\nu}^{\rm us}  \,
 \frac{1}{i n_+ D_{c}} \, i \Slash D_{\perp c}
 \right) \frac{\slash n_+}{2}
\, \xi 
\nonumber\\[0.2cm]
&&\hspace*{2cm}
- \, \frac12 \, \bar \xi \left\{
 x_\perp^\mu \left[(x_\perp \partial) \, (n_-\partial) \,
   A_{\mu,\rm us}\right]\right\} \frac{\slash n_+}{2} \, \xi.
\eeq
With this all remaining terms involving ultrasoft gauge fields 
can again be expressed in terms of field strength tensors and
(\ref{2xi}) follows. The derivation of (\ref{2q}), where 
one has to take account of (\ref{new2q}), is straightforward. The 
application of the $\xi$ equation of motion at several steps in this 
calculation is equivalent to the field redefinition
(\ref{Taylor/xinew}).


\section{Heavy-to-light form factors}

\label{sec:ffdef}

The form factors relevant to heavy-to-light $B$ decays are defined as
in \cite{Beneke:2000wa}
\begin{equation}
\langle P(p')|\bar \psi \, \gamma^\mu b |\bar{B}(p)\rangle =
f_+(q^2)\left[p^\mu+p^{\prime\,\mu}-\frac{m_B^2-m_P^2}{q^2}\,q^\mu\right]
+f_0(q^2)\,\frac{m_B^2-m_P^2}{q^2}\,q^\mu,
\label{fvector}
\end{equation}
\begin{equation}
\langle P(p')|\bar \psi \, \sigma^{\mu\nu} q_\nu b|\bar{B}(p) \rangle =
\frac{i f_T(q^2)}{m_B+m_P}\left[q^2(p^\mu+p^{\prime\,\mu})-
(m_B^2-m_P^2)\,q^\mu\right],
\label{ftensor}
\end{equation}
where $m_B$ is the $B$ meson mass, $m_P$ the mass of the pseudoscalar 
meson and $q=p-p'$. The relevant form factors for $B$ decays into vector 
mesons are defined as 
\begin{equation}
\langle V(p',\varepsilon^\ast)| \bar \psi \gamma^\mu b | \bar{B}(p) \rangle =
 \frac{2iV(q^2)}{m_B+m_V} \,\epsilon^{\mu\nu\rho\sigma}
 \varepsilon^{\ast}_\nu \, p^{\prime}_\rho p_\sigma,
\label{V}
\end{equation}
\beq
\langle V(p',\varepsilon^\ast)| \bar \psi \gamma^\mu\gamma_5 b | \bar{B}(p) 
\rangle &=&
  2m_VA_0(q^2)\,\frac{\varepsilon^\ast\cdot q}{q^2}\,q^\mu + 
  (m_B+m_V)\,A_1(q^2)\left[\varepsilon^{\ast\mu}-
  \frac{\varepsilon^\ast\cdot q}{q^2}\,q^\mu\right]
\nonumber\\[0.0cm]
  &&\hspace*{-2cm}
-\,A_2(q^2)\,\frac{\varepsilon^\ast\cdot q}{m_B+m_V}
 \left[p^\mu+p^{\prime\,\mu} -\frac{m_B^2-m_V^2}{q^2}\,q^\mu\right],
\eeq
\vskip0.2cm
\begin{equation}
\langle V(p',\varepsilon^\ast)| \bar \psi \sigma^{\mu\nu}q_\nu b | \bar{B}(p)
\rangle =
  2\,T_1(q^2)\,\epsilon^{\mu\nu\rho\sigma}\varepsilon^{\ast}_\nu\, 
  p_\rho p^{\prime}_\sigma,
\end{equation}
\begin{eqnarray}
\langle V(p',\varepsilon^\ast)| \bar \psi \sigma^{\mu\nu} \gamma_5 q_\nu b | 
\bar{B}(p) \rangle&=&
(-i)\,T_2(q^2)\left[(m_B^2-m_V^2)\,\varepsilon^{\ast\mu}-(\varepsilon^\ast\cdot
q)\,(p^\mu+p^{\prime\,\mu})\right]
\nonumber\\[0.0cm]
 && \hspace*{-2cm}
+\,(-i)\,T_3(q^2)\,(\varepsilon^\ast\cdot
q)\left[q^\mu-\frac{q^2}{m_B^2-m_V^2}(p^\mu+p^{\prime\,\mu})\right],
\label{ffdef}
\end{eqnarray}
where $m_V$ ($\varepsilon$) is the mass (polarization vector) 
of the vector meson and we use the sign convention $\epsilon^{0123}=-1$.

Calculating the various traces in (\ref{ff:general}) we can express
the form factors (\ref{fvector}) to (\ref{ffdef}) in terms of the sets
$f_P^{(kl)}$ and $f_{\perp}^{(kl)}$, $f_{\parallel}^{(kl)}$ introduced
in (\ref{APV-full-def}).  Neglecting terms suppressed by a relative
factor $m_{P,V}^2 / E^2 \sim \lambda^4$ we obtain
\beq
 f_+(q^2) &=& f_P^{(++)}(E)
   - f_P^{(+-)}(E) 
      - \frac{q^2}{m_B^2}\left(f_P^{(-+)}(E)-f_P^{(--)}(E)\right),
\nonumber \\[0.2em]
  f_0(q^2) &=& \frac{m_B^2-q^2}{m_B^2} \left\{
    f_P^{(++)}(E)
    - f_P^{(+-)}(E) 
      + \frac{q^2}{m_B^2}\left(f_P^{(-+)}(E)-f_P^{(--)}(E)\right)\right\},
\nonumber \\[0.2em]
 f_T(q^2) &=& \frac{m_B+m_P}{m_B} \left\{ f_P^{(++)}(E)
     + f_P^{(+-)}(E) 
      - f_P^{(-+)}(E)-f_P^{(--)}(E)\right\}
\eeq
for decays into light pseudoscalars, and
\beq
  V(q^2) &=& \frac{m_B+m_V}{m_B} \left\{ f_\perp^{(++)}(E) 
    + f_\perp^{(+-)}(E) 
      - f_\perp^{(-+)}(E)-f_\perp^{(--)}(E)\right\},
\nonumber \\[0.2em]
 A_0(q^2) &=&  \frac{m_B^2-q^2}{2 m_B m_V} \left\{ f_\parallel^{(++)}(E)
  - f_\parallel^{(+-)}(E) 
       + \frac{q^2}{m_B^2}
       \left(f_\parallel^{(-+)}(E)-f_\parallel^{(--)}(E)
  \right)\right\},
\nonumber \\[0.2em]
 A_1(q^2) &=& 
   \frac{m_B^2-q^2}{m_B (m_B+m_V)} \left\{ f_\perp^{(++)}(E)
   + f_\perp^{(+-)}(E)+f_\perp^{(-+)}(E) + f_\perp^{(--)}(E) \right\},
\nonumber \\[0.2em]
 A_2(q^2) &=& - \frac{m_B+m_V}{m_B} \left\{
   f_\parallel^{(++)}(E) - f_\parallel^{(+-)}(E) -
     \frac{q^2}{m_B^2}
     \left(f_\parallel^{(-+)}(E)-f_\parallel^{(--)}(E)\right)
   \right\}
\cr && + \frac{m_B+m_V}{m_B} \left\{
       f_\perp^{(++)}(E) + f_\perp^{(+-)}(E)
       +f_\perp^{(-+)}(E)+f_\perp^{(--)}(E)
   \right\},
\nonumber \\[0.2em]
 T_1(q^2) &=& f_\perp^{(++)}(E)
   - f_\perp^{(+-)}(E)- \frac{q^2}{m_B^2} \left(
       f_\perp^{(-+)}(E) -f_\perp^{(--)}(E) \right),
\nonumber \\[0.2em]
 T_2(q^2) &=& \frac{m_B^2-q^2}{m_B^2} \left\{
   f_\perp^{(++)}(E)
   - f_\perp^{(+-)}(E) + \frac{q^2}{m_B^2} \left(
       f_\perp^{(-+)}(E) -f_\perp^{(--)}(E) \right) \right\},
\nonumber \\[0.2em]
 T_3(q^2) &=& - f_\parallel^{(++)}(E) - f_\parallel^{(+-)}(E) +
     f_\parallel^{(-+)}(E)+f_\parallel^{(--)}(E)
\nonumber\\
&& + f_\perp^{(++)}(E) - f_\perp^{(+-)}(E)
       + \frac{q^2}{m_B^2} \left(f_\perp^{(-+)}(E)-f_\perp^{(--)}(E)\right)
\eeq
for decays into light vector mesons.

\end{appendix}


\begin{thebibliography}{10}

\bibitem{Bauer:2000ew}
C.~W.~Bauer, S.~Fleming and M.~E.~Luke,
\newblock Phys.\ Rev. {\bf D63} (2001) 014006 [hep-ph/0005275].

\bibitem{Bauer:2000yr}
C.~W. Bauer, S.~Fleming, D.~Pirjol, and I.~W. Stewart,
\newblock Phys. Rev. {\bf D63} (2001) 114020 [hep-ph/0011336].

\bibitem{Bauer:2001ct}
C.~W. Bauer and I.~W. Stewart,
\newblock Phys. Lett. {\bf B516} (2001) 134 [hep-ph/0107001].

\bibitem{Bauer:2001yt}
C.~W. Bauer, D.~Pirjol, and I.~W. Stewart,
\newblock Phys. Rev. {\bf D65} (2002) 054022 [hep-ph/0109045].

\bibitem{Bauer:2002nz}
C.~W. Bauer, S.~Fleming, D.~Pirjol, I.~Z. Rothstein, and I.~W. Stewart,
\newblock Phys.\ Rev. {\bf D66} (2002) 014017 [hep-ph/0202088].

\bibitem{Collins:1988gx}
J.~C.~Collins, D.~E.~Soper, and G.~Sterman,  in: \textit{Perturbative Quantum
Chromodynamics}, ed.\ A.~H.~Mueller (World Scientific, Singapore,
1989), p.~1. 

\bibitem{Brodsky:-240pv} 
S.~J.~Brodsky and G.~P.~Lepage, in: \textit{Perturbative Quantum
Chromodynamics}, ed.\ A.~H.~Mueller (World Scientific, Singapore,
1989), p.~93. 

\bibitem{Beneke:1998ui}
M.~Beneke,
\newblock Phys. Rept. {\bf 317} (1999) 1 [hep-ph/9807443].

\bibitem{Beneke:2000wa}
M.~Beneke and T.~Feldmann,
\newblock Nucl. Phys. {\bf B592} (2001) 3 [hep-ph/0008255].

\bibitem{Beneke:1998zp}
M.~Beneke and V.~A. Smirnov,
\newblock Nucl. Phys. {\bf B522} (1998) 321 [hep-ph/9711391].

\bibitem{Kogut:1970xa}
J.~B. Kogut and D.~E. Soper,
\newblock Phys. Rev. {\bf D1} (1970) 2901.

\bibitem{Brodsky:1998de}
S.~J. Brodsky, H.-C. Pauli, and S.~S. Pinsky,
\newblock Phys. Rept. {\bf 301} (1998) 299 [hep-ph/9705477].

\bibitem{Srivastava:1999gi}
P.~P. Srivastava and S.~J. Brodsky,
\newblock Phys. Rev. {\bf D61} (2000) 025013 [hep-ph/9906423].

\bibitem{Manohar:2002fd}
A.~V. Manohar, T.~Mehen, D.~Pirjol, and I.~W. Stewart,
\newblock Phys.\ Lett. {\bf B539} (2002) 59 [hep-ph/0204229].

\bibitem{Manohar:1997qy}
A.~V. Manohar,
\newblock Phys. Rev. {\bf D56} (1997) 230 [hep-ph/9701294].

\bibitem{Eichten:1990zv}
E.~Eichten and B.~Hill,
\newblock Phys. Lett. {\bf B234} (1990) 511.

\bibitem{Georgi:1990um}
H.~Georgi,
\newblock Phys. Lett. {\bf B240} (1990) 447.

\bibitem{Grinstein:1990mj}
B.~Grinstein,
\newblock Nucl. Phys. {\bf B339} (1990) 253.

\bibitem{Coleman:1965aa}
S.~Coleman and R.~Norton,
\newblock Nuov. Cim. {\bf 38} (1965) 438.

\bibitem{Chay:2002vy}
J.~Chay and C.~Kim,
\newblock Phys.\ Rev. {\bf D65} (2002) 114016 [hep-ph/0201197].

\bibitem{Chay:2002mw}
J.~Chay and C.~Kim,
\newblock [hep-ph/0205117].

\bibitem{Luke:1992cs}
M.~E. Luke and A.~V. Manohar,
\newblock Phys. Lett. {\bf B286} (1992) 348 [hep-ph/9205228].

\bibitem{Bauer:2002uv}
C.~W.~Bauer, D.~Pirjol and I.~W.~Stewart,
\newblock [hep-ph/0205289].

\bibitem{Beneke:1999br}
M.~Beneke, G.~Buchalla, M.~Neubert, and C.~T. Sachrajda,
\newblock Phys. Rev. Lett. {\bf 83} (1999) 1914 [hep-ph/9905312].

\bibitem{Beneke:2001at}
M.~Beneke, T.~Feldmann, and D.~Seidel,
\newblock Nucl. Phys. {\bf B612} (2001) 25 [hep-ph/0106067].

\bibitem{Bosch:2001gv}
S.~W. Bosch and G.~Buchalla,
\newblock Nucl. Phys. {\bf B621} (2002) 459 [hep-ph/0106081].

\bibitem{Charles:1998dr}
J.~Charles, A.~Le~Yaouanc, L.~Oliver, O.~Pene, and J.~C. Raynal,
\newblock Phys. Rev. {\bf D60} (1999) 014001 [hep-ph/9812358].

\bibitem{Burdman:2000ku}
G.~Burdman and G.~Hiller,
\newblock Phys. Rev. {\bf D63} (2001) 113008 [hep-ph/0011266].

\bibitem{Ball:1998tj}
P.~Ball,
\newblock JHEP {\bf 09} (1998) 005 [hep-ph/9802394].

\bibitem{Ball:1998kk}
P.~Ball and V.~M. Braun,
\newblock Phys. Rev. {\bf D58} (1998) 094016 [hep-ph/9805422].

\bibitem{Isgur:1990ed}
N.~Isgur and M.~B. Wise,
\newblock Phys. Lett. {\bf B237} (1990) 527.

\end{thebibliography}

\end{document}